\documentclass[twocolumn]{aastex701}

\usepackage{amsmath}
\usepackage{bm}
\shorttitle{BINDing the lightcone}
\shortauthors{M. E. Lee et al.}

\begin{document}

\title{BINDing the lightcone: A suite of astrophysical ray-traced weak lensing and SZ maps}

\author[orcid=0000-0002-2318-3087,sname='M.E. Lee']{Max E. Lee}
\affiliation{Department of Astronomy, Columbia University, MC 5246, 538 West 120th Street, New York, NY 10027, USA}
\email[show]{max.e.lee@columbia.edu}  

\author[0000-0002-3185-1540]{Shy Genel}
\affiliation{Center for Computational Astrophysics, Flatiron Institute, 162 Fifth Ave, New York, NY, 10010, USA}
\email{sgenel@flatironinstitute.org}

\author[0000-0003-3633-5403]{Zolt\'an Haiman}
\affiliation{Department of Astronomy, Columbia University, MC 5246, 538 West 120th Street, New York, NY 10027, USA}
\affiliation{Department of Physics, Columbia University, MC 5255, 538 West 120th Street, New York, NY 10027, USA}
\affiliation{Institute of Science and Technology Austria, Am Campus 1, Klosterneuburg 3400 Austria}
\email{Zoltan.Haiman@ista.ac.at}

\author[0000-0003-2630-9228]{Greg L. Bryan}
\affiliation{Department of Astronomy, Columbia University, MC 5246, 538 West 120th Street, New York, NY 10027, USA}
\affiliation{Center for Computational Astrophysics, Flatiron Institute, 162 Fifth Ave, New York, NY, 10010, USA}
\email{gbryan@columbia.edu}

\author[0000-0002-2312-3121]{Boryana Hadzhiyska}
\affiliation{Institute of Astronomy, Madingley Road, Cambridge, CB3 0HA, UK}
\affiliation{Kavli Institute for Cosmology, Cambridge, Madingley Road, Cambridge, CB3 0HA, UK}
\email{bth26@cam.ac.uk}

\begin{abstract}
Recent multiwavelength observations of galaxy group and cluster gas suggest stronger baryonic feedback than our best-calibrated hydrodynamical simulations produce, while modeling this feedback remains a primary source of uncertainty in Stage-IV weak-lensing (WL) analyses. We present a suite of ray-traced maps generated with \textsc{BIND} (Baryonic INpainting with Deep learning), a conditional flow-matching model that paints baryonic mass and gas thermodynamics onto the halos of dark-matter-only simulations, and which was developed in a companion paper. Applied to IllustrisTNG300-Dark and ray-traced, we generate convergence, optical depth, and Compton-$y$ maps at five source redshifts, each with $1000$ pseudo-independent realizations. We build lightcones across a 256-node Sobol sequence spanning the thirty-dimensional IllustrisTNG galaxy formation prior, with individual parameter variations and at the fiducial model. In validation, the maps match those built from the IllustrisTNG300 halos to within LSST-Y10-like precision for a range of WL statistics. Across the prior, the response to feedback exceeds Stage-IV statistical precision by more than an order of magnitude on small scales, and different statistics respond to different model sectors: galactic winds control the WL power spectrum and gas auto- and cross-spectra, while the stellar initial mass function slope and AGN parameters shape the morphological statistics (PDF, peaks, minima, and Minkowski functionals). Finally, we find that the response of statistics can be compressed into seven halo properties which linearly predict a range of WL and SZ statistics. We publicly release the maps, statistics, and model tables.
\end{abstract}



\section{Introduction}
\label{sec:intro}

Modeling baryonic feedback is a leading source of systematic uncertainty in Stage-IV weak lensing analyses \citep[e.g.][]{Amon-2022, Preston-2023, Wright-2025}. Unfortunately, the small scales that drive this uncertainty, where galactic winds and active galactic nuclei (AGN) redistribute gas, are furthermore the same scales that carry a wealth of cosmological information. Mismodeling these processes therefore has dire consequences for cosmological analyses. A simulation with AGN feedback tuned to reproduce the X-ray and optical properties of galaxy groups induces a $\sim40\%$ bias in $w_0$ if that feedback is left unmodeled \citep{Semboloni-2011}, and neglecting baryons biases $\Omega_m$ and $\sigma_8$ by more than $5\sigma$ for a Stage-IV survey \citep{Schneider-2020}. Avoiding small scales by removing them from analyses and relying only on large-scale modes inflates the $\Omega_m$--$\sigma_8$ contour area by more than an order of magnitude \citep{Schneider-2020}. This makes analysis pipeline requirements demanding, requiring the matter power spectrum to be predicted to roughly one percent accuracy out to $k \simeq 10\,h\,\mathrm{Mpc}^{-1}$ \citep{Huterer-2005, Hearin-2012, Chisari-2019}, the same regime where galaxy formation model decisions dominate.

The past decade of multiwavelength observations has begun to narrow down where the real universe sits in this vast galaxy formation model space, although consensus on feedback strength remains elusive. Kinetic Sunyaev-Zel'dovich (kSZ) profiles, on their own and jointly with \textit{eROSITA} X-ray gas masses, suggest that group-scale halos eject gas more efficiently and to greater radii than the weakest-feedback state-of-the-art hydrodynamical simulations, such as IllustrisTNG, predict \citep{Hadzhiyska-2024, Hadzhiyska-2025, Siegel-2025}. For example, the fiducial \textsc{flamingo} run, calibrated to pre-\textit{eROSITA} X-ray gas fractions, was shown to be $>8\sigma$ discrepant with stacked kSZ profiles of groups, while its strongest-feedback variant (fgas$-8\sigma$) agrees to within $\sim2\sigma$; likewise, \textit{eROSITA} X-ray gas fractions in groups are $\sim2$ times lower than the fiducial \textsc{flamingo} predicts \citep{Siegel-2025}. By stacking optically selected groups, recent literature finds that hot gas accounts for a small fraction ($20$--$40\%$) of the cosmic baryon budget within $R_{200}$. Every flagship high-resolution, large-volume hydrodynamical simulation overpredicted this by up to a factor of three, except \textsc{magneticum} and \textsc{simba} \citep{Popesso-2026}. For weak lensing surveys, semi-analytic baryon correction models can translate these measurements into a matter power-spectrum suppression, although the translation rests on an assumed gas profile and halo model. Under those assumptions, the measurements imply a suppression of $10\pm2\%$ at $k=1\,h\,\mathrm{Mpc}^{-1}$, double the $\sim5\%$ that most current lensing pipelines assume \citep{Siegel-2025b}. 

For thermal measurements of gas through the thermal Sunyaev-Zel'dovich (tSZ) effect, joint analyses of DES Year~3 weak lensing with ACT tSZ detect the shear$\times$tSZ cross-correlation at $21\sigma$ and find $2$--$4\sigma$ tension with mild-feedback simulations \citep{Pandey-2025}. The implication is that group-sized halos ($10^{13}\,M_{\odot}\,h^{-1} \leq M < 10^{14}\,M_{\odot}\,h^{-1}$) require significantly reduced thermal gas pressure \citep{Pandey-2022, Pandey-2025}. The newest and highest signal-to-noise kSZ measurement, DESI DR2 crossed with ACT DR6, detects the effect at $18\sigma$ around luminous red galaxies \citep{Qu-2026}, with companion measurements around bright and emission-line galaxies \citep{Hadzhiyska-2026a}, and favors gas ejection from group-scale halos more efficiently than the standard AGN gas profile \citep{Battaglia-2016} predicts. These results are consistent with earlier weak lensing plus kSZ constraints, which preferred gas fractions systematically below simulation predictions \citep{Hadzhiyska-2024, Hadzhiyska-2025, Bigwood-2024}, and with the kSZ benchmark established against DESI+ACT data, in which the fiducial feedback calibrations of \textsc{flamingo}, \textsc{antilles}, \textsc{bahamas}, and \textsc{fable} appear disfavored at $>3\sigma$, while stronger-AGN variants of \textsc{flamingo} and \textsc{bahamas}, as well as \textsc{simba}, reproduce the measured signal \citep{Bigwood-2025,McCarthy-2024}. These analyses point to stronger baryonic feedback than the weakest-feedback simulations predict, although how strong it is and whether every probe agrees on a single calibration remain open questions.

Beyond the weak lensing angular power spectrum analyses, Stage-IV surveys are designed to exploit non-Gaussian information through statistics such as peak counts, minimum counts, Minkowski functionals, and the one-point probability distribution function, each of which outperforms two-point statistics in forecast precision \citep{EuclidHOS-2023, Zuercher-2021, Martinet-2021}, and has now been demonstrated across a wide range of surveys \citep{Liu-2015, Kacprzak-2016, Martinet-2018, Shan-2018, HarnoisDeraps-2021, Gatti-2022, Lu-2022, Lu-2023}. Peak counts combined with the angular power spectrum tighten $S_8$ by $38\%$ over the power spectrum alone in DES Year~3 \citep{Zuercher-2022}, and peaks and minima combined with the power spectrum yield a $35\%$ improvement in HSC Year~1 \citep{Lu-2023,Marques-2024}. Each of these statistics is affected by baryonic physics \citep{Osato-2021,Coulton-2020,Lu-2022,Broxterman-2024,Grandon-2024,Marinichenko-2025}, however, these impacts vary across the statistic space. Ignoring baryons biases peak-count inference by ${\approx}4\sigma$ while leaving minimum counts nearly unaffected at ${\approx}0.5\sigma$ \citep{Coulton-2020}. \citet{Lee-2026a} showed that the effect of baryons on morphological statistics, such as peak counts, is sourced primarily from the cores of massive halos, while the power spectrum is sensitive to the baryonic imprint from a broad range of halo masses and radii, a separation also visible in \cite{vanDaalen-2020, Yang-2011, Liu-Haiman-2016, Broxterman-2024, Marinichenko-2025}. A correction calibrated to reproduce the power spectrum is therefore not guaranteed to be correct for other statistics, and in practice, it is not. \citet{Lee-2023} found that baryon correction models (BCMs) tuned on the power spectrum reproduce the baryonic effects in peak counts at the percent level only for peaks with ${\rm S/N} < 4$. For larger peak amplitudes, the effects of baryons are improperly modeled by BCMs, which implies they will become insufficient for use in LSST and \textit{Euclid} WL analyses.

Baryonic modeling approaches fall into three broad categories. Full hydrodynamical simulations (e.g., \textsc{flamingo}, \textsc{bahamas}, IllustrisTNG, \textsc{simba}) resolve the complex interplay and physics across scales, but are computationally prohibitive for the large volumes and wide parameter-space coverage that inference pipelines require \citep{Schaye-2023,VillaescusaNavarro-2021}. Semi-analytic baryon correction models are fast and flexible, providing $\sim2\%$ power-spectrum accuracy \citep{Schneider-2015, Schneider-2019, Arico-2021, Giri-2021, Schneider-2025}, and have recently been extended to operate directly on maps, where they have been calibrated against higher-order statistics rather than only against $P(k)$. For example, \citet{Anbajagane-2024} jointly fit the second, third, and fourth moments of the lensing and tSZ fields to within measurement uncertainties above ${\sim}1\,$Mpc, and \citet{Zhou-2025} fit two-point statistics, wavelet phase harmonics, scattering coefficients, and the third and fourth moments to within $2\%$ for $\ell < 2000$ using only three parameters. However, these semi-analytic models impose spherically symmetric parametric profiles and therefore cannot represent aspherical morphology or stochasticity in gas properties for a given halo, to which non-Gaussian statistics are sensitive. Deep-learning approaches occupy a middle ground, painting baryonic fields directly onto $N$-body simulations \citep{Troester-2019, Thiele-2020b, Dai-2021, Chadayammuri-2023, Sharma-2024, Liu-2025}, but existing models are trained on a single galaxy-formation calibration rather than conditioned on the subgrid parameter space, or are trained to reproduce the full volume, sacrificing resolution on small halo scales where the baryonic effects are sourced. Outside of these three categories, a model-free calibration of the suppression has recently been proposed, which cross-correlates shear with direct tracers of the ionized gas: fast radio burst dispersion measures \citep{Leung-2025} or the kSZ effect \citep{Hadzhiyska-2026b, Ganguly-2026}. While these do measure the two-point suppression without a galaxy formation model, they say nothing about non-Gaussian statistics and how they may respond. Further, none of the models above allow analysis of the connection between physics model parameters, their induced effects on halo populations, and the propagation of these effects into various weak lensing and CMB statistics.

This is the gap we aim to address in this and our companion work \citep{Lee-2026b} by introducing \textsc{BIND} (Baryonic INpainting with Deep learning). \textsc{BIND} is a conditional generative model trained via flow matching to paint baryonic fields onto the halos of dark-matter-only $N$-body simulations, conditioned on the full set of $\Lambda$CDM cosmological and IllustrisTNG galaxy formation parameters \citep{Lee-2026b}. \textsc{BIND} is halo-centric, allowing for high-resolution emulations of individual halos, generative so that it can sample from the distribution of effects for a given halo, and it generates simultaneous channels so the same structures, at the same positions, appear in gas, stellar, and dark matter mass distributions as well as in the resulting gas temperatures, pressures, and entropies. The companion paper \citep{Lee-2026b} introduces the model and validates it at the halo level, while here, we apply BIND to the lightcone.

In this paper, we generate ray-traced weak lensing convergence ($\kappa$), electron scattering ($\tau$), and Compton-$y$ ($y$) maps across the IllustrisTNG galaxy formation parameter space using \textsc{BIND}, and use these maps to measure how the astrophysical parameters propagate into weak lensing and SZ summary statistics. We then use these measurements to build an analytic model of weak lensing statistics as a function of measurable halo quantities. Applied to lightcones built from TNG300-Dark, \textsc{BIND} produces $(\kappa, y, \tau)$ map triplets at a fraction of the cost of a full hydrodynamical simulation, allowing us to cover the parameter space instead of sampling it at a handful of points. We generate one set of map triplets by varying the thirty astrophysical parameters one at a time while keeping all others fixed to their fiducial values, and a second set by varying all thirty parameters simultaneously across the full thirty-dimensional prior volume using a 256-node Sobol sequence. At each parameter-space location, we generate maps for five source redshifts and produce $1000$ pseudo-independent realizations. To our knowledge, this is the first systematic study of how a full galaxy formation parameter space propagates into a suite of Stage-IV weak lensing summary statistics and into the gas auto- and cross-spectra on a cosmological lightcone.

We demonstrate three principal results. 
\begin{enumerate}
    \item First, at the fiducial IllustrisTNG parameters, the generated \textsc{BIND} lightcones reproduce every WL statistic from an analogously constructed lightcone with hydrodynamical halos to within the statistical precision of an LSST-Y10-like survey. 
    \item Second, feedback responses vary by statistic and feedback type. The projected 2D WL statistics and every gas spectrum respond primarily to the galactic-wind sector feedback parameters, while the WL morphological statistics are more sensitive to the slope of the stellar initial mass function and AGN feedback parameters. 
    \item Third, the statistical response to the astrophysical parameters is compressible. A linear model can be generated from the shapes of parameter effects on statistics and a handful of measured halo properties. This implies that the imprint of feedback on the maps and statistics is carried by a few physical, observable halo properties, and provides a bridge between the multiwavelength gas measurements discussed above and Stage-IV cosmological inference \citep{Siegel-2025,Bigwood-2025,vanDaalen-2020,Schneider-2022}.
\end{enumerate}

This paper is structured as follows. Section~\ref{sec:methods} reviews \textsc{BIND}, the IllustrisTNG-300 simulation, halo conditioning and pasting procedure, ray tracing, and introduces the various lightcone sets to be generated. Section~\ref{sec:validation} validates the fiducial generation against the hydro-pasted truth, first at the halo level (scaling relations and radial profiles) and then at the map level (all WL statistics and the gas auto- and cross-spectra). \S~\ref{sec:astro} measures the response of twelve statistics across the 30-dimensional astrophysical parameter space, identifies the dominant parameters (\S~\ref{sec:correlations}), and categorizes their response into five distinct families (\S~\ref{sec:families}). Section~\ref{sec:emulator} builds an analytic linear model from the results of \S~\ref{sec:astro}. We develop a physical interpretation for our results in \S~\ref{sec:story}. Sections~\ref{sec:caveats} and \ref{sec:outlook} catalog the limits of the maps' construction and their remedies, and Section~\ref{sec:conclusions} summarizes our main conclusions. We publicly release the lightcones, measured statistics, and model tables.

\section{Methods}\label{sec:methods}
In this section, we review the \textsc{BIND} methodology, the simulation suites used, and the procedure for generating halos, lightcones, and our ray-traced maps.
 
\subsection{BIND}\label{sec:BIND}
\textsc{BIND} is a conditional generative model for creating hydrodynamical fields in and around dark-matter-only halos. More specifically, \textsc{BIND} is a trained \textit{conditional flow model} \citep{Lipman-2022, Lipman-2024, kannan-2025}, which learns to sample from the posterior
\begin{equation}\label{eq:BIND_posterior}
P(\bm{F}^H_{\rm hydro} \,|\, \rho^H_{\rm DMO}, \bm{\theta}, z),
\end{equation}
where $\bm{F}^H_{\rm hydro}$ is a set of hydrodynamical fields such as gas, stellar, and dark matter densities, $\rho^H_{\rm DMO}$ is a dark matter density field from an $N$-body simulation centered on a halo $H$, $\bm{\theta}$ is a set of cosmological and astrophysical parameters, and $z$ is the redshift.
 
\textsc{BIND} was trained on halos with $M_{200c}\geq10^{13}\,{\rm M}_\odot\,h^{-1}$ inside the $50\,{\rm Mpc}\,h^{-1}$ CAMELS suite of 1024 dark-matter-only and hydrodynamical simulation pairs \citep{Genel-2026}\footnote{Where $M_{200c}$ is the mass enclosed within a radius inside which the mean density is $200\rho_c$, with $\rho_c$ the critical density of the universe.}. Each simulation in the suite used random initial phase fluctuations, a set of five cosmological parameters, and thirty galaxy formation model subgrid parameters that control astrophysical effects such as star formation, stellar feedback, and AGN feedback. For a detailed list of the parameters and their descriptions, see \citet{Genel-2026}.
 
In \citet{Lee-2026b}, \textsc{BIND} was trained to generate hydrodynamical density fields such as gas, stellar, and dark matter masses on $128\times128$ pixel grids centered on DMO halos with pixel scales of $48\,{\rm kpc}\,h^{-1}$ such that the full patch is $6.25\times6.25\,{\rm Mpc}\,h^{-1}$. In this work, we retrain the same network with the same map resolution and scale, but also include thermodynamic channels for gas temperature, $T$, pressure, $P$, and entropy $K$, such that the total list of fields generated by \textsc{BIND} is $\bm{F}^H_{\rm hydro} = \{\rho_{\rm DM}, \rho_{\rm gas}, \rho_{\rm star}, T_{\rm gas}, P_{\rm gas}, K_{\rm gas}\}$. It is important to note that \textsc{BIND} is a 2D model, and as such, it generates a projected field of $6.25\times6.25\,{\rm Mpc}\,h^{-1}$, where each pixel contains the surface density projected over a $50\,{\rm Mpc}\,h^{-1}$ depth. This decision naturally followed from the CAMELS boxes being $(50\,{\rm Mpc}\,h^{-1})^3$ volumes. However, we will see in \S~\ref{sec:conditioning} that the lightcone generation requires simulation projections, which actually turns \textsc{BIND}'s projection mechanisms into an advantage.
 
For a detailed explanation of the \textsc{BIND} architecture, training, and validation, see \citet{Lee-2026b}. We verified that the retraining described above produces analogous results in the mass channels and will validate the thermodynamic channels in \S~\ref{sec:validation} as well.
 
\subsection{TNG simulations and conditioning generation}\label{sec:conditioning}
The goal of this work is to use \textsc{BIND} to generate weak lensing, tSZ, and $\tau$ maps with various astrophysical parameter changes, which requires a dark-matter-only simulation with halos for BIND to paste, and which is large enough to produce these maps. IllustrisTNG \citep{Weinberger-2017, Pillepich-2018, Springel-2018} is the galaxy formation model that \textsc{BIND} was trained on, and so we use the largest IllustrisTNG dark-matter-only simulation, IllustrisTNG-300 Dark (henceforth TNG300-Dark) \citep{Nelson-2019}. We also use the paired hydrodynamical simulation, IllustrisTNG300 (henceforth TNG300), to validate our procedure. Both TNG300 and TNG300-Dark cover a simulated volume $(205\,\mathrm{Mpc}\,h^{-1})^3$ with dark matter resolutions of $4.0\times 10^7\,h^{-1}\,{\rm M}_\odot$ and, $4.73\times 10^7\,h^{-1}\,{\rm M}_\odot$ respectively, and both are generated with the cosmological parameters inferred from \citet{planck-2015} ($\Omega_m = 0.3089$, $\sigma_8 = 0.8158$, $\Omega_b = 0.0486$, and $H_0 = 67.74\,{\rm km}\,{\rm s}^{-1}\,{\rm Mpc}^{-1}$). The TNG suite contains 100 simulation snapshots output between redshifts $z = 0$ and $ z = 20$, each containing halo and subhalo catalogs computed on the fly and saved using the FoF/Subfind algorithms.

This work leverages the TNG300-Dark simulation suite for pasting, but in \S~\ref{sec:validation}, we test our procedure by comparing maps generated at the fiducial TNG300 parameter space location with the true TNG300 hydrodynamical simulation. Another advantage of using the IllustrisTNG suite is that several previous works have built and validated ray-tracing pipelines \citep{Osato-2021, Lee-2023, Lee-2026, Lee-2026a}. We treat TNG300 as the ground-truth simulation; however, TNG300 is only a single hydrodynamical realization of a field generated from the fiducial set of 30 astrophysical parameters. In this work with BIND, we can arbitrarily modify any of these parameters to generate different hydrodynamical halo realizations, which we then paste back into the dark-matter-only maps and feed through the ray-tracing pipeline.

To \textsc{BIND}\footnote{We use \textsc{BIND} as a verb frequently throughout the text, as we believe it is useful and clear.} a simulation, one only requires a projection of a dark-matter-only simulation, a catalog of halo positions within the projection, the cosmological parameters of the simulation, the desired astrophysical subgrid parameters, and a random seed from which the flow matching starts its sampling. As TNG300-Dark contains the snapshots between $z=0$--$20$, and is a volume of $(205\,{\rm Mpc}\,h^{-1})^3$ per snapshot, we can turn each snapshot into four slabs projected over a depth of $\delta\chi=51.25\,{\rm Mpc}\,h^{-1}$ and, using Cloud-in-cell \citep[CIC;][]{hockney} interpolation, generate dark matter surface density maps for each, which are $4096\times 4096$ pixels. The pixel resolution and projection depth for each slab then closely match (with a mere $2.5\%$ offset) those used to train \textsc{BIND}, allowing it to be used as conditioning in the model. We find that this small discrepancy doesn't affect our emulated results, as we show in our validations in \S~\ref{sec:val_halo}. To identify the halos, once the pixelized slabs are generated, the FoF catalog positions of halos with $M_{200c}\geq 10^{13}\, M_\odot\,h^{-1}$ can be identified in pixel coordinates, and the conditioning field, $\rho^H_{\rm DMO}$, for each halo extracted as a $128\times 128$ pixelized grid.

In practice, to build a lightcone, we require only 20 snapshots between $z=0$ and $z=2.5$. From these 20 snapshots, we typically want to reduce repeated structures along a given line of sight, which can occur when we naively project along the same axis in each snapshot. To avoid this, we follow \citet{Osato-2021, Lee-2026}, using the simulation's periodic boundary conditions and applying random translations and rotations to each snapshot before projecting. This ensures that when the slabs are stacked to perform the ray tracing described in \S~\ref{sec:raytracing}, there are fewer repeated structures along a line of sight. Each of the 20 snapshots used to generate the lightcone receives a unique rotation, translation, and randomly chosen projection direction (along x, y, or z axes), from which the catalog of halo conditioning grids is extracted and used to generate with BIND.
 
\begin{figure*}
    \centering
    \includegraphics[width=\linewidth]{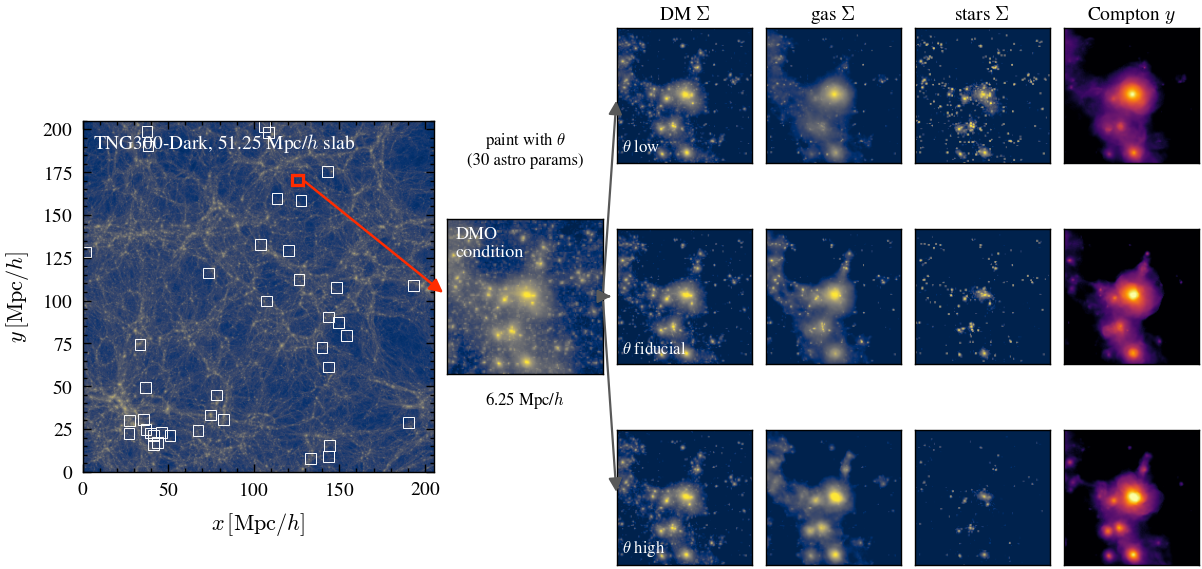}
    \caption{The lightcone \textsc{BIND}ing pipeline. \textit{Left:} a $51.25\,{\rm Mpc}\,h^{-1}$ slab of TNG300-Dark at $z=0.0337$, rotated, translated, and projected onto a $4096\times4096$ grid. Squares mark halos with $M_{200c}\geq 10^{13}\,M_\odot\,h^{-1}$ identified in the FoF catalog. \textit{Center:} the $128\times128$ pixel conditioning field $\rho^H_{\rm DMO}$ extracted around one halo. \textit{Right:} fields generated by \textsc{BIND} from this conditioning (dark matter, gas, and stellar surface masses, and the Compton-$y$ field) for low, fiducial, and high values of a chosen parameter while all other parameters are fixed to the TNG fiducial values.}
    \label{fig:pipeline_diagram}
\end{figure*}
 
Once the full halo conditioning catalog is created, we can choose a set of astrophysical parameters, the TNG300 set of cosmological parameters, and the given redshift of the snapshot and generate each of the halos. We show this procedure in Fig.~\ref{fig:pipeline_diagram}, where the leftmost field shows a slab of TNG300-Dark rotated, translated, and projected over $51.25\,{\rm Mpc}\,h^{-1}$ to a $4096\times 4096$ grid. Each small square contains a halo identified by the FoF catalog, for which a $ 128\times128$-pixel patch is extracted, as shown in the center panel. With this conditioning, we can use \textsc{BIND} to draw from the posterior and generate density fields such as dark matter, gas, and stellar density, or thermodynamic fields such as the Compton-$y$ field, for any chosen parameter combination. We show each field for low, fiducial, and high values of a single galaxy formation model parameter\footnote{While irrelevant for this discussion and Fig.~\ref{fig:pipeline_diagram}, the parameter that is varied is $A_{SN1}$. We simply wanted to present the pipeline and qualitative changes for a parameter; a full analysis of the effects due to parameters can be found in \S~\ref{sec:astro}}, while keeping all other parameters fixed to the TNG fiducial. We can see clear differences right away. The stellar density is significantly suppressed, and the gas concentration and Compton-$y$ become more centrally concentrated in the halo when this parameter is turned up. When the parameter is turned low, there are more small-scale features in all channels, including a dramatic change for the stellar field; however, in the halo core, the gas and Compton-$y$ distributions appear to be more diffuse. For a more detailed exploration of the individual effects of the parameters on the halos, we refer the reader to \citet{Lee-2026b}.

\subsection{Generating halos across the Sobol sequence}\label{sec:sobol}
After creating the halo catalog for conditioning described in the previous section, we can generate halos for each slab across all 20 snapshots. The generated halos are then pasted back into the TNG300-Dark simulation following the pasting scheme outlined in \S5 of \citet{Lee-2026b}, where each generated density field blends into the dark-matter-only density field following a cosine tapering beginning at $3R_{200c}$. For the thermodynamic fields, the painted halo gas follows the same tapering. However, because there is no gas outside the painted halo regions in the TNG300-Dark, for the $\tau$ and Compton-$y$ fields, we approximate the electron density outside halo regions by multiplying the dark matter field by the baryon fraction. In this way, we paint each slab with baryons centered on the halos and stack the slabs as in \citealt{Osato-2021} to create a lightcone stretching from $z=0$ to $ z=2.5$. We create five sets of lightcones for use in this work, which we describe below.
 
\begin{itemize}
\item \textbf{DMO set}: TNG300-Dark is used with no halo painting to create the lightcone.
\item \textbf{Hydro-pasted set}: TNG300-Dark has its halos replaced with TNG300 halos to create the lightcone.
\item \textbf{Fiducial set}: Each halo is generated with the fiducial TNG300 parameters to create the lightcone. This can then be validated against the hydro-pasted set.
\item \textbf{1P set}: Each halo is generated with all TNG300 parameters at their fiducial values except for one of the 30 parameters, which is set to one of its prior bounds from \citet{Genel-2026}. This yields 57 lightcones (rather than 60, as three parameters have their fiducial values at a prior bound, see \citet{Genel-2026} for details) at different locations in parameter space, allowing us to isolate the effects of individual parameters.
\item \textbf{Sobol set}: Each of the 30 astrophysics parameters is varied simultaneously over a 256-node Sobol sequence to generate 256 lightcones. This is a smaller version of the \citet{Genel-2026} Sobol sequence, which contains 1024 simulations over a 35-dimensional parameter space. 
\end{itemize}

In all of the sets described above, the cosmological parameters are fixed to the TNG300 fiducial values, and only the 30 astrophysical parameters are varied. This is because generating halos with DMO in a different cosmology would be out-of-distribution from the training set. BIND is not a tool for cosmological rescaling, but could in the future be used on rescaled cosmological simulations \citep{Angulo-2010, Zennaro-2019}. 

We also generate our lightcones for each set in the same way. They contain the same rotations, translations, and projections of the snapshots and the same set of replaced halos (either with the generative model's field or the true hydro field). Further, each halo has a unique random seed for flow matching, used to generate it once per parameter-space location, and that seed is shared across all lightcones, so differences between lightcones are purely parameter-driven. The result is then 314 lightcones at different parameter space locations that use BIND and a single lightcone from the TNG300-Dark simulation, and a single lightcone that has replaced the TNG300-Dark halos of $M_{200c}\geq 10^{13}\,h^{-1}\,{\rm M}_\odot$ with their true hydrodynamical field counterparts.\footnote{For more information on this replacement procedure, see \citet{Lee-2026a}.}

\subsection{Ray tracing and map generation}\label{sec:raytracing}

\begin{figure*}[!t]
    \centering
    \includegraphics[width=\linewidth]{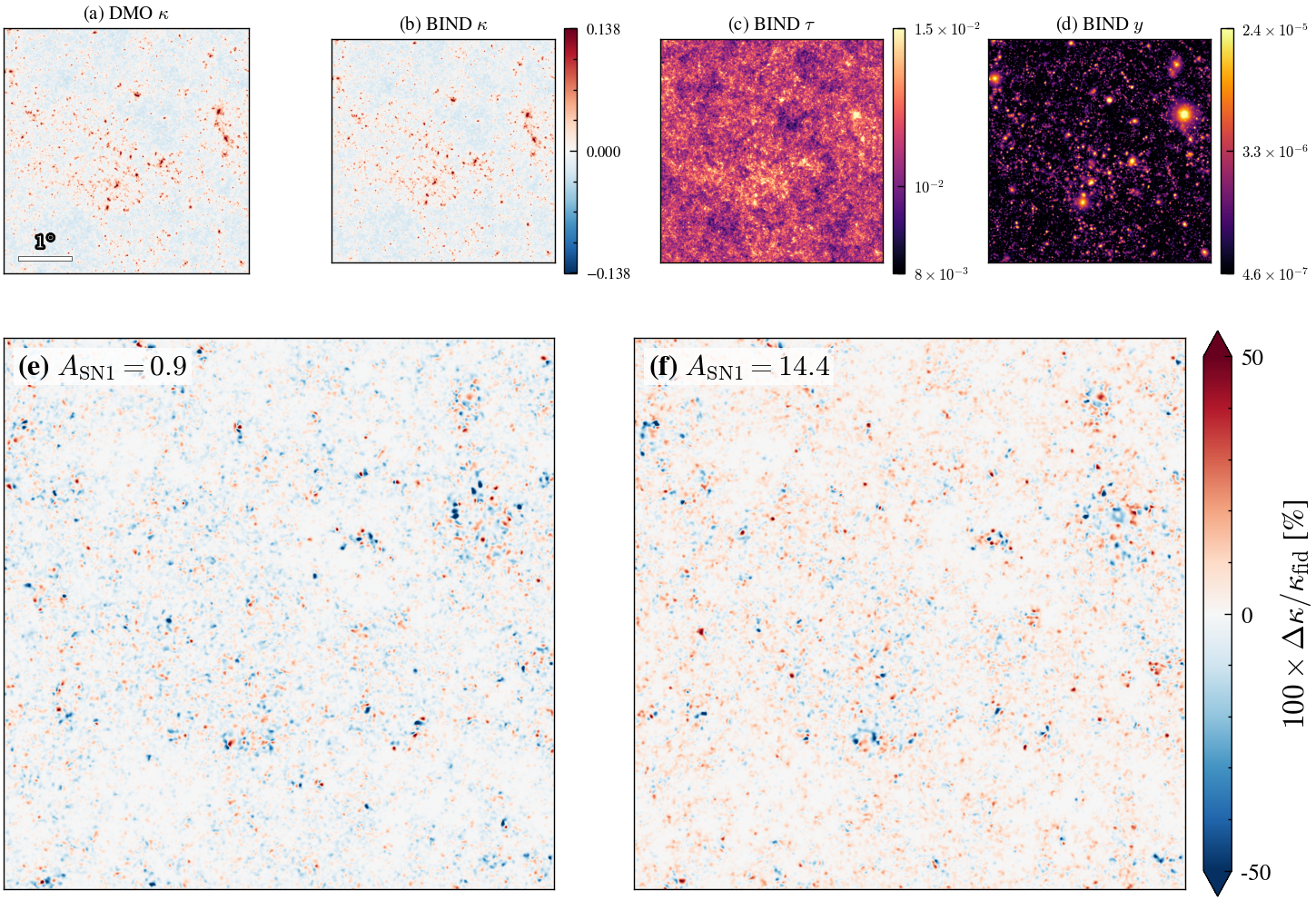}
    \caption{Example maps from the pipeline described in \S~\ref{sec:raytracing}. \emph{Top row left to right:} a dark-matter-only convergence map, and a \textsc{BIND}ed convergence $\kappa$, optical depth $\tau$ and Compton-$y$ map generated at the fiducial parameter space location for TNG300. The halos are consistently traced, with peaks appearing at the same locations across all three observables. \textit{Bottom row:} differences between convergence maps with a single parameter set to each of its two prior bounds while all others are fixed to the TNG300 fiducial, normalized by the fiducial convergence map, with clear differences appearing around the halos. All maps shown in this plot are at $z=1$}
    \label{fig:hero}
\end{figure*}
We follow the procedure of \citet{Osato-2021}, which uses the multiple lens-plane ray-tracing algorithm \citep{Jain-2000, Hilbert-2009, Petri-2016a} to account for lensing deflections and generate the convergence $\kappa$, Compton-$y$, and $\tau$ maps. We ray-trace each of the surface density stacks described in the previous section, where each slab in a stack, projected over a depth of $\delta\chi=51.25\,h^{-1}\,{\rm Mpc}$, has surface density $\Sigma$ after the pasting procedure. From each slab we compute the lensing potential at plane $k$ via

\begin{equation}
    \nabla^2_{\bm{\beta}^k} \psi^k(\bm{\beta}^k) = 2\Sigma^k(\bm{\beta}^k),
\end{equation}
where $\bm{\beta}^k$ is the angular position on the $k^{\rm th}$ plane. At each plane we then compute the deflection angle $\bm{\alpha}$,
\begin{equation}
    \nabla_{\bm{\beta}^k} \psi^k(\bm{\beta}^k) = \bm{\alpha}^k(\bm{\beta}^k),
\end{equation}
which, summed across all planes, gives the final lensed position of the ray given the original pointing angle on the sky $\phi$,
\begin{equation}\label{eq:beta}
    \bm{\beta}^k(\bm{\phi}) = \bm{\phi} - \sum_{i=1}^{k-1}\dfrac{\chi^k-\chi^i}{\chi^k}\bm{\alpha}^i(\bm{\beta}^i), \quad (k=2, 3,\dots),
\end{equation}
where $\chi^i$ is the comoving distance at the $i^{\rm th}$ lensing plane, and $\chi^{k}$ is the comoving distance of the source. The lensed position at a given pointing angle from the observer, and out to a comoving distance is then related to the convergence and shear of the field via the lensing Jacobian,
\begin{equation}
    A_{ij}(\bm{\phi}, \chi) \equiv \dfrac{\partial\beta_i(\bm{\phi}, \chi)}{\partial\bm{\phi}_j},
\end{equation}
from which the convergence follows as $\kappa = 1 - \tfrac{1}{2}\left(A_{11}+A_{22}\right)$ (for the full recursion and higher-order terms, we refer the reader to \citealt{Osato-2021} and \citealt{Hilbert-2009}).
For each surface density stack, we generate $1000$ pseudo-independent lightcones of $25\,{\rm deg}^2$ by following the same rotation convention as \citet{Osato-2021}, randomly rotating each slab by $0^\circ$, $90^\circ$, $180^\circ$, or $270^\circ$ about each of the three axes and translating the pixels in each direction. This rotation and translation process allows for a large statistical sample of map realizations, with up to $10,000$ pseudo-independent realizations demonstrated in \citet{Petri-2016}. The resulting $\kappa$, $y$, and $\tau$ maps are output on $1024^2$ pixel grids, downsampled from the $4096^2$ planes following the same scheme as \citet{Osato-2021}, at five source redshifts $z_s = 0.5$, $1.0$, $1.5$, $2.0$, and $2.44$.
 
In the ray tracing procedure, the convergence is determined by the weighted deflections accumulated by a ray along its line of sight. However, we can follow the same light ray and, instead of computing a lensing term, simply sum the gas density in each pixel the ray crosses. Accounting for the ionization state of the gas, this amounts to an integral of the ionized gas along the (deflected) path of the ray, the optical depth,
\begin{equation}\label{eq:tau}
    \tau(<z_s) = \sigma_{\rm T}\int_0^{z_s} n_e\,{\rm d}l_{\rm prop}
    = \frac{\sigma_{\rm T}\, x_e}{m_{\rm p}}\sum_{k} \Sigma^{k}_{\rm gas},
\end{equation}
where $\Sigma^k_{\rm gas}$ is the gas surface density of plane $k$ evaluated along the deflected ray, and $x_e = X_{\rm H} + Y_{\rm He}/2 = 0.88$ electrons per proton mass, assuming fully ionized gas of primordial composition ($X_{\rm H} = 0.76$, $Y_{\rm He} = 0.24$). The end result is a map of the optical depth accumulated to a given source redshift along the same line of sight, deflected by gravitational lensing. Similarly, as the Compton-$y$ parameter is the integrated electron pressure along the line of sight, we follow the same ray and sum the pressure planes,
\begin{equation}\label{eq:compy}
    y(<z_s) = \frac{\sigma_{\rm T}}{m_e c^2}\int_0^{z_s} P_e\,{\rm d}l_{\rm prop}
    = \frac{\sigma_{\rm T}}{m_e c^2}\sum_{k} \Sigma^{k}_{P},
\end{equation}
where $\Sigma^k_P$ is the integrated electron pressure of plane $k$.
In practice, the pipeline evaluates the electron pressure directly from the generated gas density and temperature channels, $P_e = n_e k_{\rm B} T_e$, with the ionization prefactors given above, i.e. $P_e = \frac{2+2X_{\rm H}}{3+5X_{\rm H}}P_{\rm th}\approx0.52\,P_{\rm th}$ for the total thermal pressure $P_{\rm th}$ of a fully ionized primordial gas.

Because $\kappa$, $\tau$, and $y$ are accumulated along the same deflected rays and output at the same source redshifts, every released map triplet of $\kappa$, $\tau$, and $y$ is mutually consistent. 
 
Of course, the optical depth $\tau$ and Compton-$y$ maps suffer from the fact that the box is not entirely filled with hydrodynamical gas. Only regions centered on the $M_{200c}\geq10^{13}\,M_\odot\,h^{-1}$ halos contain painted gas tapering to an approximation of the gas density outside by using the dark matter field and cosmological baryon fraction, and a constant temperature approximation of $10^4\,{\rm K}$, roughly the temperature to which the photoionized intergalactic medium is held by the ultraviolet background \citep{McQuinn-2016}, and consistent with the cool diffuse phase that dominates the gas outside massive halos in IllustrisTNG \citep{Martizzi-2019}. However, each of the halo regions contains the integrated gas and pressure across the full $51.25\,{\rm Mpc}\,h^{-1}$ slab, allowing a compensated aperture photometry \citep[CAP;][]{Schaan-2021, Amodeo-2021} style evaluation of the Compton-$y$ and $\tau$ signals of halos in the final maps. In future work, we plan to use these maps to perform CAP-style measurements of the halos along the lightcone.

We further note that we do not model any velocities. Our maps are of the optical depth itself, and the translation to a kSZ observation, or to a dispersion-measure observation in the case of fast radio bursts, is left to the observational side, where velocity assumptions or measurements come into play. We return to both modeling statements and their consequences for which statistics the maps support in \S~\ref{sec:caveats}.

In Fig.~\ref{fig:hero}, we show examples of the resulting maps from the procedure described in this section. The left top panels show a dark-matter-only convergence map compared to the fiducial \textsc{BIND}ed convergence map. The top right two panels show the \textsc{BIND}ed optical depth and Compton-$y$ maps. The halos are traced consistently across the maps, with peaks appearing in the convergence, $\tau$, and Compton-$y$ maps at the same locations. In the bottom two panels, we hint at the effects of changing the parameters, which are discussed in \S~\ref{sec:astro}, by showing the difference between convergence maps where a single parameter is set to each of its two prior bounds (all others at the fiducial) and the fiducial convergence map. Clear differences appear, particularly around the halos, suggesting that variations in individual parameters propagate to the full weak lensing maps and are likely observable in the resulting summary statistics.

\section{Validation}\label{sec:validation}
In this section, we perform a series of validations comparing the fiducial set to the hydro-pasted set. We begin with halo-level relations, computed at the snapshot level before ray tracing, then move to map-level validations. Because TNG300 is only run at a single parameter space location, we are limited to comparing our maps at only this fiducial parameter space location, however, in the companion paper, \citet{Lee-2026b}, we focused on applying a battery of tests and showed that \textsc{BIND} accurately captured the effects of IllustrisTNG model parameters on the generated fields. Ideally, we would compare against a full suite of parameter-varying, ray-traced maps, but this is currently computationally infeasible. 

One subtlety for the following sections is that they are out-of-distribution tests. The TNG300-Dark projections are sourced from the TNG300 box, which is $205\,{\rm Mpc}\,h^{-1}$. Breaking this into four surface density slices following \S~\ref{sec:raytracing} means that each projection is $\sim 51\,{\rm Mpc}\,h^{-1}$ containing an extra $\sim 1\,{\rm Mpc}\,h^{-1}$ of matter in the projection compared to the training set used for \textsc{BIND} in \citealt{Lee-2026b}. Before performing any ray tracing, we need to be sure that this out-of-distribution conditioning can still be used with \textsc{BIND}.
 
\subsection{Halo-level validation}\label{sec:val_halo}
We first validate the \textsc{BIND}ed halos by comparing various relations to the true TNG300 hydro halos. We illustrate results only at redshift $z=0.0337$ for clarity of presentation, although we have performed the same comparisons at higher redshifts and find similar results \citep[we also explore redshift validation and effects in the companion paper,][]{Lee-2026b}. At $z=0.0337$, we identify $2933$ halos with $M_{200c}\geq10^{13}\,{\rm M}_\odot\,h^{-1}$, which are used in the following analyses.
 
\begin{figure}
    \centering
    \includegraphics[width=0.9\linewidth]{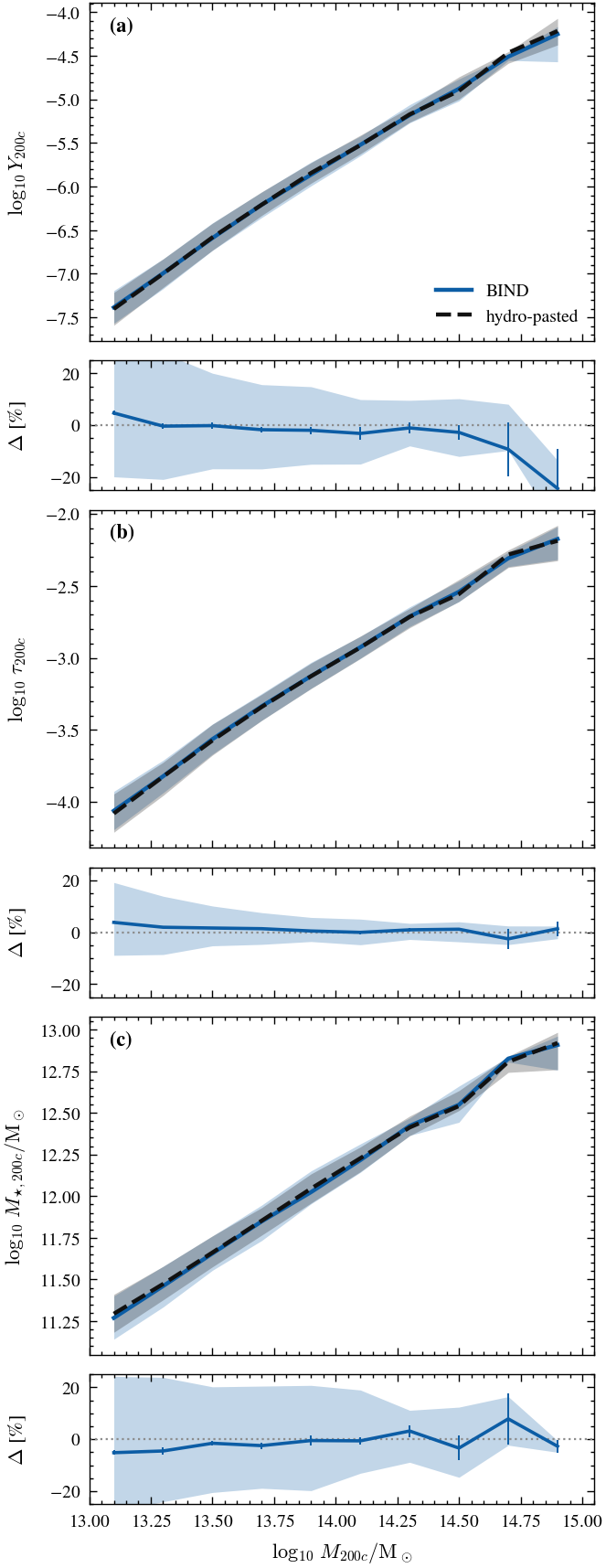}
    \caption{the $Y_{200c}-M$, $\tau_{200c}-M$, and $M_{\star,200c}-M$ relations for the \textsc{BIND}ed halos (blue) and the TNG300 halos (black dashed) computed within $R_{200c}$ at redshift $z=0.0337$. The lines are computed as the median across all halos in ten evenly log-spaced halo mass bins; the shaded bands show the $16^{\rm th}$--$84^{\rm th}$ percentile halo-to-halo scatter. Below each panel, we show the relative difference, $\Delta = ({\rm BIND} - \text{hydro-pasted})/\text{hydro-pasted}$, with error bars computed as bootstrap uncertainty on the median. Masses on the horizontal axis are in ${\rm M}_\odot\,h^{-1}$, as in the text.}
    \label{fig:halo_validation}
\end{figure}
 
We first show the $Y-M$, $\tau-M$, and $M_{*}-M$ scaling relations for both the \textsc{BIND}ed and TNG300 halos in Fig.~\ref{fig:halo_validation}. Each quantity is computed as a sum over the projected $R_{200c}$ of each halo. Of course, these relations include foreground and background material generated in the full projection, making them ``contaminated'' halo scaling relations. We paste these patches into the multi-lensplane algorithm of \S~\ref{sec:raytracing}, which is itself a projection, so we want to validate the full projection; these contaminated relations are a useful metric for that validation.

Each panel shows halos in ten bins evenly log-spaced between the mass threshold of $M_{200c} = 10^{13}\,{\rm M}_{\odot}\,h^{-1}$ and the most massive halo in the sample at $M_{200c}=10^{14.98}\,{\rm M}_\odot\,h^{-1}$. The solid line represents the median in a given halo mass bin, and the shaded regions, representing the scatter, are computed as the $16^{\rm th}$--$84^{\rm th}$ percentiles. Below each scaling relation panel, we show the relative difference between the \textsc{BIND}ed halos and TNG300 halos $\Delta = ({\rm BIND} - \text{TNG300})/\text{TNG300}$. Error bars in the lower panels are computed with bootstrap resampling to estimate median uncertainty in each bin.
 
\begin{figure}
    \centering
    \includegraphics[width=0.9\linewidth]{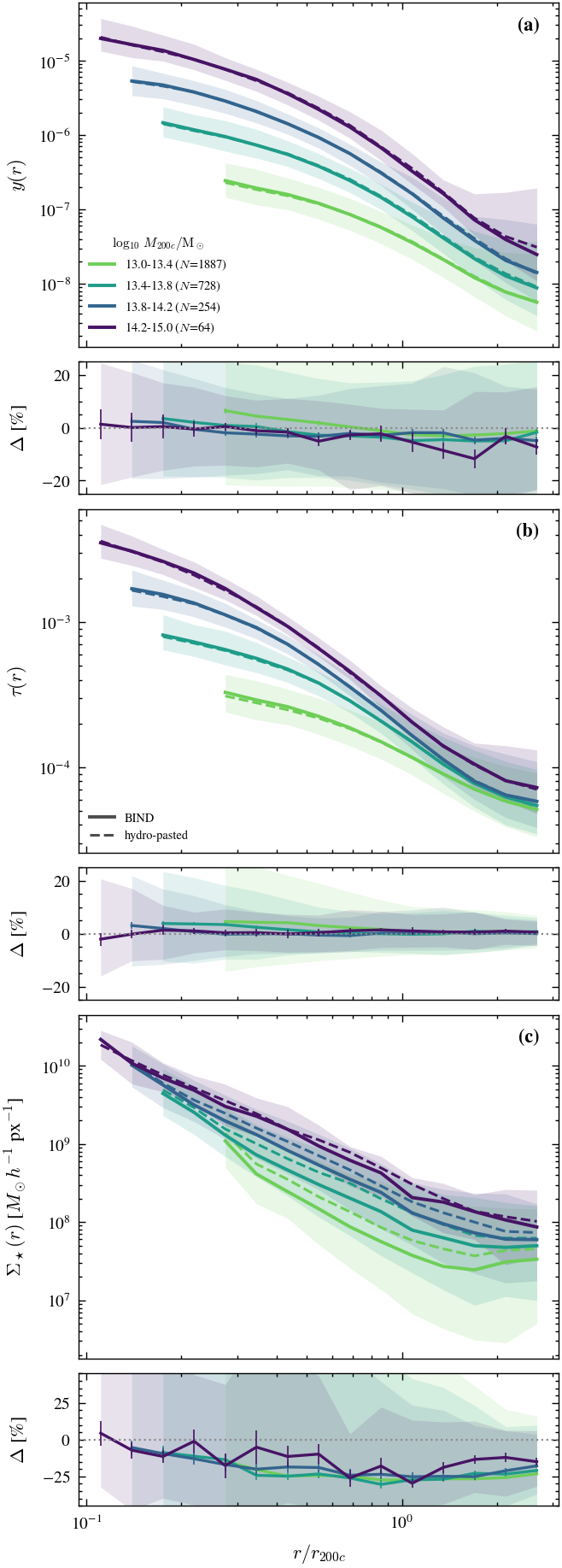}
    \caption{Radial profiles for $y$, $\tau$, and $\Sigma_\star$ for the \textsc{BIND}ed (solid) and TNG300 (dashed) halos at $z=0.0337$, in four mass bins shown as different colors. Similar to Fig.~\ref{fig:halo_validation}, the bands show the $16^{\rm th}$--$84^{\rm th}$ percentile halo-to-halo scatter, and the error bars are computed with bootstrap resampling. Each curve begins at the pixel scale of its mass bin and extends to $3R_{200c}$, where the taper for pasting into the dark-matter-only simulation begins (see \S~\ref{sec:sobol} for more details). Masses in the legend are in ${\rm M}_\odot\,h^{-1}$, as in the text.}
    \label{fig:radial_validation}
\end{figure}

In each panel, we see that the bootstrap uncertainty increases at the highest masses; however, these bins also contain few halos: 149 above $10^{14}\,{\rm M}_\odot\,h^{-1}$, 14 above $3\times10^{14}\,{\rm M}_\odot\,h^{-1}$, and only 6 above $4\times10^{14}\,{\rm M}_\odot\,h^{-1}$. \textsc{BIND} is a generative model that draws from a conditional distribution, and the resulting draw may differ by up to the scatter associated with a given halo mass. Most bins show that BIND captures this halo-to-halo scatter well, with small biases; however, it clearly underpredicts the Compton-$y$ parameter in the highest-mass bin. We suspect this is a training-set effect rather than a ray-tracing one. The CAMELS boxes are $(50\,{\rm Mpc}\,h^{-1})^3$, so halos above a few $\times10^{14}\,{\rm M}_\odot\,h^{-1}$ are rare in training, and the $Y$--$M$ relation is the steepest of the three ($Y\propto M^{5/3}$ for self-similar halos, against roughly linear for $\tau$ and $M_\star$), so it is the relation where a draw that leans toward the better-sampled, lower-mass part of the conditional distribution costs the most. Fig.~\ref{fig:radial_validation} shows where in the halo this deficit sits.

In Fig.~\ref{fig:radial_validation}, we turn to halo profiles in four mass bins and compare them between the \textsc{BIND}ed and TNG300 halos. We compute radial profiles for each halo by integrating the masses or pressures within fifteen radial bins from $0.1R_{200c}$ to $3R_{200c}$. Each profile begins at the pixel scale, so the curves for the different mass bins start at different fractional $R_{200c}$ radii. Each line again is the median in the mass bin, and the scatter is the $16^{\rm th}$--$84^{\rm th}$ percentiles for a given quantity.
 
Similar to Fig.~\ref{fig:halo_validation}, we find a close agreement across the profile quantities and mass bins. The gas-related profiles (the top two quantities) agree to within a few percent inside $R_{200c}$ in every mass bin. The one exception is the $y$ profile of the highest-mass bin, which falls $5$--$10\%$ low between $R_{200c}$ and $3R_{200c}$. This is the $Y$--$M$ deficit discussed in Fig.~\ref{fig:halo_validation}, where \textsc{BIND} reproduces the pressure in the cores of the most massive halos and underproduces it in their outskirts, where the projected pressure is carried by extended, low-density gas that the model sees least often in training. The stellar channel shows a clear bias, particularly at intermediate radii and out beyond $R_{200c}$.  Although its integrated mass is only $\sim6\%$ low for the lowest mass bins, the $\Sigma_\star$ profiles sit $25$--$40\%$ low in individual annuli, meaning \textsc{BIND} misplaces stellar mass radially more than it underproduces it. This is consistent with the stellar channel being the hardest to reconstruct because of its sparsity \citep{Lee-2026b}. As stars contribute only a small fraction of the total projected mass, this does not appreciably affect the $\kappa$ maps.
 
\subsection{Map-level validation}\label{sec:val_map}
We now validate the \textsc{BIND}ed fiducial maps against the hydro-pasted maps. Recall that, as described in \S~\ref{sec:raytracing}, the fiducial and hydro-pasted sets contain 1000 pseudo-independent maps at five redshifts. For each map and redshift, we compute several weak-lensing statistics, including both Gaussian and non-Gaussian statistics. No smoothing is applied to any map, and all statistics are computed on the native $1024^2$ pixel grid ($0.293$ arcmin pixels).
 
For each convergence map, we calculate the angular power spectrum,
\begin{equation}\label{eq:C_l}
C^{\kappa\kappa}(\ell_i) = \langle|\tilde{\kappa}(\bm{\ell})|^2\rangle_{\bm{\ell} \in [\ell_i^{\rm min}, \ell_i^{\rm max}]},
\end{equation}
where $\tilde{\kappa}(\bm{\ell})$ is the two-dimensional Fourier transform of the convergence field $\kappa(\bm{\phi})$ in the flat-sky approximation, so that $\bm{\ell}$ is the two-dimensional wavevector conjugate to $\bm{\phi}$ rather than a spherical-harmonic index $(\ell, m)$, and the angle brackets denote averaging over all Fourier modes $\bm{\ell}$ within the multipole bin $[\ell_i^{\rm min}, \ell_i^{\rm max}]$. From the TNG300-Dark set of maps, we define the power spectrum suppression,
\begin{equation}\label{eq:suppression}
S(\ell) = \frac{C^{\kappa\kappa}_{\rm BIND}(\ell)}{C^{\kappa\kappa}_{\rm DMO}(\ell)},
\end{equation}
computed realization by realization. Because each map between simulation sets contains matched rotations, translations, and generative seeds, the power-spectrum suppression division cancels realization noise in the ratio.
 
We use the \emph{LensTools} package \citep{Petri-2016a}\footnote{\url{lenstools.readthedocs.io}} to compute non-Gaussian statistics from the sets of convergence maps. All are computed as functions of the signal-to-noise ratio $\nu = \kappa/\sigma_0$, where $\sigma_0$ is the root-mean-square convergence of the \emph{fiducial} map set, held fixed across all map sets so that bins remain directly comparable between runs. We use the same set of statistics as in \citet{Lee-2026a},
\begin{enumerate}
\item \textbf{Peak counts} $N_{\rm pk}(\nu)$: The number of local maxima in the convergence field, defined as pixels with greater values than their $8$ nearest neighbors \citep{Kratochvil-2010, Maturi-2010}.
\item \textbf{Minimum counts} $N_{\rm min}(\nu)$: The number of local minima in the convergence field, defined as pixels with smaller values than their $8$ nearest neighbors \citep{Coulton-2020}.
\item \textbf{One-point PDF} $p(\nu)$: The probability distribution of convergence values, representing the full histogram of $\nu$ values \citep{Thiele-2020}.
\item \textbf{Minkowski functionals} (MFs): Three descriptors of convergence excursion sets \citep{Kratochvil-2012, Petri-2013} (area $V_0(\nu)$, boundary length $V_1(\nu)$, and genus $V_2(\nu)$) that encode shape information.
\end{enumerate}
 
For the Compton-$y$ and $\tau$ maps, we compute the two-point auto- and cross power spectra with the same estimator,
\begin{equation}\label{eq:cross_spectra}
\begin{split}
C^{AB}(\ell_i) = \langle {\rm Re}[\tilde{A}(\bm{\ell})\tilde{B}^*(\bm{\ell})]\rangle_{\bm{\ell} \in [\ell_i^{\rm min}, \ell_i^{\rm max}]}, \\
 A, B \in \{\kappa, y, \tau\},
 \end{split}
\end{equation}
giving $C^{yy}$, $C^{\tau\tau}$, $C^{\kappa y}$, $C^{\kappa\tau}$, and $C^{y\tau}$ in addition to $C^{\kappa\kappa}$.
 
Each $\ell$-statistic is computed in log-spaced bins between $\ell=10^2$ and $4\times10^4$. The Nyquist mode of the $1024^2$ maps is $\ell_{\rm Ny} = \pi/\theta_{\rm pix} \approx 3.7\times10^4$, and we only present results up to this Nyquist mode. For the non-Gaussian statistics, we compute all of them in 22 bins between $-2\leq\nu\leq8$.
 
\begin{figure*}
    \centering
    \includegraphics[width=\linewidth]{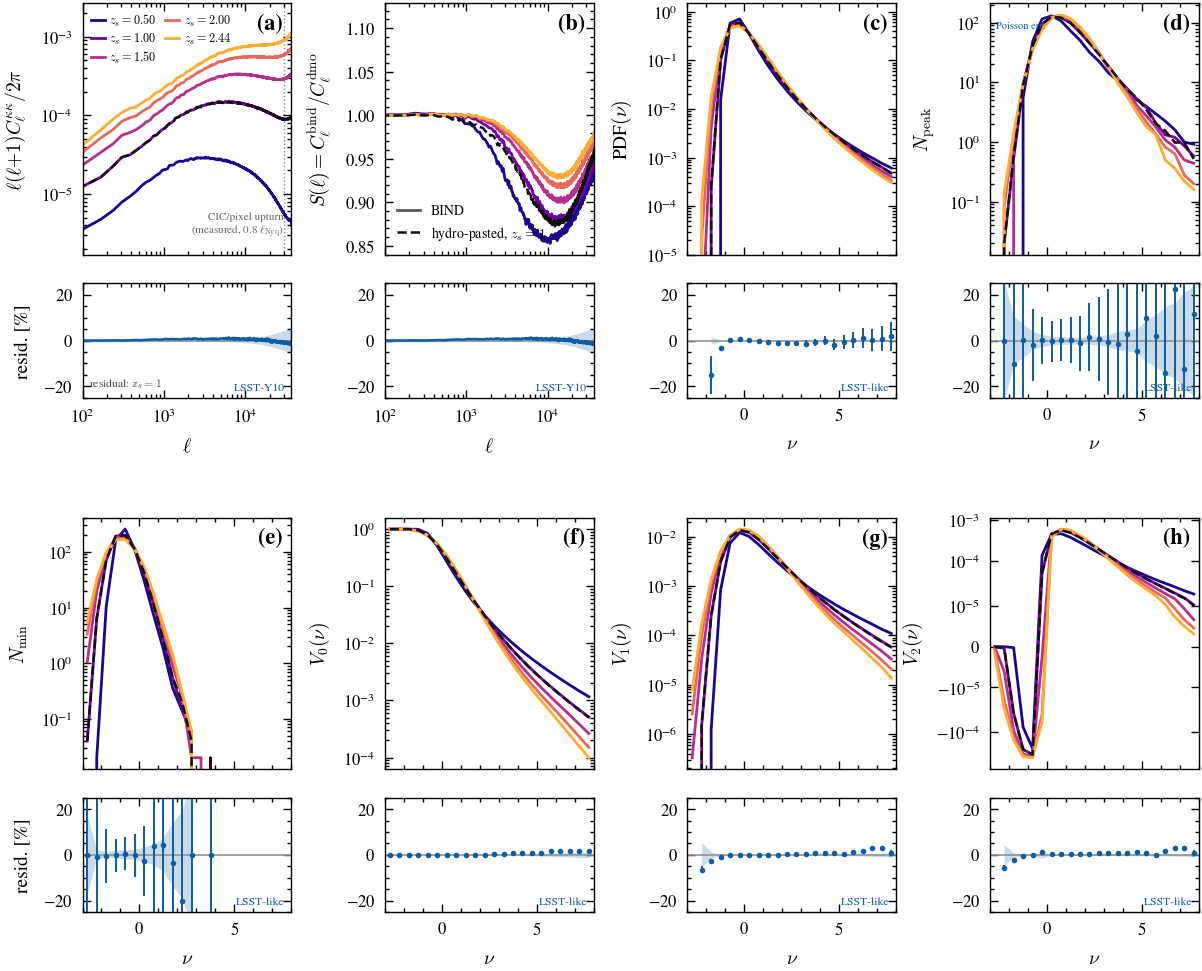}
    \caption{Weak lensing statistics for the \textsc{BIND}ed (solid) and hydro-pasted (dashed) maps at source redshift $z_s=1$. The weak lensing angular power spectrum (\textbf{a}), the power spectrum suppression $S(\ell)$ (\textbf{b}), the one-point convergence PDF (\textbf{c}), weak lensing peak counts (\textbf{d}), minimum counts (\textbf{e}), and the three Minkowski functionals (\textbf{f-h}). Panels (\textbf{a}, \textbf{b}) are functions of multipole $\ell$; panels (\textbf{c}--\textbf{h}) are functions of the signal-to-noise $\nu=\kappa/\sigma_0$. Bottom panels show the mean residual between \textsc{BIND} and hydro-pasted at the same $\ell$ or $\nu$, with error bars scaled to an LSST-Y10-like survey area of $\sim18,000\,{\rm deg}^2$ following Eq.~\ref{eq:area_scaling}.}
    \label{fig:field_validation}
\end{figure*}

Note that the hydro-pasted maps are \textbf{not} the full TNG300 hydro ray-traced maps. Instead, they are TNG300-Dark maps where each halo with $M_{200c}\geq 10^{13}\,{\rm M}_\odot\,h^{-1}$ has been replaced exactly with the dark matter, stellar, and gas particles of the TNG300 simulation. This is the same procedure that was followed in \citet{Lee-2026a}, where the authors found that replacing halos of $10^{13}\,{\rm M}_\odot\,h^{-1}$ and above accounts for $\sim60$--$65\%$ of the response between dark-matter-only and hydrodynamical simulations at the level of the power spectrum, but virtually all of the response for peak counts and Minkowski functionals. We return to the consequences of this construction in \S~\ref{sec:caveats}. The comparison here is also noiseless, meaning no shape noise, beam, or survey systematics enter at any stage. By comparing \textsc{BIND}ed with the hydro-pasted maps, we are able to directly observe this halo-centric and field-level generative model approach in an apples-to-apples comparison.
 
In Fig.~\ref{fig:field_validation}, we compare the weak lensing statistics between the hydro-pasted and \textsc{BIND} maps at source redshift $z_s=1.0$. We generate maps at all five source redshifts but limit the comparison to $z_s=1$ for clarity. However, the results are comparable at the other redshifts. Each figure shows the given statistic in the top panels, while the bottom panels show the residual at $z_s=1$.
 
In the bottom panels, we assess the statistical distinguishability of the statistics given an LSST-like covariance matrix. To compute this, we follow a similar procedure to \citet{Lee-2023}, and use the hydro-pasted convergence maps to estimate, for a given statistic,
\begin{equation}\label{eq:cov}
    C_{ij} = \frac{1}{N-1} \sum_{k=1}^{N}
    \left(n_{i}^{(k)} - \langle n_i\rangle\right)\left(n_{j}^{(k)} - \langle n_j\rangle\right),
\end{equation}
where $n_i^{(k)}$ is the statistic measured in the $k^{\rm th}$ hydro-pasted realization, the sum runs over the $N=1000$ pseudo-independent realizations, and the subscripts refer to the $\ell$ or $\nu$ bins of the data vector.
When a precision matrix is required, we debias it as
\begin{equation}\label{eq:inv_cov}
    \hat{\bm{C}}^{-1} = \frac{N-d-2}{N-1}\bm{C}^{-1},
\end{equation}
where $d$ is the length of the data vector \citep{Hartlap-2007}. The lightcones are limited to $25\,{\rm deg}^2$, but we can test statistical distinguishability for LSST-like areas $A$ by scaling the covariance matrix as
\begin{equation}\label{eq:area_scaling}
    \bm{C}(A) = \left(\dfrac{A}{25\,{\rm deg}^2}\right)^{-1}\bm{C}(25\,{\rm deg}^2),
\end{equation}
with $A = 18,000\,{\rm deg}^2$ for LSST-Y10. This should not be read as a realistic prediction for LSST, which would require super-sample covariance and the inclusion of many other systematics; however, it is a good test of statistical distinguishability against the hydro-pasted model with an LSST-like covariance. 
 
The bottom panels of each plot show the mean residual with error bars associated with an LSST-Y10-like survey, with Poisson error bars for the number counts in a given bin. We can read off by eye the statistical distinguishability relative to an LSST-like survey by comparing the residual lines to the error contours in the bottom panels, and we clearly see that the \textsc{BIND}ed statistics match the hydro-pasted maps to within LSST-like precision for all weak lensing statistics.

\begin{figure*}
    \centering
    \includegraphics[width=\linewidth]{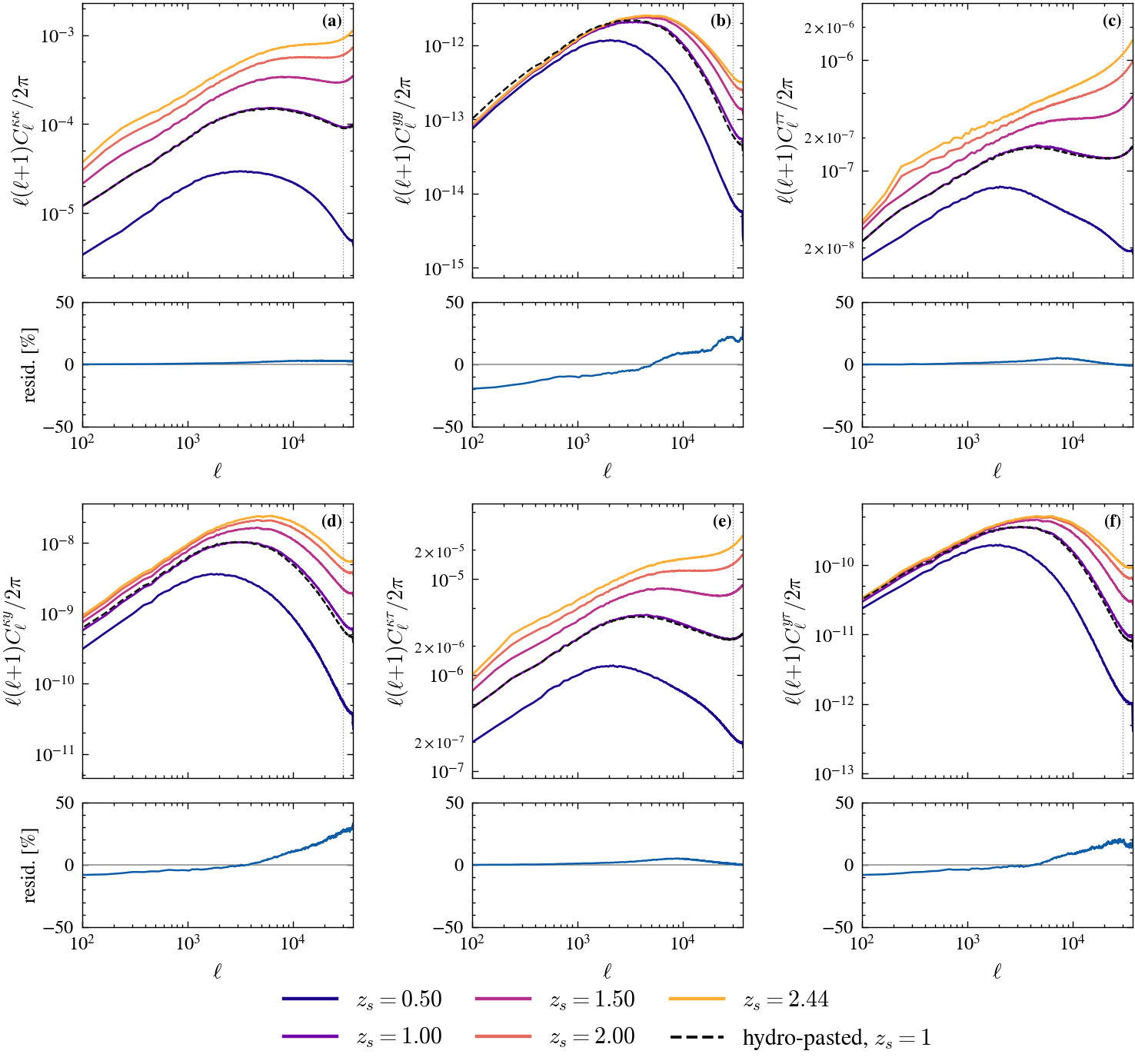}
    \caption{Auto and cross power spectra of the \textsc{BIND}ed maps at all five source redshifts (colors), compared with the hydro-pasted maps at $z_s=1$ (black dashed), though each panel shows all five source redshifts. The $\kappa\kappa$ (\textbf{a}), $yy$ (\textbf{b}), and $\tau\tau$ (\textbf{c}) auto spectra, and the $\kappa y$ (\textbf{d}), $\kappa\tau$ (\textbf{e}), and $y\tau$ (\textbf{f})  cross spectra. Residual panels show the $z_s=1$ comparison. Unlike Fig.~\ref{fig:field_validation}, no survey scaling is applied, as no LSST-like covariance analog exists for the gas observables.}
    \label{fig:spectra_validation}
\end{figure*}

Beyond weak lensing statistics, we are interested in the trend in baryon correction models used to generate thermodynamic predictions on the lightcone, as in \citet{Anbajagane-2024} and \citet{Schneider-2025}. We compare our generated $\tau$ and Compton-$y$ maps with respect to the hydro-pasted set at $z_s=1$ in Fig.~\ref{fig:spectra_validation}. We follow the same format as above and show \textsc{BIND}ed maps at each source redshift, with comparisons at $z_s=1$ to the hydro-pasted maps. The top panels show the three auto power spectra, while the bottom three panels show the cross spectra. In the absence of an LSST-like analog for distinguishability, we show the percent error of the prediction relative to the truth in the bottom panels, with the standard error on the median curve. Unlike the weak lensing statistics, the thermodynamic spectra, both auto and cross, are more biased. The $yy$ spectrum is the worst case: it sits $\sim20\%$ low at $\ell\simeq10^2$, crosses zero near $\ell\simeq5\times10^3$, and rises to $\sim20\%$ high at the Nyquist scale. The tSZ power at low $\ell$ is dominated by the most massive halos, and $C_\ell^{yy}$ scales as $Y^2$, so the $10$--$15\%$ underprediction of $Y$ in the highest-mass bin, concentrated in the halo outskirts, is enough to account for the $\sim20\%$ deficit. The $\tau$ spectra trace gas mass rather than pressure and have no such deficit at the halo level, and are correspondingly less biased.

Given these biases, it is fair to ask what the $\tau$ and Compton-$y$ maps are useful for. We are primarily interested in how the statistics respond to parameter variations, and the results in \S\S~\ref{sec:astro}--\ref{sec:emulator} are largely insensitive to these offsets. The rank correlations of \S~\ref{sec:correlations} are invariant to any bin-wise bias constant across the parameters. Moreover, the biases are an order of magnitude below the actual feedback response found in \S~\ref{sec:astro}, so the maps resolve feedback responses at high contrast. However, predictions of the absolute $C_\ell^{yy}$ or the mean Compton-$y$ inherit these offsets, as well as the approximated diffuse gas discussed in \S~\ref{sec:caveats}.

Finally, we emphasize again that the validation above is performed only at the fiducial parameters because TNG300 provides the only available hydrodynamical truth and is available only at this parameter-space location. The parameter response of \textsc{BIND} itself, that the generated halos move correctly as the parameters vary, was validated against the CAMELS simulations in \citet{Lee-2026b}, where held-out parameter locations were recovered across the prior volume, and correlations between generated quantities and the parameters were matched almost perfectly. Validation in both this section and the companion paper supports the validity of the parameter variations we now turn to. In the following section, we use the Sobol and 1P sets to measure how weak lensing and gas statistics respond across the full 30-dimensional astrophysical parameter space.

\section{Astrophysical effects}\label{sec:astro}

In the previous sections, we showed that \textsc{BIND} generates lightcones comparable to TNG300 at the fiducial parameter-space location; we now turn to the effects on weak lensing and CMB statistics as functions of IllustrisTNG galaxy formation model parameters. 

In this work, we do not explore varying cosmologies or correlations between astrophysics and cosmological parameters. TNG300-Dark was run with a single set of cosmological parameters, so we only explore astrophysical parameter variations independently of cosmology here. In future work, we will consider the effect of cosmology by applying cosmological rescaling methods or by running additional dark matter simulations with varying cosmologies to use as conditioning. 

Our first question is whether, at a fixed cosmology, the IllustrisTNG galaxy formation model parameters induce statistically distinguishable effects for an LSST-like survey. Current gas and lensing data already prefer stronger feedback than the TNG300 fiducial calibration \citep{Bigwood-2024, Hadzhiyska-2024, Siegel-2025, Bigwood-2025b}, but with the generated Sobol set of lensing maps we can directly test whether parameter changes induce a detectable shift in weak lensing statistics given LSST-like uncertainty. As a reminder, the amplitude of the power spectrum suppression is limited by the incomplete pasting, which only includes more massive halos of $M\geq10^{13}\,{\rm M}_\odot\,h^{-1}$, but the relative variation of statistics with respect to the parameters is accurate. 

For this measurement, we assume that the TNG300 fiducial parameter-space location is the ``true'' observed data, and we build the survey precision exactly as in \S~\ref{sec:val_map} with the covariance of $C_\ell^{\kappa\kappa}$ measured across the $1,000$ hydro-pasted realizations and scaled to the area of an LSST-like survey (exactly as in Eq.~\ref{eq:area_scaling}). 

\begin{figure}
    \centering
    \includegraphics[width=\linewidth]{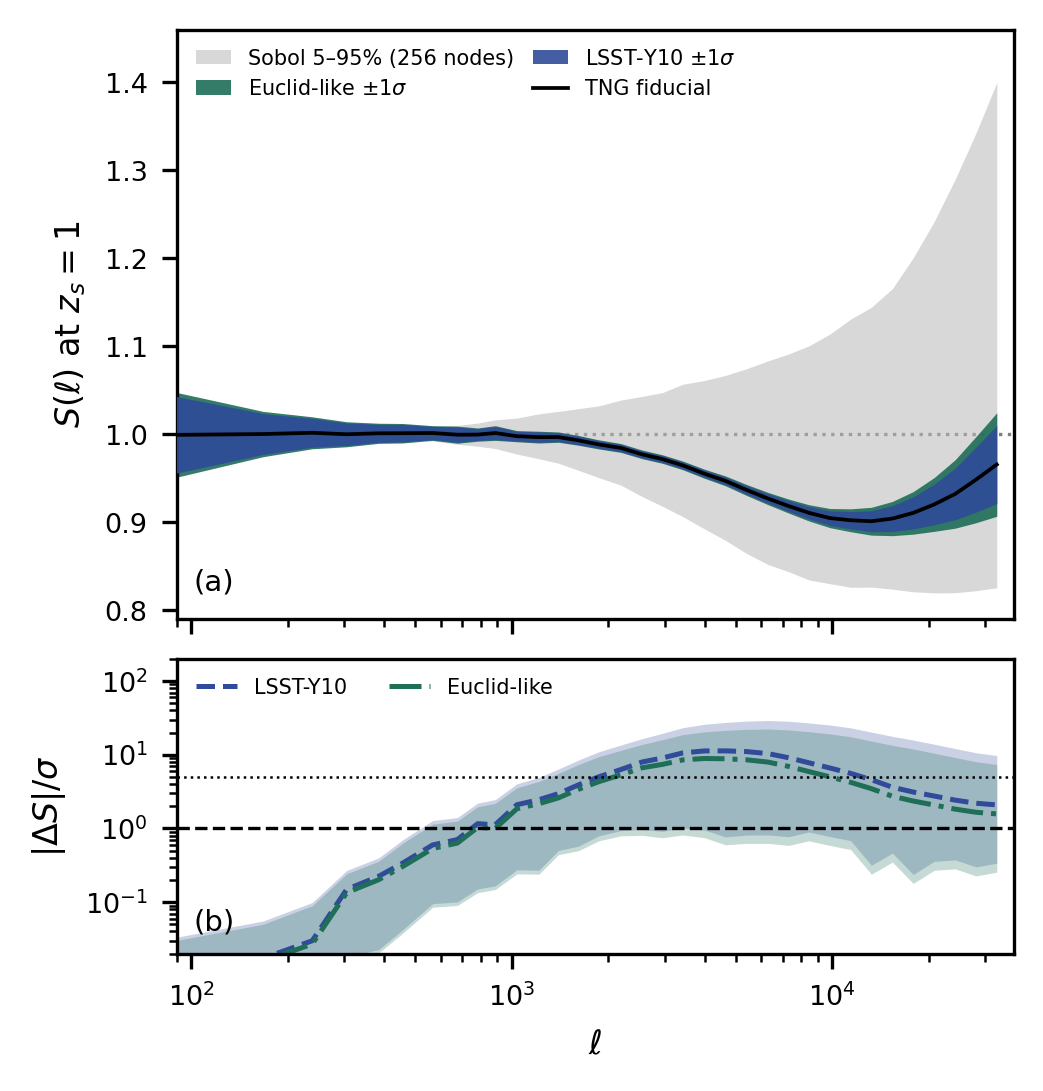}
    \caption{The power spectrum suppression, $S(\ell)$, across 256 IllustrisTNG model parameter space locations at redshift $z_s=1$ with respect to LSST-Y10-like uncertainty. \textbf{(a)} The $5$--$95\%$ spread of $S(\ell)$ for the Sobol sequence (gray), the fiducial (black), and the $\pm1\sigma$ LSST-Y10 and \textit{Euclid}-like precisions drawn on the fiducial. \textbf{(b)} The same Sobol sequence spread, but shown as a measure of significance of statistical distinguishability $|S_{\rm node}-S_{\rm fid}|/\sigma$. The bands are the $5$--$95\%$ range across nodes, while the lines represent the medians.}
    \label{fig:s3_opener}
\end{figure}

Fig.~\ref{fig:s3_opener} panel (a) shows the extent of the power spectrum suppressions from the entire Sobol set in gray, the fiducial suppression in black, and the two survey precisions as envelopes drawn on the fiducial. Panel (b) shows the same information but rewritten as a detection significance per band-power for each of the Sobol nodes, $|S_{\rm node}-S_{\rm fid}|/\sigma$. We take the uncertainty, $\sigma$, to be the square root of the diagonal of the covariance matrix (Eq.~\ref{eq:cov}). 

Over $\ell\simeq2.2\times10^3$--$1.8\times10^4$, the Sobol set prior spread exceeds the LSST-Y10 precision by more than $10\sigma$, peaking at $22\sigma$ near $\ell\simeq5\times10^3$. For \textit{Euclid} the window is narrower owing to the smaller area of the survey, with $\ell\simeq3\times10^3$--$1.3\times10^4$ and a peak of $17\sigma$. All 256 parameter space locations of the Sobol set vary from the fiducial by more than $1\sigma$ somewhere in the range of $300<\ell<0.8\,\ell_{\rm Ny}$, and $84\%$ ($73\%$) of the prior is statistically distinguishable by more than $5\sigma$ for LSST-Y10 (\textit{Euclid}). Ultimately, this plot shows that the galaxy formation model parameter space yields two-point statistics that are distinguishable from the ground truth for an LSST- or Euclid-like survey. This holds beyond power-spectrum suppression; in fact, we find the same for all statistics we study in this work. We show the power-spectrum suppression to connect with previous literature \citep{vanDaalen-2011, vanLoon-2024, vanDaalen-2025}.

In the following sections, we examine which TNG300 parameters drive the effects we see in the various weak lensing statistics. We start by looking for correlations between the statistics and parameter changes, as in \citet{Lee-2026b}. We then consider if individual parameter variations fall into similar families of effects and hypothesize about the physical reasoning behind this.

\subsection{Correlations across the Sobol set}\label{sec:correlations}

\begin{figure*}
\centering
\includegraphics[width=\linewidth]{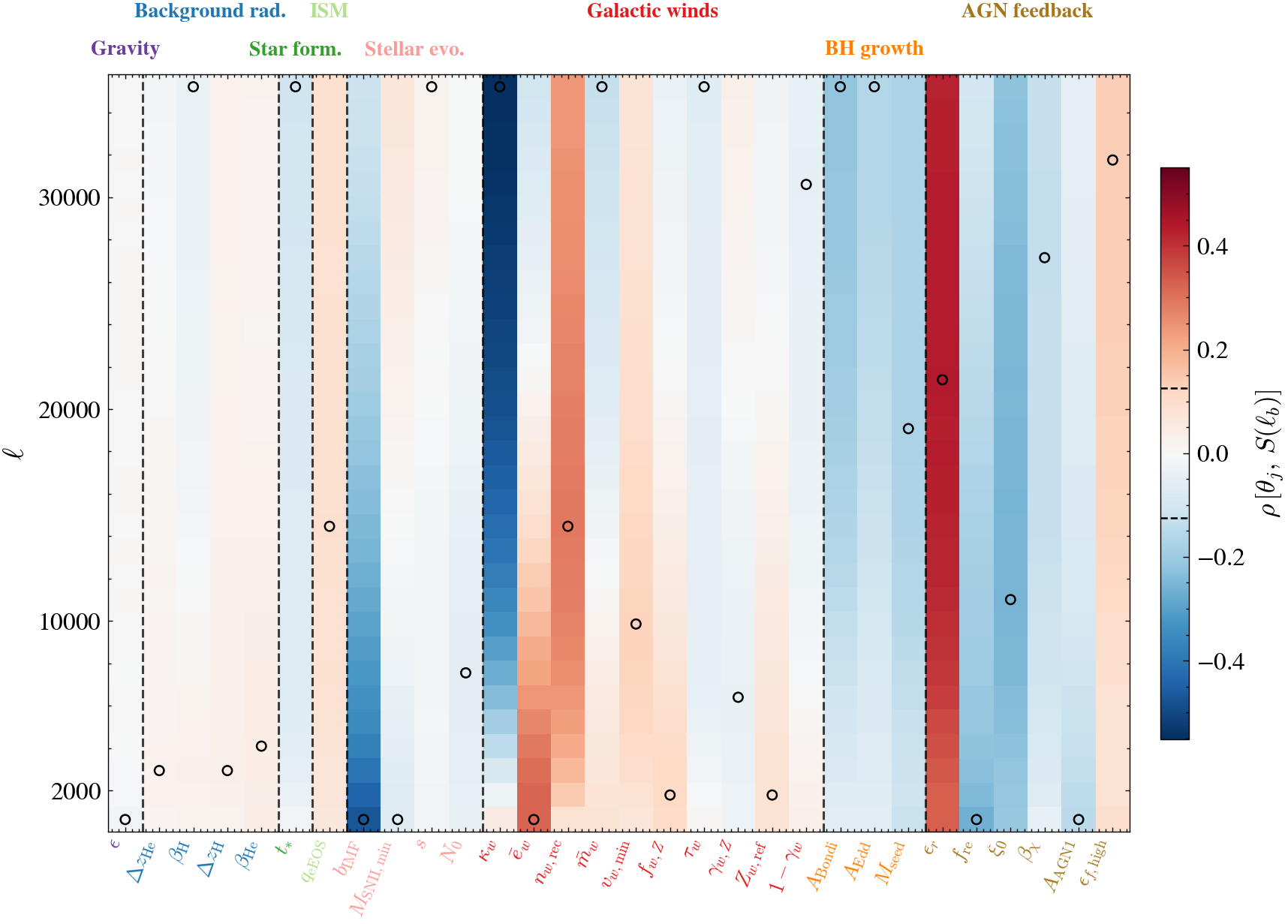}
\caption{The peak-correlation construction of \S~\ref{sec:correlations} for the power spectrum suppression alone. Each column is one of the thirty galaxy formation parameters, each row an $\ell$ bin, and the color shows the signed Spearman rank correlation $\rho[\theta_j, S(\ell_b)]$ measured across the 256 Sobol nodes at $z_s=1$. Circles mark the bin with the strongest correlation for each parameter, whose peak-$|\rho|$ values are the entries that feed the statistic-parameter grid of Fig.~\ref{fig:corr_matrix}. Dashed lines on the colorbar mark the chance-correlation level for a single pre-chosen bin, $|\rho| = 2/\sqrt{256} = 0.125$ (\S~\ref{sec:correlations}). The wind sector (VarWindVelFactor, WindFreeTravelDens) drives the strongest and most scale-coherent responses.} 
\label{fig:srow}
\end{figure*}
 
Our goal is to address a few concrete questions: which galaxy model parameters impact which statistics? Is this consistent across all statistics, or does each have unique sensitivities? For hypotheses on the physical origins of these effects, see \S~\ref{sec:story}.

Following a similar routine to \citet{Lee-2026b}, we compute the signed Spearman rank correlation $\rho[\theta_j, n_i]$ between each of the 30 parameters and each bin of each measured statistic across the 256 Sobol nodes, at $z_s=1$.  For each (statistic, parameter) pair, we then extract the peak $|\rho|$ across that statistic's bins, restricted to the trusted scale range. Fig.~\ref{fig:srow} shows this construction for the power spectrum suppression alone. Each column is one of the thirty parameters, each row an $\ell$ bin, and the circles mark the bin with the strongest correlation for each parameter. Some parameters clearly induce effects more prominently on smaller scales, while others on larger angular scales. Some parameters also appear to have little to no effect across the full spectrum. 
 
However, we now want to consider the 12 commonly explored statistics\footnote{Note that we show $S(\ell)$, which has an identical response to $C_\ell^{\kappa\kappa}$. Hence, the 12 statistics are the auto-correlations in Fig.~\ref{fig:spectra_validation}, and the weak lensing statistics in Fig.~\ref{fig:field_validation}} introduced in \S~\ref{sec:val_map}, each with its own bins in $\ell$ or $\nu$, its own noisiness, and its own correlations across bins. One could imagine eleven more versions of Fig.~\ref{fig:spectra_validation}, each for a different statistic. 

A maximum as represented by the open circles in Fig.~\ref{fig:srow} taken over many bins is biased high. Even a parameter that affects nothing still draws one chance correlation per bin, and by taking the peak, we report its luckiest. A search of this size can therefore be fooled by peaks produced by random fluctuations or a shared systematic effect. What we are really after is a map of parameters against statistics in which a peak correlation appears only if chance alone would not have produced it. To this end, we subject the peak correlations to three levels of statistical rigor.

\begin{enumerate}

\item \textit{Tile level, from the scatter across Sobol nodes.} With $N=256$ nodes, a single \emph{pre-chosen} bin of a statistic that responds to nothing still shows chance correlations of order $1/\sqrt{N}$. We take the $2\sigma$ level, $|\rho| = 2/\sqrt{256} = 0.125$. This level applies before any bin search.

\item \textit{Row level, from the look-elsewhere across a statistic's own bins.} Neighboring bins in either $\ell$ or $\nu$ for a statistic are correlated, so the number of effectively independent bins matters when considering the bin with peak correlation. We measure this empirically by randomly assigning the Sobol parameter vectors to the measured statistics, breaking the connection between parameters and statistics while preserving correlations among bins. Any peak correlation in these shuffled sets arises solely from bin covariance and noise. We set each statistic's threshold at the 95th percentile of its shuffled peaks, with $|\rho| $ ranging from 0.14 to 0.182 across the twelve statistics.

\item \textit{Column level, from the look-elsewhere across all statistics.} The full grid of twelve statistics and thirty parameters shares the same random assignments as in the row level, so correlations are shared not only across a statistic's bins but between the statistics themselves. From this, we obtain the per-parameter threshold across statistics at the 95th percentile, $|\rho| = 0.200$, and the whole-grid null, $|\rho| = 0.253$, which is the level the single largest cell of a pure-noise grid would reach only $5\%$ of the time.
\end{enumerate}
A larger search always yields a luckier peak, so a parameter matters for a statistic only when it clears the level appropriate to the claim being made.
 
\begin{figure*}
\centering
\includegraphics[width=\linewidth]{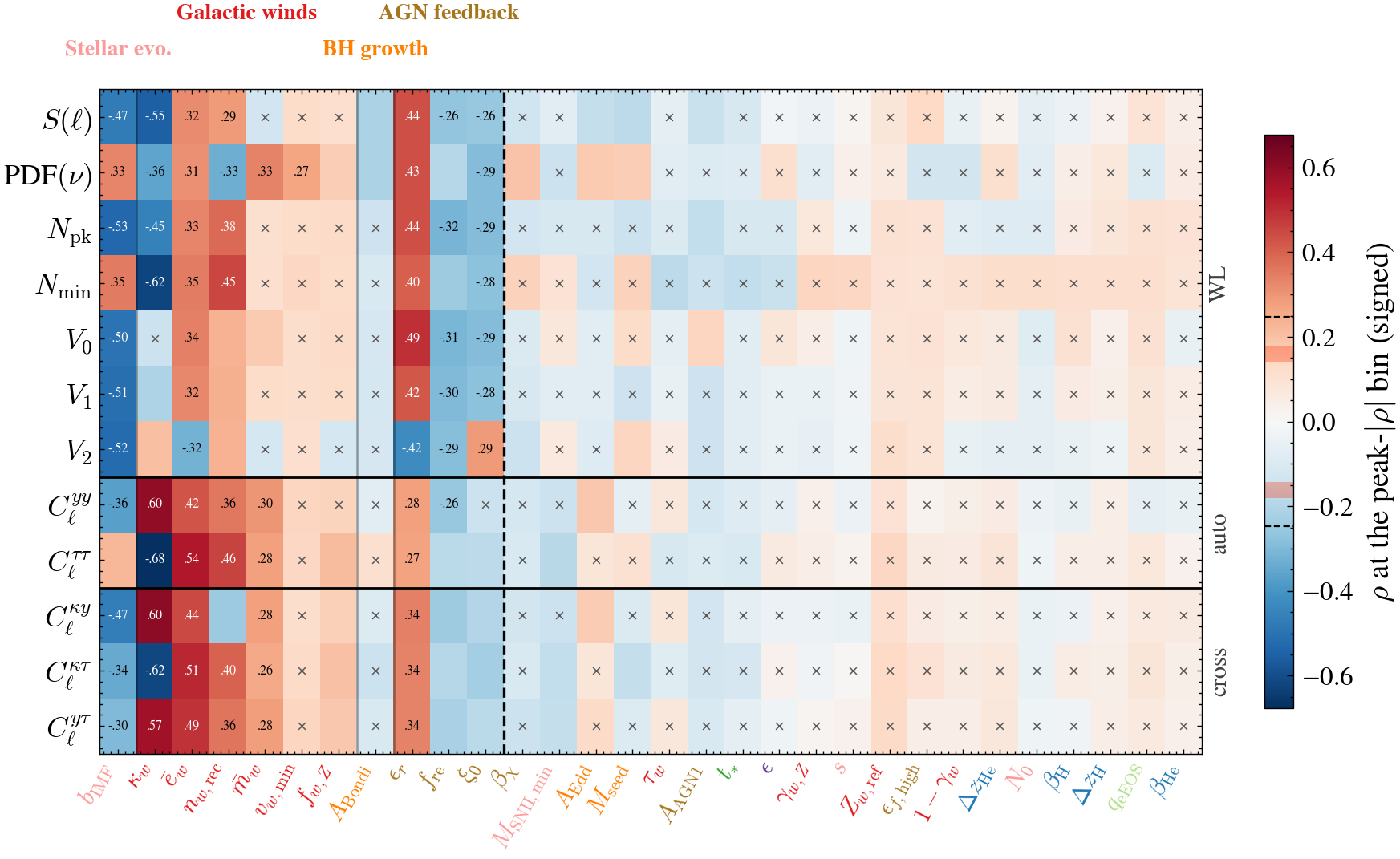}
\caption{The signed Spearman correlation $\rho$ evaluated at each pair's peak-$|\rho|$ bin, for the 12 statistics (rows) against the 30 parameters (columns, sorted by peak $|\rho|$). The null levels described in the text of \S~\ref{sec:correlations} are marked on the colorbar. Printed values mark cells clearing the whole-grid null ($|\rho| \geq 0.253$), and cells marked $\times$ fall below that statistic's null. Blank, uncrossed cells lie between the two. The vertical dashed line marks the per-parameter look-elsewhere null ($0.200$) such that parameters to its right never clear it for any statistic.}
\label{fig:corr_matrix}
\end{figure*}

In Fig.~\ref{fig:corr_matrix} we show the peak correlations for each of the 12 statistics across all 30 parameters, with parameters sorted by their strongest effect on any statistic. Each cell shows the signed correlation at the peak-$|\rho|$ bin for that pair. A parameter matters for a statistic only when it crosses the appropriate threshold. We find 11 parameters to the left of the dashed line clear the per-parameter look-elsewhere null ($0.200$) for at least one statistic. Of these, a few contain at least one cell beyond the whole-grid null ($0.253$), an effect stronger than $5\%$ of the entire null space. Cells marked $\times$ fall below their statistic's individual null and are consistent with bin covariance rather than parameter influence, and blank cells without numerals lie between their own-statistic null and the whole-grid level.
 
Fig.~\ref{fig:corr_matrix} compresses all statistics and parameter effects into a concise image. A handful of parameters dominate the correlations across every statistic, while the majority never clear their look-elsewhere null (19/30 deemed insignificant with $|\rho|\leq0.2$). 

The row structure provides more insight into the connection between the galaxy formation model and the statistics. The $\ell$-domain statistics and every gas spectrum appear to be dominated by the wind parameters, while the $\nu$-domain morphological statistics (peaks, minima, and Minkowski functionals) are sourced primarily by the IMF slope $b_{\rm IMF}$ and the black hole radiative efficiency $\epsilon_r$. Different statistic families listen to different feedback channels, and in \S~\ref{sec:story}, we theorize more about the physical causes.

\subsection{Response families}\label{sec:families}
In the previous section, we saw that 11 parameters (those to the left of the dashed line in Fig.~\ref{fig:corr_matrix}) control the effects on the statistics, based on peak correlations across bins, and that the wind parameters dominate the $\ell$-domain statistics while the $\nu$-domain statistics are shaped primarily by the AGN parameters and the IMF slope. Here, instead of the amplitude of the parameter effects, we explore their scale dependence. Our goal is to determine whether the parameters act in distinct ways or whether separate families of parameters drive the statistics to respond similarly.

\begin{figure*}
\centering
\includegraphics[width=\linewidth]{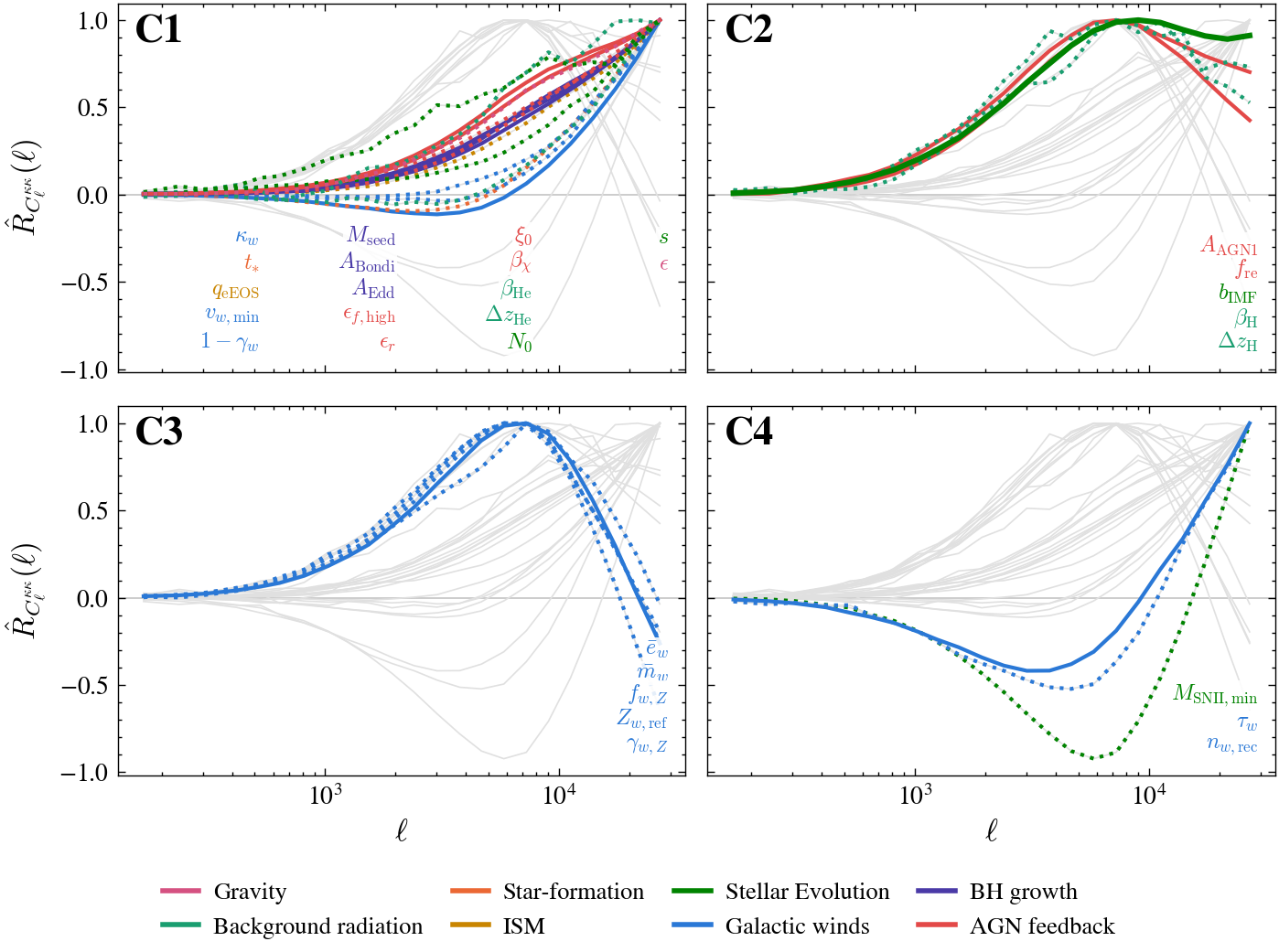}
\caption{Response families of the convergence power spectrum. Each panel shows all thirty normalized 1P responses $\hat{R}_{C_\ell^{\kappa\kappa}}(\ell)$ (gray), with the panel's family members colored by their astrophysical sector from Table~1 of \citet{Genel-2026} (legend). Solid lines mark parameters passing the statistic-level noise floor of Fig.~\ref{fig:corr_matrix}, while dotted lines do not.}
\label{fig:families_cl}
\end{figure*}
 
We start with the $\ell$-domain lensing power spectrum $C_\ell^{\kappa\kappa}$, computed across the 57 1P lightcones at $z_s=1$. For each parameter, we compute the fractional response of the statistic,
\begin{equation}\label{eq:response}
R_i(\ell) = \dfrac{C_{\ell}^{\kappa\kappa}\big|_{\theta_{i,\rm max}} - C_{\ell}^{\kappa\kappa}\big|_{\theta_{i,\rm min}}}{C_{\ell}^{\kappa\kappa}\big|_{\theta_{\rm fid}}},
\end{equation}
where, for the three parameters whose fiducial value sits at a prior bound, the response is one-sided, $(C_\ell|_{\theta_{\rm bound}} - C_\ell|_{\theta_{\rm fid}})/C_\ell|_{\theta_{\rm fid}}$. We compute this for each of the 50 realizations of every 1P lightcone and take the median. Since we are interested here in the shape of each parameter's effect rather than its amplitude, we normalize every curve by its extremal value,
\begin{equation}\label{eq:rhat}
\hat{R}_i(\ell) = \dfrac{R_i(\ell)}{\max_\ell\,[R_i(\ell)]}.
\end{equation}
Finally, to identify connections between the responses, we compute the correlation between each normalized curve and every other curve, and consider curves with $r \geq 0.85$ to belong to the same family.
 
In Fig.~\ref{fig:families_cl}, we show the resulting four distinct families. Each panel shows all thirty normalized response curves, with the panel's family members colored by their astrophysical sector, taken exactly from Table~1 of \citet{Genel-2026}: gravity, background radiation, star formation, ISM, stellar evolution, galactic winds, black hole growth, and AGN feedback. Solid lines mark parameters that pass the statistical noise floor of Fig.~\ref{fig:corr_matrix}, and dotted lines do not. Because Eq.~\ref{eq:rhat} normalizes each curve by its value at the scale of its largest difference, every curve reaches $+1$ at that scale regardless of the sign of the underlying response, so the families group parameters by the \emph{shape} of their scale dependence and not by its sign. The signs and amplitudes are those of the $S(\ell)$ row of Fig.~\ref{fig:corr_matrix}.
 
The four families sort the galaxy formation parameters by the radius at which they act, which we discuss in more detail in \S~\ref{sec:story}. C1 controls clustering on the core scales, C2 on the halo volume, C3 on the halo radius scale, and C4 on both the core and halo radius scales. The unique responses of the families are the structure that the analytic model will exploit in S~\ref{sec:emulator}, and it is also why a single suppression amplitude cannot describe the parameter space. Two parameters can produce the same suppression at $\ell \simeq 6\times10^3$ while moving the core in opposite directions, and the peaks and Minkowski functionals, which \citet{Lee-2026a} showed baryonic effects are dominated by the cores of massive halos, will tell them apart where the power spectrum cannot.

\section{An analytic model of the statistics}\label{sec:emulator}
Fig.~\ref{fig:families_cl} showed that the full IllustrisTNG parameter space affects the power spectrum with roughly four unique shapes as a function of $\ell$. Low-dimensional descriptions of baryonic effects have precedent, such as in previous works that investigated principal-component-based compressions of feedback's imprint on shear data vectors \citep{Eifler-2015, Huang-2019}, and single-amplitude models that capture much of the two-point response \citep{Mead-2021, Amon-2022, Preston-2023}. Here, the basis instead emerges from the measured parameter-response families and extends to every statistic considered. With that knowledge, we can build a simple linear model by expanding these families as basis functions,
\begin{equation}
\tilde{C}_\ell^{\kappa\kappa} = \overline{C}_\ell^{\kappa\kappa} + \sum_k a_k\, \hat{B}_k(\ell),
\end{equation}
with the $k = 1$--$4$ shape families as basis functions $\hat{B}_k(\ell)$, an amplitude $a_k$ for each, and a reference power spectrum $\overline{C}_\ell^{\kappa\kappa}$ which we describe below.

We are not limited to the power spectrum here. In fact, in Appendix~\ref{app:figs} we show that the parameter effects on the convergence PDF also decompose into correlated families from the 1P set, in the same way as in \S~\ref{sec:families}. In general, then, for each statistic we can define
\begin{equation}\label{eq:analytic_model}
\tilde{S}(x) = \overline{S}(x) + \sum_k a_k\, \hat{B}_k(x),
\end{equation}
where $x$ denotes the statistic's bin coordinate ($\ell$ or $\nu$). Moving forward requires three choices to define in Eq.~\ref{eq:analytic_model} -- the reference statistic $\overline{S}(x)$, the combination of a family's individual parameter responses into its basis $\hat{B}_k(x)$, and the amplitudes $a_k$.

\subsection{Computing the basis functions}\label{sec:model_parts}
\begin{figure}
    \centering
    \includegraphics[width=\linewidth]{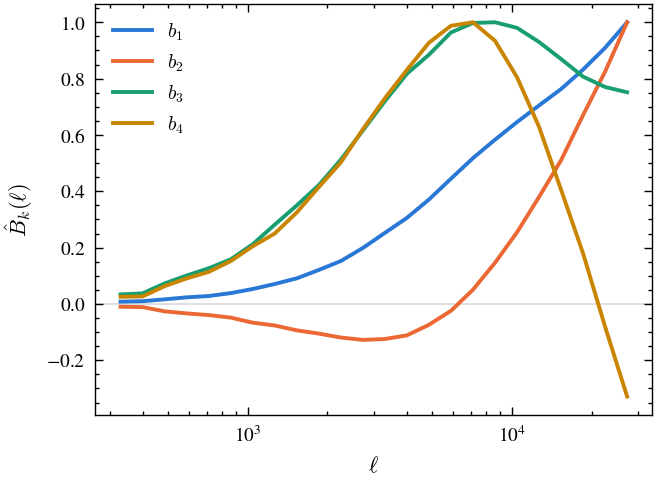}
    \caption{The basis functions $\hat{B}_k(\ell)$ for $S(\ell)$, built as the correlation-weighted mean of each family's 1P responses shown in Fig.~\ref{fig:families_cl} with weights computed from Fig.~\ref{fig:corr_matrix}. }
    \label{fig:basis}
\end{figure}
 
The basis functions are computed from the individual parameter responses of the 1P set as the weighted mean over a family's members, normalized as in Eq.~\ref{eq:rhat}:
\begin{equation}
B_k(x) = \frac{\sum_{j\in F_k} w_j\, R_j(x)}{\sum_{j\in F_k} w_j}, \qquad
\hat{B}_k(x) = \frac{B_k(x)}{\max_x\,[B_k(x)]},
\end{equation}
where $F_k$ is the member set of the family $k$ and the weights are the peak correlation magnitudes of \S~\ref{sec:correlations}, $w_j = |\rho_j|$, evaluated for the statistic being modeled. This lets the family's basis be shaped most by its statistically strongest members, while low-amplitude members are naturally down-weighted. One can simply compute a weighted mean for each panel of Fig.~\ref{fig:families_cl}.

We show in Fig.~\ref{fig:basis} the resulting basis functions for the power spectrum obtained from the same lines in Fig.~\ref{fig:families_cl}. From the curves in Fig.~\ref{fig:basis}, we argue that we can generate any weak lensing power spectrum (or power spectrum suppression, or any other statistic for that matter, as long as we have computed the basis functions from the 1P set) in the entire Sobol set given the correct set of amplitude factors and reference suppression.

The reference statistic we take to be the mean over the full Sobol set of $N = 256$ runs,
\begin{equation}
\overline{S}(x) = \frac{1}{N}\sum_{r=1}^N S_r(x),
\end{equation}
though the choice is somewhat arbitrary, as any reference within the suite, such as the fiducial or an arbitrary prior point, would only shift the amplitudes and be absorbed. 
 
For the amplitudes, we consider the deviation of each Sobol run's statistic from the reference, $\bm{d}_r = \bm{S}_r - \overline{\bm{S}}$, written as a vector over the statistic's bins, and solve the ordinary least squares problem per run,
\begin{equation}
\bm{a}_r = \arg\min_{\bm{a}} \Bigl\lVert\, \bm{d}_r - \sum_k a_k \bm{b}_k \,\Bigr\rVert^2
= (\bm{B}^T \bm{B})^{-1}\bm{B}^T \bm{d}_r,
\label{eq:amp-objective}
\end{equation}
with $\bm{B} = [\,\bm{b}_1 \cdots \bm{b}_4\,]$ the basis matrix.

\begin{figure}
    \centering
    \includegraphics[width=\linewidth]{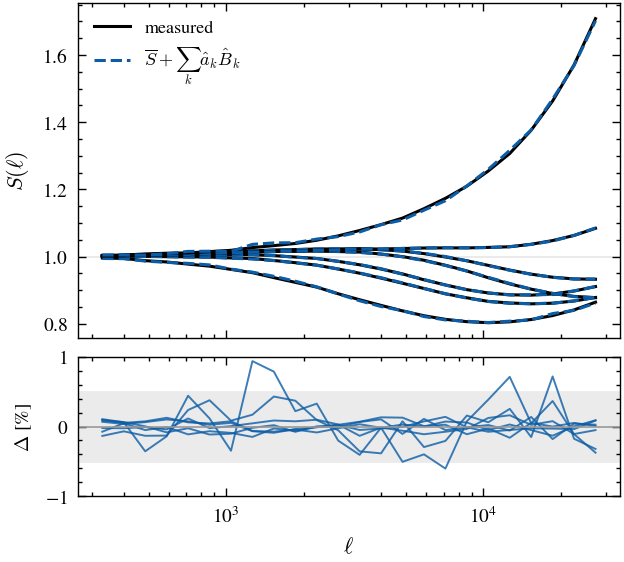}
    \caption{\textit{Top:} the measured suppression $S(\ell)$ (solid) for a set of Sobol nodes spanning the parameter priors, against the linear model $\overline{S}+\sum_k \hat{a}_k \hat{B}_k$ (dashed), with the amplitudes fit to each run directly via Eq.~\ref{eq:amp-objective}. \textit{Bottom:} the fractional residual, with the shaded band marking $\pm0.5\%$.}
    \label{fig:freeamp}
\end{figure}

Solving for the amplitudes at all Sobol points gives a matrix of 256 fitted vectors $\bm{a}_r$, from which any Sobol-set statistic can be regenerated. In Fig.~\ref{fig:freeamp} we show the power spectrum suppression for a set of Sobol lightcones along with the fit model of Eq.~\ref{eq:analytic_model}. With the above linear prescription, we achieve sub-percent-level accuracy at the tested locations, a trend that holds across the entire Sobol set. This is in and of itself a non-trivial result. We computed the basis vectors from parameter variations in the 1P set, while the amplitudes came from the Sobol set. Recovering the statistics here means the statistic can be written as a linear decomposition of individual parameter variations, with an amplitude factor. We caution, though, that this does not imply a linear mapping between the model parameters and the statistics themselves. If this were the case, then there would be a perfect mapping between the thirty model parameters and the amplitudes which were fit in the above ordinary least squares. We do not find this to be the case. Instead, \textit{the resulting field statistics}, which arise from nonlinear effects induced by the parameters, can be written as a linear model.

To generalize the model across the entire prior, however, we want the amplitudes as functions of measurable quantities or simulation parameters rather than a lookup table. To do this, we need a transformation that converts simulation information into amplitude factors. A neural network is one option; however, here we opt for the simpler and more interpretable approach of a linear model
\begin{equation}\label{eq:amp_latent}
\alpha_k(\bm{\lambda}) = \bm{M}_k\cdot(\bm{\lambda}-\bar{\bm{\lambda}}),
\end{equation}
where $\bm{\lambda}$ is a latent vector of simulation-level properties extracted from each simulation in the Sobol set (discussed in depth in \S~\ref{sec:latent_search}), $\bar{\bm{\lambda}}$ is the Sobol-mean latent vector of $\lambda$, and $\bm{M}_k$ is a transformation matrix. where k indexes the components of the latent vector to be discussed in the following section. For $\bm{M}$, we stack all latent vectors $\lambda$ and fitted amplitudes $a_r$ from each simulation to generate $\bm{\Lambda}$ and $\bm{A}$, respectively, and again use least squares to solve for the linear transformation,  

\begin{equation}\label{eq:Mfit}
\bm{M} = (\bm{\Lambda}^\top\bm{\Lambda})^{-1}\bm{\Lambda}^\top \bm{A}.
\end{equation}

The full model is then a three-stage chain. We first extract latent quantities from each simulation (see the following sections for details). We then use these latents to find amplitude values, which we multiply into our basis vectors to obtain our statistics.

\subsection{Searching for the latents}\label{sec:latent_search}
Now the question becomes, \textit{what properties should we extract from each simulation to form the $\lambda$ vector?} The argument above rests on the idea that there exists some linear mapping between the simulations and the resulting statistics we care about. In \citet{vanDaalen-2020}, the authors find strong correlations between power-spectrum suppression and the baryon fraction of group-scale halos, and the trend in weak-lensing two-point statistics also suggests a potential linear mapping between halo quantities and weak-lensing statistics. Subsequent work has made this mapping quantitative, either by predicting the suppression directly from measured baryon fractions and cluster gas properties \citep{Salcido-2023, vanLoon-2024, Grandis-2024, vanDaalen-2025} or by learning it from halo gas observables in the CAMELS suites \citep{Delgado-2023, Pandey-2023}. 

Because of this, we explore whether a linear mapping exists between some set of halo-level quantities and our weak lensing statistics. However, instead of looking only for the set of quantities that work for the power spectrum, we aim to find the full set that works for all statistics. In our suite, we have access to various halo-level observable proxies. For example, X-ray and tSZ/kSZ campaigns can provide access to halo group and cluster temperatures, pressures, Compton $Y$--$M$ relations, gas masses, and baryon fractions, and photometric surveys can probe stellar masses for massive halos. We then seek the best combination of observable halo-level properties to define each simulation's $\lambda$.

Rather than cherry-picking halo properties and potentially introducing selection bias, we let the data from the generated halos choose. From each simulation's halo catalog, we build a library of candidate summaries, listed in Table~\ref{tab:library}. The library covers the gas budget (the baryon, stellar, and gas fractions), its arrangement (the gas concentration), its thermal state (temperature, entropy, pressure, and integrated Compton parameter, each also in self-similar-scaled form), and its halo-to-halo scatter, every one measured in seven evenly log-spaced mass bins from groups to clusters, together with the cross-bin mass-trend slopes of the fractions and of $T$, $P_e$, and $Y$, for ninety candidates in total.

\begin{deluxetable*}{llll}
\tabletypesize{\footnotesize}
\tablecaption{The candidate library used for connecting amplitudes in Eq.~\ref{eq:analytic_model} to halo quantities measured from generated halos. Twelve quantities are measured within $R_{500c}$ for every halo in seven bins of $\log_{10} M_{500c}/{\rm M}_\odot$ from $13.0$ to $14.9$, and six cross-bin slopes. The first four groups are computed as the median over halos in a mass bin, the scatter group as the 16th--84th percentile half-width across halos in the mass bin, and the slopes as linear fits of the bin medians against $\log_{10} M_{500c}$ across the seven bins. Subscripts ${\rm mw}$ denote mass weighting within the aperture, and $Y$ is integrated within it. The last column gives the bin at which the search of \S~\ref{sec:latent_search} selected the quantity (Fig.~\ref{fig:fm_search_path}). \label{tab:library}}
\tablewidth{0pt}
\tablehead{\colhead{Symbol} & \colhead{Quantity} & \colhead{Definition} &
            \colhead{Picked}}
\startdata
\sidehead{\textit{Mass fractions}}
$f_{\rm bar}$ & baryon fraction & $(M_{\rm bar}/M_{\rm tot})/(\Omega_b/\Omega_m)$ &  $[13.0]$ \\
$f_\star$ & stellar fraction & $(M_\star/M_{\rm tot})/(\Omega_b/\Omega_m)$ &  -- \\
$f_{\rm g}$ & gas fraction & $(M_{\rm gas}/M_{\rm tot})/(\Omega_b/\Omega_m)$ &  -- \\
\sidehead{\textit{Structure}}
$c_{\rm gas}$ & gas concentration & $M_{{\rm gas},500c}/M_{{\rm gas},200c}$ &  -- \\
\sidehead{\textit{Thermodynamics}}
$T$ & temperature & $\log_{10} T_{\rm mw}$ &  $[13.2]$ \\
$K$ & entropy & $\log_{10} K_{\rm mw}$ &  -- \\
$P_e$ & electron pressure & $\log_{10} P_{e,{\rm mw}}$ &  $[14.0]$ \\
$Y$ & Compton parameter & $\log_{10} Y$ &  -- \\
\sidehead{\textit{Self-similar scaled}}
$T_{\rm ss}$ & scaled temperature & $\log_{10}\,(T_{\rm mw}/M_{500c}^{2/3})$ &  -- \\
$Y_{\rm ss}$ & scaled Compton parameter & $\log_{10}\,(Y/M_{500c}^{5/3})$ &  $[13.4]$ \\
\sidehead{\textit{Scatter}}
$\sigma(f_{\rm bar})$ & scatter of baryon fraction & half-width of $f_{\rm bar}$ &  $[13.2]$, $[14.0]$ \\
$\sigma(Y)$ & scatter of Compton parameter & half-width of $\log_{10} Y$ &  -- \\
\sidehead{\textit{Cross-bin slopes}}
${\rm d}f_{\rm bar}/{\rm d}\log M$ & slope of baryon fraction & ${\rm d}f_{\rm bar}/{\rm d}\log_{10} M_{500c}$ &  -- \\
${\rm d}f_\star/{\rm d}\log M$ & slope of stellar fraction & ${\rm d}f_\star/{\rm d}\log_{10} M_{500c}$ & yes \\
${\rm d}f_{\rm g}/{\rm d}\log M$ & slope of gas fraction & ${\rm d}f_{\rm g}/{\rm d}\log_{10} M_{500c}$ &  -- \\
${\rm d}\log T/{\rm d}\log M$ & slope of temperature & ${\rm d}\log_{10} T_{\rm mw}/{\rm d}\log_{10} M_{500c}$ & -- \\
${\rm d}\log P_e/{\rm d}\log M$ & slope of electron pressure & ${\rm d}\log_{10} P_{e,{\rm mw}}/{\rm d}\log_{10} M_{500c}$ & -- \\
${\rm d}\log Y/{\rm d}\log M$ & slope of Compton parameter & ${\rm d}\log_{10} Y/{\rm d}\log_{10} M_{500c}$ & -- \\
\enddata
\end{deluxetable*}

We then try to build the candidate set of latents from this catalog. To determine whether a set of latents is actually good, we need a test set the model has not seen during training. We randomly assign the 256 Sobol simulations to five groups (five-fold cross-validation). For each group in turn, the latent map $\bm{M}$ of Eq.~\ref{eq:Mfit} is fit on the other four groups' latent vectors, and the held-out group's statistics are predicted from its measured latents alone. (The basis and the per-run amplitude fits of Eq.~\ref{eq:amp-objective} involve no latents, so only $\bm{M}$ needs to be held out.) The score of a candidate set is then built from these held-out predictions via an $R^2$ per statistic. The set's total score is the average $R^2$ over all statistics.

This procedure produces a noisy score because it depends on how the simulations were assigned to the five groups. We measure that noise by re-drawing the random fold partition twenty times and recomputing the score each time. We find that the resulting scatter in the score is $6\times10^{-4}$, which we treat as one standard deviation ($\sigma$). We use this level to set the acceptance criterion for our search, admitting a candidate latent into the group only if it raises the pooled score by more than roughly $3\sigma$, or $2\times10^{-3}$. An accepted candidate then confidently increases the group score beyond what chance would expect at the $3\sigma$ level.

We start the latent search with an empty set, which grows one candidate at a time. At each step, we try every remaining candidate, score it as a trial addition, and keep the best candidate if it clears the acceptance criterion. After each addition, we also allow the reverse move. If later picks make an earlier pick redundant (the group without it outscores the best group of that size found so far), we kick it out. This removal can temporarily lower the score, but it allows the search to grow from a better base. The search ends when no remaining candidate clears the threshold. We score it on its own five-fold grouping of simulations, independent of the folds used for every accuracy we report. This search score covers all the statistics we wish to emulate, namely, all the weak lensing statistics in \S~\ref{sec:val_map}. So, latents accepted into the group must truly influence all associated statistics.
\begin{figure}
    \centering
    \includegraphics[width=\linewidth]{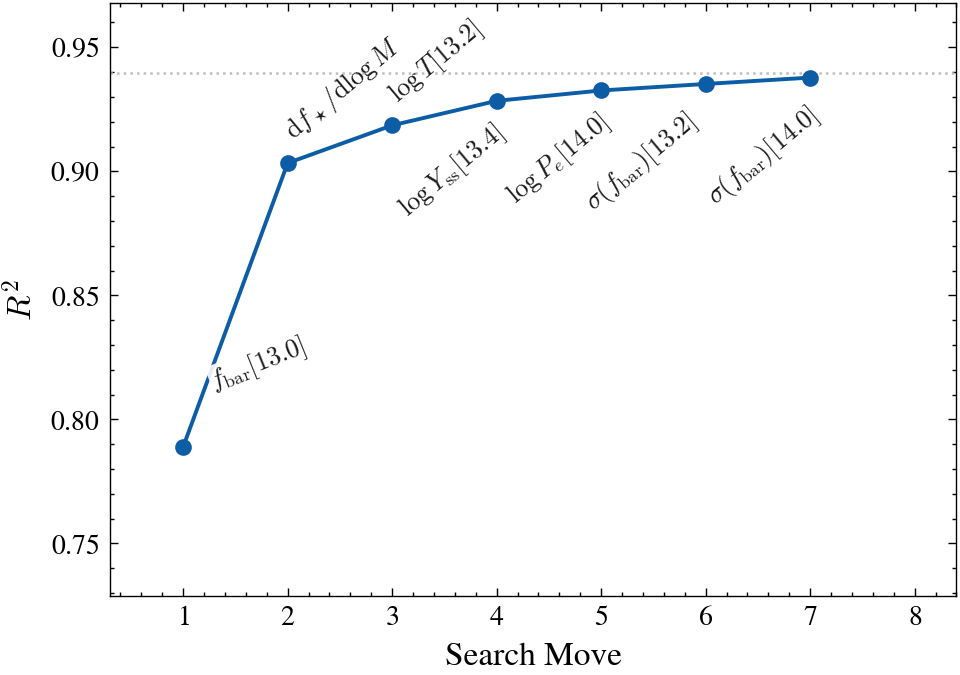}
\caption{The path taken by the search to find halo quantities in Table~\ref{tab:library} that map to the amplitudes of Eq.~\ref{eq:analytic_model}. The x-axis is the search step, and the y-axis is the mean value of $R^2$ over all WL statistics after adding a given member to the set. These members are listed next to the points along the lines. The search terminates at the dotted horizontal line, where the best remaining candidate would improve the $R^2$ below the $3\sigma$ acceptance threshold of $2\times10^{-3}$.}
    \label{fig:fm_search_path}
\end{figure}

After this procedure, we find a set of seven accepted latent halo quantities (Fig.~\ref{fig:fm_search_path}). In order, their choices tell a physical story in concordance with \S\S~\ref{sec:correlations}--\ref{sec:families}. The first pick is the baryon fraction of low-mass groups ($13.0 \le \log_{10} M < 13.2$). With no guidance and ninety candidates, the search rediscovers the same variable as \citet{vanDaalen-2020} as the single most informative halo property for the lensing statistics. This mass budget controls how much gas feedback has left $R_{500c}$ at the group scale. This is also in agreement with where the redistribution families C3 and C4 move gas across and where the halo family C2 acts, as discussed in \S~\ref{sec:families} and in Fig.~\ref{fig:families_cl}. 

The second pick is the stellar fraction mass-trend slope, ${\rm d}f_\star/{\rm d}\log M$. This quantity records whether feedback has preferentially stripped low-mass halos or high-mass ones, which makes sense, as the shape of the stellar-to-halo mass function is controlled by how well stellar winds quench the lower mass range and the AGN sector the higher masses. Between these two numbers, $f_{\rm bar}$ and ${\rm d}f_\star/{\rm d}\log M$ capture $0.90$ of the variance in every weak lensing statistic.

The next three picks are thermodynamic quantities at increasing masses. The group temperature ($13.2 \le \log_{10} M < 13.4$), the self-similar-scaled Compton $Y$ one mass bin higher, and the cluster electron pressure ($14.0 \le \log_{10} M < 14.3$) together add only $+0.03$ to the $R^2$. Once the baryon fraction and the stellar partition are determined, the gas temperature remains, and these three thermodynamic quantities sample it from groups to clusters. $Y_{\rm ss} = Y_{500c}/M_{500c}^{5/3}$ is the informative form here rather than $Y$ itself, because dividing out the self-similar mass scaling leaves only the departure from gravitational heating, which is more reflective of the thermal energy that feedback has added. 

The final two picks are the halo-to-halo scatter of the baryon fraction at fixed mass, $\sigma(f_{\rm bar})$, at the group scale and the cluster scale, each adding an even smaller $+0.003$ to the $R^2$ score. The median gas content of halos does not capture the full picture, and apparently, how these quantities vary halo to halo at fixed mass also controls the suppression.

The final set (rightmost column of Table~\ref{tab:library}) then spans four physical axes, in order of importance: the baryon budget, the stellar partition across halo mass, the thermal state from groups to clusters, and the halo-to-halo scatter.

\subsection{Halo properties, not parameters}\label{sec:why_latents}
At this point, it is fair to ask why we went through the latent search at all and why we have not just built a mapping between the 30 IllustrisTNG parameters and the amplitude vectors $a_r$. Nothing in the math prevents this, as Eq.~\ref{eq:Mfit} could take in the 30 simulation parameters for the latents (such that $\bm{\Lambda}$ becomes the $256\times30$ design matrix) and the model would become an emulator that goes from parameters as input to the statistics as an output. 

We avoid this for a few reasons. First, the 30 parameters used in this work exist only inside IllustrisTNG. So a model that requires the user to define 30 subgrid parameters to generate a weak lensing statistic is a statement about one code and offers little physical insight, whereas halo quantities can be measured in any simulation and, in principle, in data. So, from the physical reasoning introduced with the family basis functions to the measured halo quantities for the amplitudes, the model presented here is directly connected to the physics driving changes in the lensing statistics, rather than statements about the IllustrisTNG subgrid prescription (see \S~\ref{sec:story} for more details on our hypotheses of the underlying physics). The second, more pragmatic reason is that using the 30 parameters as latents will not work, and we tested this. A linear map from the parameters to the amplitudes following Eq.~\ref{eq:amp_latent} reaches only $R^2 \simeq 0.6$ for $S(\ell)$, and replacing the least squares approach with progressively more flexible regressions (gradient-boosted trees, Gaussian processes, deep-network ensembles) moves it no further. In comparison, as shown in \S~\ref{sec:latent_search} the mapping from the seven measured halo numbers to the amplitudes with least squares reaches $R^2=0.96$ under a purely linear map.

It is worth considering why we can achieve this result. The linear map between parameters and the amplitudes fails because the mapping between the 30 parameters and the statistics is non-linear. The attempt at building flexible regressions fails because training on 256 parameter locations in a 30-dimensional parameter space provides too few samples to learn a non-linear map. \textsc{BIND} is the tool that was created to capture this mapping, and it can do so because it learns it at the halo level, from roughly $3.6\times10^{5}$ projected halo maps drawn from over the set of SB35 simulations \citep{Lee-2026b}, rather than from 256 lightcone summaries. The halo quantities are the output of that non-linear step, and what remains between them and the statistics is a seven-dimensional linear map, which is why ordinary least squares works at such high fidelity. Because $\bm{\lambda}$ is measured rather than looked up, the model carries no reference to the simulation parameters. Measure seven numbers in any simulation (or observations!) whose halos fall inside the design, and Eqs.~\ref{eq:amp_latent}--\ref{eq:Mfit} return every statistic. Constraining the same seven numbers from X-ray and SZ or other multiwavelength observations is a natural application \citep[cf.][]{Schneider-2022, Grandis-2024, Pandey-2025}, and we believe this is a promising direction, though real data would add hydrostatic mass bias, projection effects, and many other considerations that we save for future work.

\subsection{Accuracy and validation}\label{sec:family_validation}

\begin{figure*}
    \centering
    \includegraphics[width=\linewidth]{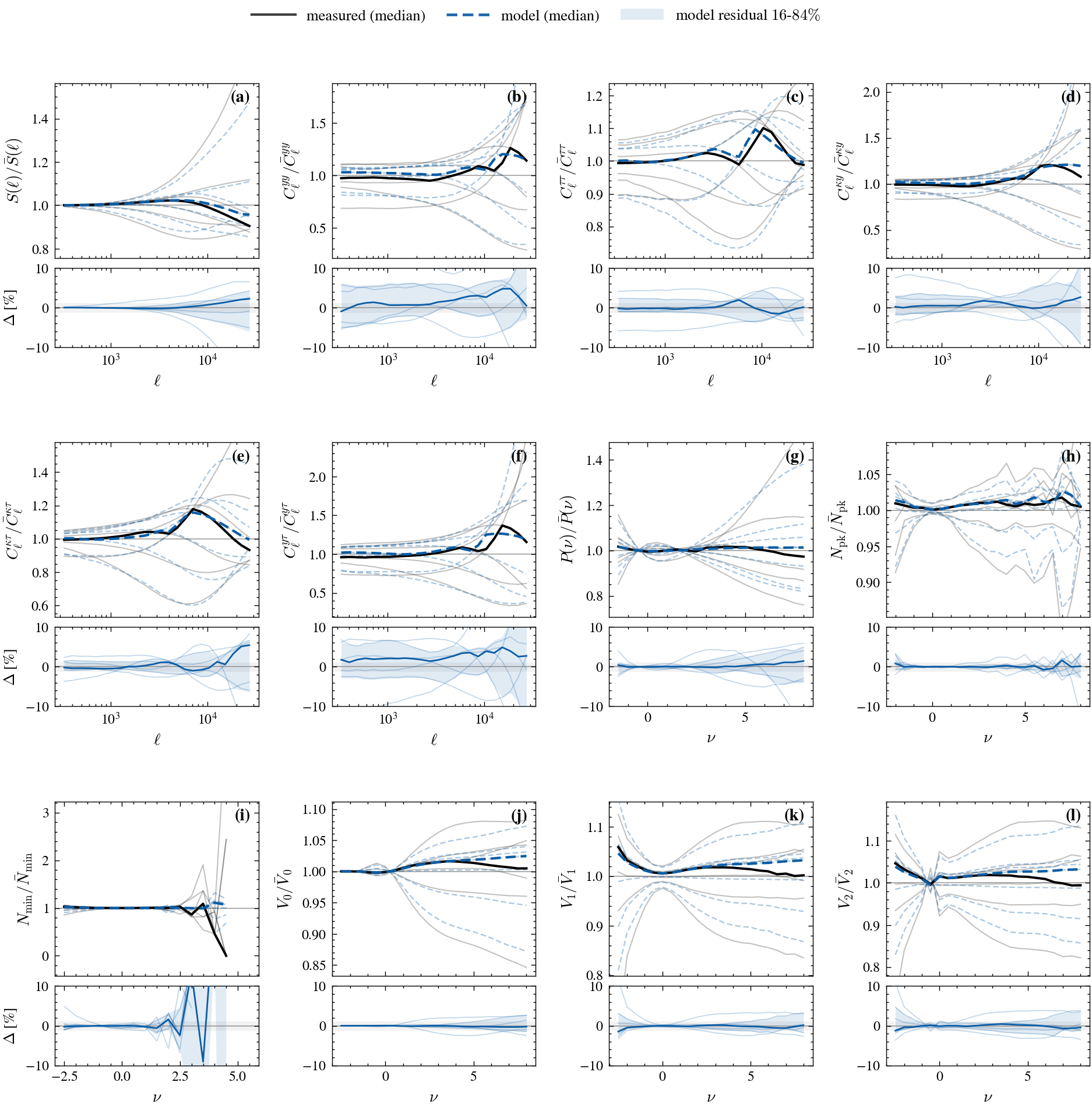}
\caption{The linear model of Eq.~\ref{eq:analytic_model} with halo quantities of Table~\ref{tab:library} (rightmost column) used to compute the linear amplitudes for all twelve statistics. We compare the model predictions against a set of seven held-out simulations normalized by the mean to highlight typical deviations for a given statistic. Thin lines are the individual held-out nodes (gray, measured; blue dashed, model), heavy lines are the median over the seven nodes (black, measured; blue dashed, model), and the shaded band in each strip is the 16th--84th percentile range of the per-node model residuals. \textit{(a)} $S(\ell)$; \textit{(b)--(f)} the gas auto- and cross-spectra $C_\ell^{yy}$, $C_\ell^{\tau\tau}$, $C_\ell^{\kappa y}$, $C_\ell^{\kappa\tau}$, and $C_\ell^{y\tau}$; \textit{(g)} the convergence PDF; \textit{(h)} peak counts; \textit{(i)} minimum counts; \textit{(j)--(l)} the Minkowski functionals $V_0$, $V_1$, and $V_2$.}
    \label{fig:family_model_curves}
\end{figure*}

Table~\ref{tab:family_model} collects the model's accuracy for the twelve statistics, the seven weak-lensing statistics, and the five gas auto- and cross-spectra. The ceiling $R^2$ is the accuracy ceiling indicating how much of the 256-simulation variation the basis can absorb when Eq.~\ref{eq:amp-objective} fits each run's amplitudes freely. At $0.86$--$0.9997$ for the weak-lensing statistics, the handful of family shapes is not the limitation for $S(\ell)$, the PDF, or the Minkowski functionals. The end-to-end ${\rm CV}\,R^2$ corresponds to the models where the amplitudes are computed by fitting to the latents following Eq.~\ref{eq:Mfit}. This procedure involves measuring $\bm{\lambda}$ from the halos themselves, applying the $\bm{\lambda}\to\bm{a}$ map fit with Eq.~\ref{eq:Mfit}, and comparing the predicted curve to the held-out test set of Sobol nodes. For the weak-lensing statistics, this reaches $0.79$--$0.96$, or $92$--$98\%$ of each statistic's ceiling, so the gap between those two columns corresponding to the information the $\bm{\lambda}\to\bm{a}$ map fails to capture is small everywhere. The count statistics sit at the bottom of both columns, so their limitation lies in the family basis itself: some residual nonlinearity or rare-pixel sensitivity that the basis does not span. The latents are not the bottleneck there. The last column repeats the end-to-end fit with the 30 simulation parameters in place of the halo quantity latents. It reaches $R^2=0.39$--$0.54$ for the weak-lensing statistics, roughly half of the halo quantity latent column for every statistic, and peaks and minima sit at the bottom of it as well. 

\begin{table}
\centering
\caption{Family-basis model accuracy per statistic. $K$ is the number of basis
families and ceiling $R^2$ is the free-amplitude ceiling. The last two columns
are end-to-end five-fold cross-validated accuracies, with the seven measured
halo latents $\bm{\lambda}$ and, on the same folds, with the 30 simulation
parameters $\bm{\theta}$ in their place. Each statistic uses its own family
basis, so $K$ varies across statistics. The pooled rows are computed over all
bins of the statistics in the block at once.}
\label{tab:family_model}
\begin{tabular}{l|cccc}
\hline
 & & ceiling & \multicolumn{2}{c}{end-to-end CV $R^2$} \\
\cline{4-5}
statistic & $K$ & $R^2$ & $\bm{\lambda}$ & $\bm{\theta}$ \\
\hline
\hline
\multicolumn{5}{l}{\textit{weak lensing}} \\
\hline
$S(\ell)$          & 4 & 0.9997 & 0.963 & 0.528 \\
PDF       & 4 & 0.985  & 0.939 & 0.498 \\
peaks              & 5 & 0.857  & 0.787 & 0.385 \\
minima             & 4 & 0.908  & 0.838 & 0.415 \\
$V_0$              & 3 & 0.9994 & 0.960 & 0.535 \\
$V_1$              & 2 & 0.986  & 0.946 & 0.457 \\
$V_2$              & 1 & 0.985  & 0.964 & 0.481 \\
\hline
pooled             &   &        & 0.914 & 0.471 \\
\hline
\hline
\multicolumn{5}{l}{\textit{gas auto- and cross-spectra}} \\
\hline
$C_\ell^{yy}$          & 2 & 0.991  & 0.862 & 0.494 \\
$C_\ell^{\tau\tau}$    & 3 & 0.997  & 0.907 & 0.564 \\
$C_\ell^{\kappa y}$    & 2 & 0.996  & 0.946 & 0.551 \\
$C_\ell^{\kappa\tau}$  & 3 & 0.990  & 0.936 & 0.535 \\
$C_\ell^{y\tau}$       & 2 & 0.992  & 0.865 & 0.526 \\
\hline
pooled                 &   &        & 0.903 & 0.534 \\
\hline
\end{tabular}
\end{table}

Fig.~\ref{fig:family_model_curves} shows the model directly against held-out measurements for all twelve statistics. For seven Sobol simulations spanning the design space, including the strongest suppression and strongest enhancement, we refit the $\bm{\lambda}\to\bm{a}$ map excluding the shown simulations and predict every statistic using only their measured halo numbers. Because the raw statistics span several decades and a few-percent feedback response would be invisible, every panel divides both the measurement and the prediction by a common reference, the Sobol design mean, so the panels display the response each node carries and the residual strips read in the same units. The thin lines are the individual held-out nodes, the heavy lines are the median over the seven, and the band in each residual strip is the 16th--84th percentile range of the per-node residuals. The five gas spectra were not part of the search objective in \S~\ref{sec:latent_search}, so their panels are truly a test of the latents rather than a fit to them.

Fig.~\ref{fig:family_model_curves} shows that the model tracks the sign, the shape, and the ordering of the held-out responses in every panel. For the weak-lensing statistics, the residual band sits within $\pm1$--$3\%$ wherever the reference is well measured, against node responses of up to $10$--$50\%$, and grows only at the edges: to about $\pm5\%$ for $S(\ell)$ at $\ell\gtrsim10^4$, to a few percent at $\nu\gtrsim5$ for the PDF and the Minkowski functionals, and at $\nu\lesssim-2$ for $V_1$ and $V_2$, where the reference itself is small. The minima are the one place the comparison breaks down, beyond $\nu\simeq2.5$, where the reference falls to a handful of counts per map and the ratio stops being meaningful. The growth of the $S(\ell)$ residual at $\ell\gtrsim10^4$ is where $b_3$ and $b_4$ finally separate (Fig.~\ref{fig:basis}); over the rest of the range the two are nearly degenerate, so the combination of amplitudes that distinguishes them is set by a handful of small-scale bins and is the hardest thing in the model to predict.

The gas spectra are a harder test. Their responses across the design are of order unity rather than tens of percent (\S~\ref{sec:astro}), and the residual band is correspondingly two to three times wider, $\pm2$--$5\%$ at $\ell\lesssim3\times10^3$ and $\pm5$--$10\%$ at $\ell\gtrsim10^4$, widest for the three $y$ spectra ($C_\ell^{yy}$, $C_\ell^{\kappa y}$, $C_\ell^{y\tau}$) that the rarest, most massive halos dominate. The optical-depth spectra carry the most scale structure in the set, with a rise to the one-halo peak near $\ell\simeq10^4$ and the return to the mean beyond it, and the model reproduces that shape in $C_\ell^{\tau\tau}$ and $C_\ell^{\kappa\tau}$ with the peak height recovered to within $2$--$5\%$. Across all twelve panels, the largest residuals occur at the most extreme nodes, and in nearly every case the model pulls the extreme node toward the design mean.

The model mapping is linear, and the design is the SB35 astrophysical prior (the CAMELS-IllustrisTNG prior of \citealt{Genel-2026}, for which we vary the 30 galaxy formation parameters), so $\bm{\lambda}$ values far outside the design ranges extrapolate unphysically. The cosmology is fixed to that of TNG300, and the basis inherits the noise of the 1P suite from which it is built. We ship the model as a single table of curves and matrices ($\overline{S}$, $\hat B_k$, $\bm{M}$, $\bar{\bm{\lambda}}$, and the per-bin predictive scatter $\sigma_{\rm pred}$) with a numpy-only loader.\footnote{Release DOI to be inserted at acceptance; see \S~\ref{sec:conclusions}.}

\section{The story of the statistics as told by the parameters}\label{sec:story}

The preceding sections presented three distinct descriptions of the lensing and CMB maps, along with their dependence on the IllustrisTNG model parameters, by posing three questions.
\begin{enumerate}
\item \S~\ref{sec:correlations} asked which parameters move which statistics. The peak-correlation grid in Fig.~\ref{fig:corr_matrix} answered that the $\ell$-domain statistics are more sensitive to the wind parameters, while the $\nu$-domain morphological statistics are more sensitive to the IMF slope and the AGN parameters.
\item \S~\ref{sec:families} asked what shape each statistic's response to the parameters takes, and found four families of shapes describing these responses in Fig.~\ref{fig:families_cl}.
\item \S~\ref{sec:latent_search} asked which halo properties carry the amplitudes of the response, and found that seven halo quantities, chosen by a search that knew nothing about the parameters, suffice.
\end{enumerate}
In this section, we attempt to tell a cohesive story that ties these three descriptions together and propose a physical narrative to connect them.

We want to be upfront about these next sections. The physics of galaxy formation, and the IllustrisTNG model used in this work, is complex, and its effects on the summary statistics are nonlinear and coupled (as shown by the $R^2$ score in the rightmost column of Table~\ref{tab:family_model}). A complete account of the physical effects and statistical intricacies lies beyond the scope of this work and will require targeted investigation. But the three constructions above provide empirical constraints on which physical processes could drive the effects. The following subsections present a physical narrative we have found that satisfies these empirical constraints.

\subsection{A gas budget and a thermal state}\label{sec:story_winds}

We start by examining the gas fields, $\tau$ and Compton-$y$, because they respond differently to gas density. $\tau$ maps are formed simply by the electron density traced along the line of sight (Eq.~\ref{eq:tau}), but pixels in the Compton-$y$ maps are proportional to the temperature-weighted density (the integrated pressure, Eq.~\ref{eq:compy}). Therefore, a parameter that changes only the \emph{amount} of gas would shift the two spectra in the same direction by moving the density up or down.

In Fig.~\ref{fig:corr_matrix}, many parameters do exactly this. For example, the wind energy $\bar{e}_w$ (WindEnergyIn1e51erg) moves both spectra in the same direction and with roughly the same magnitude, $\rho=+0.54$ and $\rho=+0.42$. The wind velocity factor $\kappa_w$ (VarWindVelFactor) does not. Instead, it produces the strongest peak correlation in the grid, $\rho=-0.68$ for $C_\ell^{\tau\tau}$, and a nearly equal and opposite peak correlation for $C_\ell^{yy}$, $\rho=+0.60$. The only thermodynamic quantity that can separate a density-weighted field from a pressure-weighted one is the temperature, so we take this to mean that $\bar{e}_w$ sets how much gas the halo holds while $\kappa_w$ sets how hot that gas is.

This is what we would expect from the IllustrisTNG wind model, in which the mass loading factor scales as \citep{Pillepich-2018}
\begin{equation}\label{eq:mass_loading}
\eta_w \propto \bar{e}_w/\kappa_w^{2}.
\end{equation}
So the two wind parameters change the mass loading in different ways. Raising $\bar{e}_w$ loads more gas into winds at the same speed, lifting gas out of the galaxy but not out of the halo. As a result, the reservoir feeding the central black hole shrinks, so the black hole never grows large enough to drive feedback that escapes the halo, and the halo keeps its gas. Both $\tau$ and $y$ grow, which is reflected in the pair of positive correlation cells in Fig.~\ref{fig:corr_matrix}.

Raising $\kappa_w$ instead lowers the mass loading and makes the winds faster and lighter. What they launch is carried out of the halo core to large radii, and when the wind recouples to the gas, it thermalizes. The density-weighted $\tau$ loses power where the gas used to be, while the pressure-weighted $y$, which is also sensitive to the thermal state of the gas, gains power wherever the gas ends up. The gas-spectrum families in Appendix~\ref{app:figs} show the same thing as a function of $\ell$: the Compton-$y$ response to $\kappa_w$ is effectively scale-independent, because the gas moved out of the core has been balanced by heating (Fig.~\ref{fig:gas_families}).

The recoupling density $n_{w,\rm rec}$ (WindFreeTravelDens) offers a similar story. A wind particle recouples with the medium once the gas density falls below the threshold set by $n_{w,\rm rec}$, so raising it makes the wind recouple sooner and deposit its mass closer to where it was launched. Every peak correlation between the statistics and $n_{w,\rm rec}$ in Fig.~\ref{fig:corr_matrix} is positive, indicating that gas is retained as the recoupling density is raised.

In this interpretation, the halo properties needed to predict all the statistics should include quantities that reflect the gas budget and thermal state of halos. Indeed, this is exactly what the search in \S~\ref{sec:latent_search} chooses. The first pick was the baryon fraction of low-mass groups, the gas budget, which alone accounts for $R^2=0.79$ of the pooled variance of the statistics and rediscovers the quantity \citet{vanDaalen-2020} identified for the power spectrum. The search's third, fourth, and fifth picks were thermal-state quantities: the group temperature, the self-similar-scaled Compton $Y$, and the cluster electron pressure.

\subsection{Where the gas ends up}\label{sec:story_radius}

The gas budget and its thermal state tell us how much gas there is and how hot it is, but the response families in Fig.~\ref{fig:families_cl} show where this gas has been redistributed. Viewed this way, the wind-sector parameters split into three families.

The wind energy $\bar{e}_w$ and every parameter that modulates it (the specific momentum and the three energy-reduction parameters) form C3 by themselves. C3 peaks at the halo scale, crosses zero near $\ell\simeq1.5\times10^4$, and is negative at the smallest scales. With $\rho=+0.32$ for $\bar{e}_w$ on $S(\ell)$ in Fig.~\ref{fig:corr_matrix}, we read this to mean that gas is lifted out of the galaxy and deposited in the circumgalactic medium, and the zero crossing in the C3 panel marks the radius that separates where the gas is taken from and where it is put.

The recoupling density $n_{w,\rm rec}$ sits in C4 with the thermal wind fraction $\tau_w$ and the SNII minimum mass $M_{\rm SNII,\,min}$. C4 is nearly the inverse of C3: its peak is at the smallest scales rather than the halo scale, and it crosses zero near $\ell\simeq10^4$. Winds that recouple at higher density deposit their mass closer to where they were launched, so power moves from the halo scale back toward the core. A larger thermal fraction makes the wind slower and more extended, so it recouples sooner; a higher SNII minimum mass means fewer supernovae per unit stellar mass and less wind energy, the inverse of raising $\bar{e}_w$. So C3 acts mainly on the halo scale with a smaller, opposite effect in the core, while C4 acts mainly on the core with a smaller, opposite effect at the halo scale.

The wind velocity factor belongs to C1, whose response vanishes at $\ell\lesssim10^3$, grows through the halo scales, and continues to grow up to the Nyquist limit. So the three wind mechanisms are: how much gas is moved (C3), how far it gets (C4), and how fast it leaves the core (C1). Each has its own characteristic shape, and the family that sets the thermal state shares its shape with the sector that heats the same gas from the inside. That is the family-space version of what the $(y,\tau)$ pair showed in the grid. The winds are not the full story, however; next, we turn to the AGN parameters and how they fit into this interpretation.

\subsection{The engine in the core}\label{sec:story_agn}

We find that the AGN parameters act on the core in two distinct ways. The first is the quasar-mode threshold $\xi_0$ (QuasarThreshold), which moves the peaks and the Minkowski functionals in Fig.~\ref{fig:corr_matrix} at $\rho\simeq-0.3$ and $S(\ell)$ at $-0.26$ and $-0.28$, while sitting at or below the null for every gas spectrum. In IllustrisTNG, this parameter sets the Eddington ratio below which a black hole switches from the thermal quasar mode to the kinetic mode, $\chi_{\rm thresh} = \min[\xi_0\,(M_{\rm BH}/10^{8}\,{\rm M}_\odot)^{\beta_X}, 0.1]$ \citep{Weinberger-2017}, and it is the kinetic mode that quenches star formation and ejects gas from halo centers \citep{Zinger-2020}. Raising the threshold shifts the transition to kinetic-mode feedback to lower black hole masses, allowing more efficient feedback from a halo core at any given time. The expectation is that the stronger feedback suppresses power, peaks, and the other statistics, which is what the negative peak correlations in Fig.~\ref{fig:corr_matrix} show.

The second lever is black hole growth, for example through the radiative efficiency $\epsilon_r$, which lowers the Eddington limit and slows black hole mass growth. The upper bound of $\epsilon_r$ corresponds to smaller black holes in massive halos and weaker integrated AGN feedback. The $\xi_0$ threshold and the $\epsilon_r$ efficiency therefore have nearly opposite effects on the feedback, and they correlate in opposite directions with the statistics in Fig.~\ref{fig:corr_matrix}.

The families do not show the sign of the effects, because in defining them in \S~\ref{sec:families}, each response is normalized by its extremal value so that every curve reaches $+1$ at its peak regardless of sign (Eq.~\ref{eq:rhat}), and the opposite effects of $\xi_0$ and $\epsilon_r$ have the same shape in Fig.~\ref{fig:families_cl}. Both, together with $\beta_X$, reside in C1 alongside $\kappa_w$, with the black hole growth and SNIa parameters as low-amplitude companions. So all four significant members of the C1 core family affect clustering only on the smallest scales, not outside the halo regions.

The two radio-mode AGN parameters, $A_{\rm AGN1}$ and $f_{\rm re}$, land in C2, the halo family, to which the clustering responds most strongly at the halo scale and at a somewhat lower level toward smaller scales.

Parameters that determine whether and how strongly the kinetic mode fires act on the core in C1, while the mode itself reshapes the gas out to larger scales in C2. We think this is also why the thermal latents of \S~\ref{sec:latent_search} climb the mass range, from the group temperature at $10^{13.2}\,{\rm M}_\odot$ to the cluster electron pressure at $10^{14}\,{\rm M}_\odot$: winds set the thermal state of the groups, and black holes set it for the clusters. The same mass split is, we believe, what separates the rows of Fig.~\ref{fig:corr_matrix}. The $\ell$-domain statistics integrate over every halo above $10^{13}\,{\rm M}_\odot\,h^{-1}$ and are dominated by the numerous groups just above that threshold, where the winds set the budget. The high-$\nu$ peaks and the deepest minima are produced by rare massive halos, whose cores, as \citealt{Lee-2026a} showed, source the baryonic response in the morphological statistics, and those cores are where the black hole engine acts. Different statistics listen to different sectors because different sectors act on different halos. The one parameter that shares the radio-mode shape but is not an AGN parameter is the IMF slope, which we discuss next.

\subsection{The stellar mass}\label{sec:story_imf}

The IllustrisTNG model \ citep {Lee-2024} has previously shown that the IMF slope behaves like an AGN parameter \citep{Lee-2024}. The IMF slope controls the number of massive stars, such that a shallower IMF contains more massive stars and produces more supernovae, and one might expect it to act as another wind parameter, excising the core or redistributing matter out to the halo scale. While this is partly true, the IMF slope is not just another wind-energy dial, because it also sets the metallicity of the halo gas, which, through enrichment and cooling, drives the growth of the central black holes \citep{Pillepich-2018, Lee-2024}.

The grid hints at the same thing with the peak correlations of $b_{\rm IMF}$ with the WL peak and minimum counts carrying opposite signs, $-0.53$ and $+0.35$, both above the whole-grid null, whereas the wind parameters move the two counts together, though the minima are the noisiest statistic in the suite, and we do not rest the argument on this.

The C2 family in Fig.~\ref{fig:families_cl} agrees that the IMF slope acts like the radio-mode AGN parameters, on the halo scale and inward, while staying responsive at the smallest scales like the wind parameters. Ahead of every thermodynamic quantity, the latent search of \S~\ref{sec:latent_search} chose the slope of the stellar fraction across halo mass, which records whether feedback has quenched the low-mass halos or the high-mass ones. A parameter that moves the stellar mass of cores through a channel that switches on with black hole mass would show up first in exactly this quantity.

\subsection{What this interpretation explains}\label{sec:story_synthesis}

Taken together, the four families in Fig.~\ref{fig:families_cl} sort the galaxy formation parameters by the radius at which they act: the core (C1), the halo (C2), and the two directions of transfer between them, outward from the core to the halo scale (C3) and back from the halo scale toward the core (C4).

The wind sector supplies one family that determines how much gas moves and one that sets how far it travels before it stops. The AGN sector sets when its engine switches on and how large it grows, and the IMF slope reaches the same core through stellar mass. This is the same structure the analytic model of \S~\ref{sec:emulator} exploits, and it is why the seven halo-quantity latents in Table~\ref{tab:family_model} present themselves as four physical axes, a baryon budget, a stellar partition across mass, a thermal state, and a scatter, rather than as an arbitrary list.

We emphasize that this reading is a hypothesis about IllustrisTNG, assembled from rank correlations, response shapes, and a halo-variable search, and that we have not traced any of its mechanisms in the simulations themselves; doing so is a natural next step. This interpretation explains why a single suppression amplitude cannot describe the parameter space. Two parameters can produce the same suppression while moving the core in opposite directions, and the peaks and Minkowski functionals, whose baryonic response \citet{Lee-2026a} showed is dominated by the cores of massive halos, tell them apart where the power spectrum cannot.

\section{Caveats}\label{sec:caveats}

Throughout this manuscript, we make several approximations, which we state explicitly here. We group these approximations and caveats by whether they concern what the maps physically contain, what the training covers, or what the validation actually establishes.

\subsection{What the maps contain}\label{sec:caveats_maps}

The most important approximation is that we replace halos via our generated model rather than an entire volume. So only regions centered on halos with $M_{200c}\geq10^{13}\,M_\odot\,h^{-1}$ carry generated baryonic fields, and below that threshold, and in the space between halos, the matter field is dark-matter-only from TNG300-Dark. The construction also omits the back-reaction of baryons on the dark matter outside of the painted regions, which can alter the dark matter power spectrum by up to tens of percent at $k\sim10\,h\,{\rm Mpc}^{-1}$ in extreme feedback models \citep{Gebhardt-2026}. As noted in \S~\ref{sec:val_map}, \citet{Lee-2026a} found that replacing halos at and above our threshold of $10^{13}\,M_\odot\,h^{-1}$ accounts for roughly $60$--$65\%$ of the power-spectrum response and for virtually all of the response in the peak counts and Minkowski functionals. By construction, our maps lack the missing $\sim35$--$40\%$ of the two-point response carried by halos below the threshold and by the diffuse field. This is why we compare every claim in this paper against the hydro-pasted control rather than a hydrodynamical simulation. The control shares the same missing physics, so the comparison isolates the generative model's error rather than the architecture's. It is also why the $\tau$ and $y$ maps should be used for responses and aperture measurements centered on halos, rather than for absolute amplitudes on large scales, where the diffuse component we do not model contributes a significant fraction of the signal.

Two smaller geometric approximations follow from the same architecture. Halos whose centers fall near a slab boundary are, in essence, cut off, so part of their gas is assigned to the neighboring shell. This affects a small fraction of halos and enters as noise, since the assignment is uncorrelated with the parameters. Second, \textsc{BIND} is trained on projections of depth $50\,{\rm Mpc}\,h^{-1}$ while the TNG300-Dark slabs are $51.25\,{\rm Mpc}\,h^{-1}$, so the conditioning fields carry about $1\,{\rm Mpc}\,h^{-1}$ more line-of-sight material than the model saw in training. We show in \S~\ref{sec:val_halo} that this does not measurably degrade the generated halos, but it is a mismatch and implies an out-of-distribution usage in this paper.

Finally, we model no velocities. The maps are of the optical depth itself, not of the kinematic Sunyaev-Zel'dovich signal, and converting between the two requires a peculiar velocity field that our pipeline does not supply. Users who need either must supply the velocity assumptions when reducing their data.

\subsection{What the training covers}\label{sec:caveats_training}

This paper only uses the IllustrisTNG galaxy formation model and its parameter space. The four response families, the seven latents, and the linear map between them are properties of that TNG model. A different subgrid family, such as \textsc{simba} \citep{Dave-2019} or \textsc{astrid}, can potentially have a completely different response family and different latents \citep{Delgado-2023, Yang-2026}. We are not claiming a universal law, but instead a model for IllustrisTNG. Expanding BIND to other subgrid parameterizations is an interesting open question, because a compression that survives changes in subgrid prescriptions would be a statement about feedback, whereas in this work we can only make a statement about IllustrisTNG.

The cosmology is also fixed at cosmological parameter constraints from \citealt{planck-2015}. \textsc{BIND} itself is conditioned on cosmological parameters through the SB35 training suite, but every lightcone here is built on TNG300-Dark, which was run at a single cosmology. This is what makes the responses in \S~\ref{sec:astro} independent of cosmology and unambiguous, but we cannot say anything about degeneracies between feedback and cosmological parameters. Those degeneracies are the reason feedback is a systematic in the first place, and quantifying them requires painting across a suite of cosmologies rather than one.

The analytic model inherits a related limit. The map from latents to amplitudes is linear and is fit to the SB35 design, so it is valid only within the prior bounds and nowhere else. A simulation whose measured $\bm{\lambda}$ falls outside the design ranges will extrapolate out of distribution. However, since the Sobol sequence used in this work extends over the full reasonable prior space of IllustrisTNG's model parameter space, we believe extrapolation is already unphysical.

\subsection{What the validation establishes}\label{sec:caveats_validation}

Our fiducial validation in \S~\ref{sec:validation} compares generated halos with hydro-pasted halos and ray-traced maps, measuring \textsc{BIND}'s error rather than the halo-replacement architecture itself, which we characterize separately in \citealt{Lee-2026a}. A full-hydro anchor, in which the same rays are traced through the full TNG300 hydrodynamical simulation, would fold in the halo-replacement error as well. \citealt{Lee-2026a} characterized that error, so we omit the full-hydro comparison here. 

Finally, the validation and the detectability comparison in \S~\ref{sec:val_map} are noiseless. We apply no shape noise, survey mask, photometric redshift uncertainty, or intrinsic alignment contamination, and we measure the covariance across realizations that are rotations and translations of a single lightcone rather than independent volumes. The precision envelopes of \S~\ref{sec:astro} are then statements about statistical power, not forecasts from a systematics-marginalized analysis.

\section{Future outlook}\label{sec:outlook}

The limits above sort naturally into three groups, distinguished by whether they require new analysis, a new model, or new science.

The first group needs neither a new model nor new science. We can add survey realism to the maps we have already released by drawing shape-noise realizations onto the convergence maps, applying a mask, and repeating the detectability comparison in \S~\ref{sec:astro} under tomography. This would convert the statistical-power statement into a forecast, which is a plan for upcoming work. 

The second group requires retraining \textsc{BIND}. The most important next step for \textsc{BIND} is to allow lower masses, which \citet{Lee-2026a} show are relevant for power spectrum measurements. Lowering the mass threshold toward $10^{12}\,M_\odot\,h^{-1}$ would recover most of the suppression that our maps currently miss, but requires thought on how to apply this to $N$-body simulations, since the halo mass function requires a significantly larger population of halos to be generated per snapshot. We envision BIND applied to lower-mass halos as the next iteration. Training the model on alternative subgrid parameter sets individually, such as those from \textsc{simba} or \textsc{astrid}, would test whether the family decomposition and the seven latents are properties of feedback in general or are just specific to IllustrisTNG. This requires a training set of halos from CAMELS at the $50\,{\rm Mpc}\,h^{-1}$ resolution for these subgrid prescriptions. Once available, we will use them to train \textsc{BIND} in this way. Painting onto dark-matter-only simulations run at varied cosmologies would open the feedback-cosmology degeneracies that this work holds fixed and allow for use in cosmological parameter analyses. Finally, moving from projected patches to three-dimensional volumes would remove the projection approximation entirely, allowing for 3D halo shapes, capturing the anisotropic ejection of matter around halos, and enabling a higher-fidelity model, but at a cost of memory and generation time.

The third group is the science the maps were built for. The compression result of \S~\ref{sec:latent_search} identifies seven halo properties that carry feedback's imprint on the lensing sky, and each of them is a measurable quantity from X-ray, SZ, and spectroscopic campaigns. The natural next step is to try to constrain those halo measurables from the existing multiwavelength observations of gas, and propagate them through the analytic model into a prediction for weak lensing statistics \citep[for steps in this direction, see][]{Schneider-2022, Pandey-2023, Ganguly-2026}. Doing this correctly requires many observational considerations, such as hydrostatic mass bias, projection effects, and the possibility that the real universe's subgrid calibration does not match TNG's. For these reasons, we believe this is out of the scope of the current work. However, the machinery is in place, and the maps, statistics, and model tables have been released, so it can be attempted.

\section{Conclusions}\label{sec:conclusions}

We set out to ask what a full galaxy formation parameter space does to the Stage-IV weak lensing and SZ statistics, and if that response could be compressed into measurable quantities. Using \textsc{BIND}, we generated $314$ lightcones spanning the 30-dimensional IllustrisTNG astrophysical prior, comprising the 256-node Sobol sequence, the 57-run 1P atlas, and the fiducial, alongside the DMO and hydro-pasted references. Every lightcone carries $\kappa$, Compton-$y$, and $\tau$ maps at five source redshifts, built from the same halos with the same flow-matching seeds, so that any difference between two lightcones is driven by the parameters alone. To our knowledge, this is the first set of lightcones in which the lensing and the gas observables of every halo respond consistently to a full galaxy formation parameter space, and it is what lets us study feedback effects in a controlled way, and with respect to observational quantities such as lensing power spectra and peak counts.

At the fiducial parameters, the \textsc{BIND}ed maps reproduce the hydro-pasted control set to within the statistical precision of an LSST-Y10-like survey for every weak lensing statistic we measure. The gas spectra are more biased. The optical-depth spectra agree with the control to within about $5\%$, while the spectra that involve Compton-$y$ run from $5$--$20\%$ low at $\ell \lesssim 10^3$ to about $25\%$ high at the smallest scales. The weak lensing maps are therefore usable as they stand, and the gas maps are usable for responses rather than for absolute amplitudes.

Across the prior, Stage-IV surveys can detect feedback, but only on small scales. Over $\ell \simeq 2.2\times10^3$--$1.8\times10^4$ the spread in $S(\ell)$ across the Sobol nodes exceeds the noiseless LSST-Y10 statistical precision by more than $10\sigma$, peaking at $22\sigma$ near $\ell \simeq 5\times10^3$, and $84\%$ of nodes depart from the fiducial by more than $5\sigma$ ($73\%$ for \textit{Euclid}). We also find that the statistics do not all respond in the same ways to the same galaxy formation model parameters. Eleven of the thirty parameters clear their look-elsewhere null for at least one statistic, and the row structure separates, with the $\ell$-domain statistics and every gas spectrum dominated by the galactic wind parameters, while the $\nu$-domain morphological statistics respond instead to the IMF slope and the AGN parameters. Within the wind sector, the wind velocity sets the thermal state of the gas and moves $\tau$ and $y$ in opposite directions, while the wind energy sets the gas budget and moves them together. The thirty 1P response curves of the power spectrum collapse into four families, sorted by the radius at which they act: a core family, a halo family, and two redistribution families that move gas outward or inward across the halo scale.

The central result of this work is that the response is compressible. Each statistic can be modeled as a linear combination of its response families, and a data-driven search over ninety candidate halo summaries then selects seven measured halo properties as the latent variables that set the amplitudes. They sort into four physical categories in order of importance: the baryon fraction of low-mass groups, its partition across halo mass, the thermal state of the gas from groups to clusters, and its halo-to-halo scatter. The search was unguided, and its first pick, the baryon fraction of low-mass groups, essentially rediscovers the \citet{vanDaalen-2020} conclusion, which also anchors observationally calibrated suppression models \citep{Salcido-2023, vanLoon-2024, vanDaalen-2025}. Under a purely linear map, these seven numbers predict every held-out statistic -- the seven weak lensing statistics and the five gas spectra alike -- at ${\rm CV}\,R^2 = 0.79$--$0.96$ (Table~\ref{tab:family_model}). The same linear map from the thirty IllustrisTNG parameters reaches only $0.39$--$0.56$, and more flexible regressions do no better, because 256 nodes in thirty dimensions cannot resolve a non-linear map. \textsc{BIND} is what absorbs that non-linearity, and once it has, what remains is linear in quantities that X-ray and SZ campaigns already measure.

We publicly release the lightcones, the measured statistics, and the model tables, the latter as a single set of curves and matrices ($\overline{S}$, $\hat{B}_k$, $\bm{M}$, $\bar{\bm{\lambda}}$, and the per-bin predictive scatter $\sigma_{\rm pred}$) with a numpy-only loader. We intend for them to be used to test gas-calibrated baryon corrections against a field-level reference, to forecast joint lensing and SZ analyses, and to ask, for any weak lensing statistic, which halo property it is really measuring.

\begin{acknowledgments}
We thank Colin Hill, Carolina Cuesta-Lazaro, Daisuke Nagai, Erwin Lau, and Amanda Lue for useful discussions during this project. MEL is supported by NSF grant DGE-2036197. ZH acknowledges financial support from NASA ATP grant 80NSSC24K1093. The Flatiron Institute is supported by the Simons Foundation.  GLB acknowledges support from the NSF (AST-2307419) and NASA (80NSSC21K1053), as well as support from the Simons Foundation through the Learning the Universe Collaboration. MEL also thanks Erin Walter for helpful comments and editing. The authors used Claude Opus and Sonnet to refine sections of code and text, and Grammarly was used to refine portions of the draft with grammatical edits. The authors take full responsibility for the final content.

\end{acknowledgments}

\begin{contribution}
MEL performed the analysis, code generation, testing, and text writing under the mentorship and support of SG, ZH, GLB and BH.

\end{contribution}

%

\software{astropy \citep{2013A&A...558A..33A,2018AJ....156..123A,2022ApJ...935..167A},
          LensTools \citep{Petri-2016a}
          }


\appendix

\section{Per-statistic response families and model overlays}\label{app:figs}

The main text develops the response-family construction for $C_\ell^{\kappa\kappa}$ (\S~\ref{sec:families}) and validates model in Fig.~\ref{fig:family_model_curves}. This appendix shows alternative per-statistic counterparts for completeness. Showing all for each statistic would be overwhelming for the text, but are available upon request. 

Here we show the family decomposition of the convergence PDF and the response families of the gas spectra, which provide the basis for the model of \S~\ref{sec:emulator} and the held-out model overlays for the six $\ell$-spectra.

\begin{figure*}
\centering
\includegraphics[width=\linewidth]{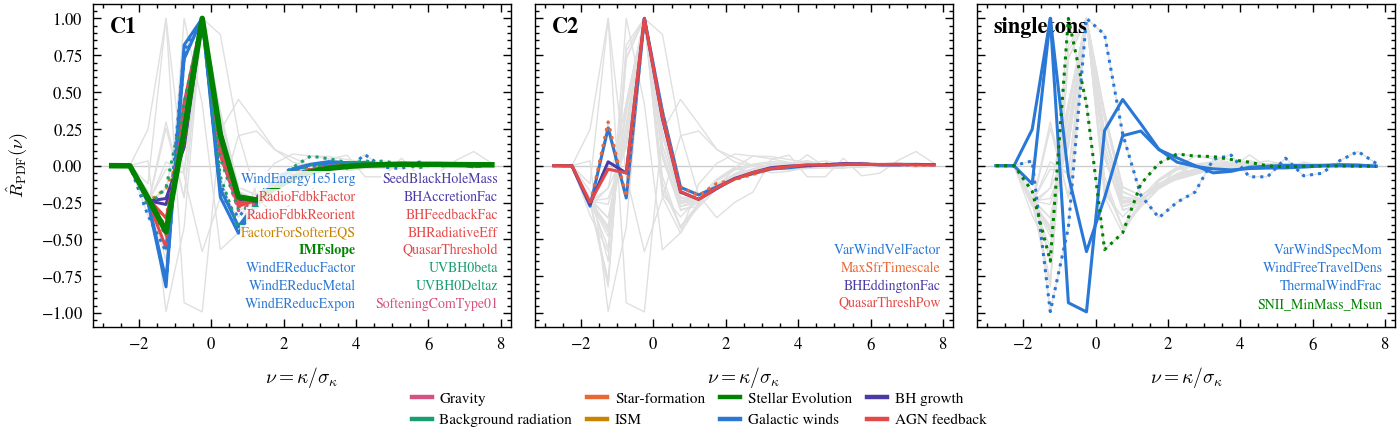}
\caption{Response families of the convergence PDF, constructed exactly as in \S~\ref{sec:families}, where the thirty normalized 1P responses $\hat{R}_{\rm PDF}(\nu)$ (gray in every panel), clustered at $r \geq 0.85$, with each panel's family members colored by their astrophysical sector and solid lines marking parameters that pass the statistic-level noise floor of Fig.~\ref{fig:corr_matrix}. The PDF collapses into two families and four singletons. C1 gathers sixteen parameters spanning AGN feedback, black hole growth, the IMF slope, and the wind energy-reduction sector into a single shape, which provides a sharpening of the peak just below $\nu = 0$ fed by a deficit near $\nu \simeq -1.3$, and C2, containing VarWindVelFactor among four members, carries a variant with a second node in the underdense wing. The singletons are the momentum-driven wind parameters and SNII\_MinMass\_Msun, neither of which clears its noise floor with a coherent shape. That thirty-parameter responses reduce to essentially two shapes is what makes the $K = 4$ PDF basis of Table~\ref{tab:family_model} sufficient.}
\label{fig:pdf_families}
\end{figure*}

Fig.~\ref{fig:pdf_families} shows the families of the convergence PDF. The procedure is identical to that of Fig.~\ref{fig:families_cl} where the responses of Eq.~\ref{eq:response} on the 1P set, normalized through Eq.~\ref{eq:rhat} and clustered at $r \geq 0.85$. The only difference here is the change in statistics. The membership rearranges relative to the power spectrum in the way Fig.~\ref{fig:corr_matrix} anticipates. The AGN, black hole growth, and IMF-slope parameters, which dominate the $\nu$-domain rows of the correlation grid, merge into one large family rather than splitting across several, and the wind sector contributes the second family and the singletons. The shapes say that essentially all of the effective parameters act on the PDF the same way, by narrowing the distribution about its peak, and differ mainly in where in the underdense wing the compensating probability is drawn from.

\begin{figure*}
\centering
\includegraphics[width=\linewidth]{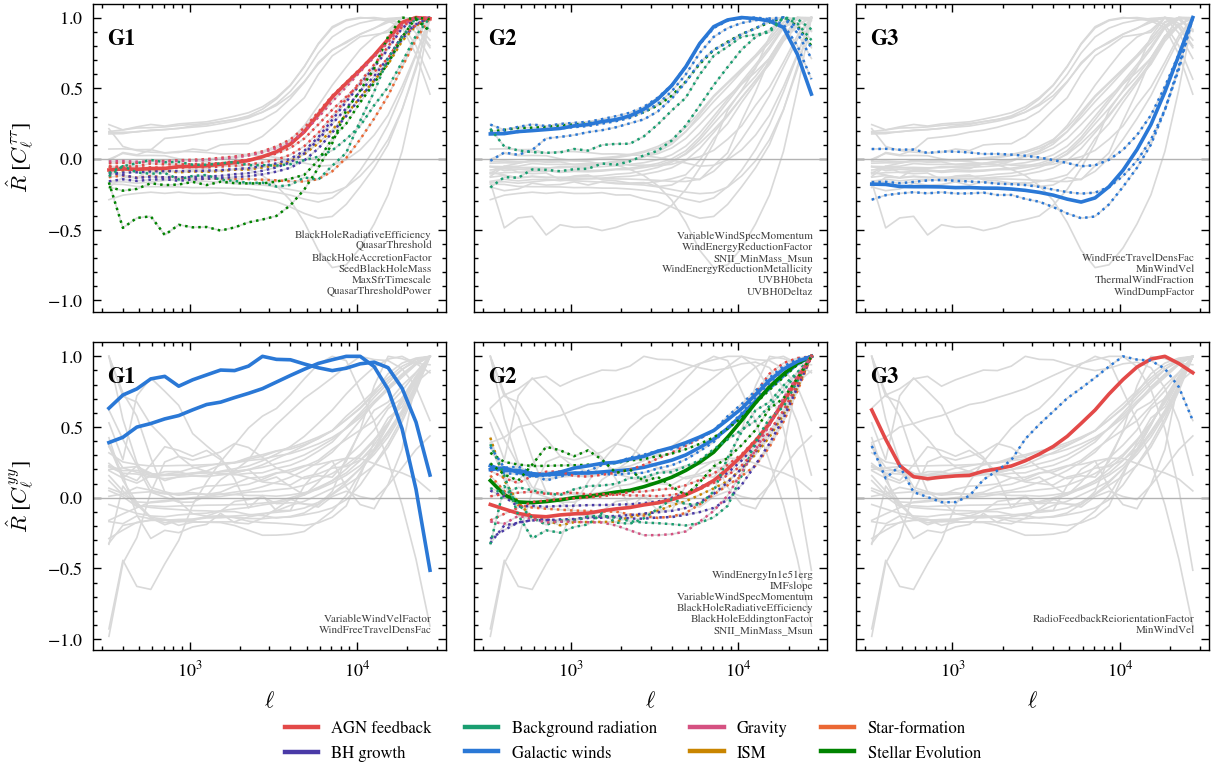}
\caption{Response families of the gas auto-spectra, built in $\ln C_\ell$ from the 1P atlas as described in the text of this appendix. $C_\ell^{\tau\tau}$ (\textit{top row}) and $C_\ell^{yy}$ (\textit{bottom row}), with the same normalization, clustering threshold, sector colors, and line conventions as Fig.~\ref{fig:families_cl}. The $\tau\tau$ families separate by feedback channel. G1 is the black hole growth and AGN sector, G2 is the wind energy-reduction and stellar-evolution parameters together with the UV background, and G3 is the wind travel and velocity parameters, whose response stays flat to $\ell \simeq 10^4$ before rising steeply. For $C_\ell^{yy}$, the two strongest correlates of Fig.~\ref{fig:corr_matrix}, VarWindVelFactor and WindFreeTravelDensFac, form their own family (G1), rising from large scales onward rather than only at high $\ell$. The $\tau\tau$ families are built from the 26 parameters whose 1P $\tau$ planes exist (see text). The cross-spectrum families are constructed identically and are not shown.}
\label{fig:gas_families}
\end{figure*}

Fig.~\ref{fig:gas_families} shows the response families of the two gas auto-spectra that enter the model of \S~\ref{sec:emulator}. Unlike the lensing statistics, these are built in the log, since the responses are multiplicative and of order unity. In all other ways, the construction is that of \S~\ref{sec:families}. The family structure again follows the correlation grid, with the optical-depth spectrum cleanly separating into an AGN family and two wind families, while the Compton-$y$ spectrum promotes the wind-velocity pair that dominates its correlations into a family of its own.

\bibliography{biblio}{}

@ARTICLE{Lu-2023,
       author = {{Lu}, Tianhuan and {Haiman}, Zolt{\'a}n and {Li}, Xiangchong},
        title = "{Cosmological constraints from HSC survey first-year data using deep learning}",
      journal = {\mnras},
         year = 2023,
        month = may,
       volume = {521},
       number = {2},
        pages = {2050-2066},
          doi = {10.1093/mnras/stad686},
archivePrefix = {arXiv},
       eprint = {2301.01354},
 primaryClass = {astro-ph.CO},
       adsurl = {https://ui.adsabs.harvard.edu/abs/2023MNRAS.521.2050L}
}

@ARTICLE{Zinger-2020,
       author = {{Zinger}, Elad and {Pillepich}, Annalisa and {Nelson}, Dylan and {Weinberger}, Rainer and {Pakmor}, R{\"u}diger and {Springel}, Volker and {Hernquist}, Lars and {Marinacci}, Federico and {Vogelsberger}, Mark},
        title = "{Ejective and preventative: the IllustrisTNG black hole feedback and its effects on the thermodynamics of the gas within and around galaxies}",
      journal = {\mnras},
         year = 2020,
        month = nov,
       volume = {499},
       number = {1},
        pages = {768-792},
          doi = {10.1093/mnras/staa2607},
archivePrefix = {arXiv},
       eprint = {2004.06132},
 primaryClass = {astro-ph.GA}
}

@ARTICLE{Pillepich-2018,
        author = {{Pillepich}, Annalisa and {Springel}, Volker and {Nelson}, Dylan and {Genel}, Shy and {Naiman}, Jill and {Pakmor}, R{\"u}diger and {Hernquist}, Lars and {Torrey}, Paul and {Vogelsberger}, Mark and {Weinberger}, Rainer and {Marinacci}, Federico},
         title = "{Simulating galaxy formation with the IllustrisTNG model}",
       journal = {\mnras},
          year = 2018,
         month = jan,
        volume = {473},
        number = {3},
         pages = {4077-4106},
           doi = {10.1093/mnras/stx2656},
 archivePrefix = {arXiv},
        eprint = {1703.02970},
  primaryClass = {astro-ph.GA},
        adsurl = {https://ui.adsabs.harvard.edu/abs/2018MNRAS.473.4077P}
}

@ARTICLE{vanLoon-2024,
       author = {{van Loon}, Maria L. and {van Daalen}, Marcel P.},
        title = "{The contribution of massive haloes to the matter power spectrum in the presence of AGN feedback}",
      journal = {\mnras},
         year = 2024,
       volume = {528},
       number = {3},
        pages = {4623-4642},
          doi = {10.1093/mnras/stae285},
archivePrefix = {arXiv},
       eprint = {2309.06478},
 primaryClass = {astro-ph.CO}
}

@ARTICLE{vanDaalen-2025,
       author = {{van Daalen}, Marcel P. and {Koutalios}, Ioannis and {Broxterman}, Jeger C. and others},
        title = "{The resummation model in FLAMINGO: precisely predicting matter power suppression from observed halo baryon fractions}",
      journal = {arXiv e-prints},
         year = 2025,
        month = sep,
          eid = {arXiv:2509.04552},
        pages = {arXiv:2509.04552},
          doi = {10.48550/arXiv.2509.04552},
archivePrefix = {arXiv},
       eprint = {2509.04552},
 primaryClass = {astro-ph.CO}
}

@ARTICLE{Salcido-2023,
       author = {{Salcido}, Jaime and {McCarthy}, Ian G. and {Kwan}, Juliana and {Upadhye}, Amol and {Font}, Andreea S.},
        title = "{SP(k) -- a hydrodynamical simulation-based model for the impact of baryon physics on the non-linear matter power spectrum}",
      journal = {\mnras},
         year = 2023,
       volume = {523},
       number = {2},
        pages = {2247-2262},
          doi = {10.1093/mnras/stad1474},
archivePrefix = {arXiv},
       eprint = {2305.09710},
 primaryClass = {astro-ph.CO}
}

@ARTICLE{Delgado-2023,
       author = {{Delgado}, Ana Maria and {Angl{\'e}s-Alc{\'a}zar}, Daniel and {Thiele}, Leander and others},
        title = "{Predicting the impact of feedback on matter clustering with machine learning in CAMELS}",
      journal = {\mnras},
         year = 2023,
       volume = {526},
       number = {4},
        pages = {5306-5325},
          doi = {10.1093/mnras/stad2992},
archivePrefix = {arXiv},
       eprint = {2301.02231},
 primaryClass = {astro-ph.CO}
}

@ARTICLE{Pandey-2023,
       author = {{Pandey}, Shivam and {Lehman}, Kai and {Baxter}, Eric J. and others},
        title = "{Inferring the impact of feedback on the matter distribution using the Sunyaev-Zel'dovich effect: insights from CAMELS simulations and ACT+DES data}",
      journal = {\mnras},
         year = 2023,
          doi = {10.1093/mnras/stad2268},
archivePrefix = {arXiv},
       eprint = {2301.02186},
 primaryClass = {astro-ph.CO}
}

@ARTICLE{Grandis-2024,
       author = {{Grandis}, Sebastian and {Aric{\`o}}, Giovanni and {Schneider}, Aurel and {Linke}, Laila},
        title = "{Determining the baryon impact on the matter power spectrum with galaxy clusters}",
      journal = {\mnras},
         year = 2024,
       volume = {528},
       number = {3},
        pages = {4379-4392},
          doi = {10.1093/mnras/stae259},
archivePrefix = {arXiv},
       eprint = {2309.02920},
 primaryClass = {astro-ph.CO}
}

@ARTICLE{Eifler-2015,
       author = {{Eifler}, Tim and {Krause}, Elisabeth and {Dodelson}, Scott and {Zentner}, Andrew R. and {Hearin}, Andrew P. and {Gnedin}, Nickolay Y.},
        title = "{Accounting for baryonic effects in cosmic shear tomography: determining a minimal set of nuisance parameters using PCA}",
      journal = {\mnras},
         year = 2015,
       volume = {454},
       number = {3},
        pages = {2451-2471},
          doi = {10.1093/mnras/stv2000},
archivePrefix = {arXiv},
       eprint = {1405.7423},
 primaryClass = {astro-ph.CO}
}

@ARTICLE{Huang-2019,
       author = {{Huang}, Hung-Jin and {Eifler}, Tim and {Mandelbaum}, Rachel and {Dodelson}, Scott},
        title = "{Modelling baryonic physics in future weak lensing surveys}",
      journal = {\mnras},
         year = 2019,
       volume = {488},
       number = {2},
        pages = {1652-1678},
          doi = {10.1093/mnras/stz1714},
archivePrefix = {arXiv},
       eprint = {1809.01146},
 primaryClass = {astro-ph.CO}
}

@ARTICLE{Mead-2021,
       author = {{Mead}, Alexander J. and {Brieden}, Samuel and {Tr{\"o}ster}, Tilman and {Heymans}, Catherine},
        title = "{HMcode-2020: improved modelling of non-linear cosmological power spectra with baryonic feedback}",
      journal = {\mnras},
         year = 2021,
       volume = {502},
       number = {1},
        pages = {1401-1422},
          doi = {10.1093/mnras/stab082},
archivePrefix = {arXiv},
       eprint = {2009.01858},
 primaryClass = {astro-ph.CO}
}

@ARTICLE{Giri-2021,
       author = {{Giri}, Sambit K. and {Schneider}, Aurel},
        title = "{Emulation of baryonic effects on the matter power spectrum and constraints from galaxy cluster data}",
      journal = {\jcap},
         year = 2021,
       volume = {2021},
       number = {12},
          eid = {046},
        pages = {046},
          doi = {10.1088/1475-7516/2021/12/046},
archivePrefix = {arXiv},
       eprint = {2108.08863},
 primaryClass = {astro-ph.CO}
}

@ARTICLE{Ganguly-2026,
       author = {{Ganguly}, Anoma and {Schaan}, Emmanuel and {Krause}, Elisabeth and others},
        title = "{Direct shear $\times$ kSZ correlation: controlling baryons without modeling galaxies}",
      journal = {arXiv e-prints},
         year = 2026,
        month = jul,
          eid = {arXiv:2607.29091},
        pages = {arXiv:2607.29091},
          doi = {10.48550/arXiv.2607.29091},
archivePrefix = {arXiv},
       eprint = {2607.29091},
 primaryClass = {astro-ph.CO}
}

@ARTICLE{Gebhardt-2026,
       author = {{Gebhardt}, Matthew and {Angl{\'e}s-Alc{\'a}zar}, Daniel and {Genel}, Shy and others},
        title = "{Cosmological back-reaction of baryons on dark matter in the CAMELS simulations}",
      journal = {arXiv e-prints},
         year = 2026,
        month = jan,
          eid = {arXiv:2601.06258},
        pages = {arXiv:2601.06258},
          doi = {10.48550/arXiv.2601.06258},
archivePrefix = {arXiv},
       eprint = {2601.06258},
 primaryClass = {astro-ph.CO}
}

@ARTICLE{Yang-2026,
       author = {{Yang}, Yanhui and {Bird}, Simeon and {Zhou}, Yihao and others},
        title = "{Matter clustering in Astrid: reduced baryonic suppression from realistic black hole dynamics}",
      journal = {arXiv e-prints},
         year = 2026,
        month = may,
          eid = {arXiv:2605.04176},
        pages = {arXiv:2605.04176},
          doi = {10.48550/arXiv.2605.04176},
archivePrefix = {arXiv},
       eprint = {2605.04176},
 primaryClass = {astro-ph.CO}
}

@ARTICLE{Schaan-2021,
       author = {{Schaan}, Emmanuel and {Ferraro}, Simone and {Amodeo}, Stefania and others},
        title = "{Atacama Cosmology Telescope: Combined kinematic and thermal Sunyaev-Zel'dovich measurements from BOSS CMASS and LOWZ halos}",
      journal = {\prd},
         year = 2021,
       volume = {103},
       number = {6},
          eid = {063513},
        pages = {063513},
          doi = {10.1103/PhysRevD.103.063513},
archivePrefix = {arXiv},
       eprint = {2009.05557},
 primaryClass = {astro-ph.CO}
}

@ARTICLE{Amodeo-2021,
       author = {{Amodeo}, Stefania and {Battaglia}, Nicholas and {Schaan}, Emmanuel and others},
        title = "{Atacama Cosmology Telescope: Modeling the gas thermodynamics in BOSS CMASS galaxies from kinematic and thermal Sunyaev-Zel'dovich measurements}",
      journal = {\prd},
         year = 2021,
       volume = {103},
       number = {6},
          eid = {063514},
        pages = {063514},
          doi = {10.1103/PhysRevD.103.063514},
archivePrefix = {arXiv},
       eprint = {2009.05558},
 primaryClass = {astro-ph.CO}
}

@ARTICLE{Battaglia-2016,
       author = {{Battaglia}, Nicholas},
        title = "{The tau of galaxy clusters}",
      journal = {\jcap},
         year = 2016,
       volume = {2016},
       number = {8},
          eid = {058},
        pages = {058},
          doi = {10.1088/1475-7516/2016/08/058},
archivePrefix = {arXiv},
       eprint = {1607.02442},
 primaryClass = {astro-ph.CO}
}

@ARTICLE{Weinberger-2017,
       author = {{Weinberger}, Rainer and {Springel}, Volker and {Hernquist}, Lars and others},
        title = "{Simulating galaxy formation with black hole driven thermal and kinetic feedback}",
      journal = {\mnras},
         year = 2017,
       volume = {465},
       number = {3},
        pages = {3291-3308},
          doi = {10.1093/mnras/stw2944},
archivePrefix = {arXiv},
       eprint = {1607.03486},
 primaryClass = {astro-ph.GA}
}

@ARTICLE{Nelson-2019,
       author = {{Nelson}, Dylan and {Springel}, Volker and {Pillepich}, Annalisa and others},
        title = "{The IllustrisTNG simulations: public data release}",
      journal = {Computational Astrophysics and Cosmology},
         year = 2019,
       volume = {6},
       number = {1},
          eid = {2},
        pages = {2},
          doi = {10.1186/s40668-019-0028-x},
archivePrefix = {arXiv},
       eprint = {1812.05609},
 primaryClass = {astro-ph.GA}
}

@ARTICLE{Lipman-2022,
       author = {{Lipman}, Yaron and {Chen}, Ricky T.~Q. and {Ben-Hamu}, Heli and {Nickel}, Maximilian and {Le}, Matthew},
        title = "{Flow matching for generative modeling}",
      journal = {arXiv e-prints},
         year = 2022,
        month = oct,
          eid = {arXiv:2210.02747},
        pages = {arXiv:2210.02747},
          doi = {10.48550/arXiv.2210.02747},
archivePrefix = {arXiv},
       eprint = {2210.02747},
 primaryClass = {cs.LG},
         note = {ICLR 2023}
}

@ARTICLE{Kratochvil-2010,
       author = {{Kratochvil}, Jan M. and {Haiman}, Zolt{\'a}n and {May}, Morgan},
        title = "{Probing cosmology with weak lensing peak counts}",
      journal = {\prd},
         year = 2010,
       volume = {81},
       number = {4},
          eid = {043519},
        pages = {043519},
          doi = {10.1103/PhysRevD.81.043519},
archivePrefix = {arXiv},
       eprint = {0907.0486},
 primaryClass = {astro-ph.CO}
}

@ARTICLE{Petri-2013,
       author = {{Petri}, Andrea and {Haiman}, Zolt{\'a}n and {Hui}, Lam and {May}, Morgan and {Kratochvil}, Jan M.},
        title = "{Cosmology with Minkowski functionals and moments of the weak lensing convergence field}",
      journal = {\prd},
         year = 2013,
       volume = {88},
       number = {12},
          eid = {123002},
        pages = {123002},
          doi = {10.1103/PhysRevD.88.123002},
archivePrefix = {arXiv},
       eprint = {1309.4460},
 primaryClass = {astro-ph.CO}
}

@ARTICLE{Troester-2019,
       author = {{Tr{\"o}ster}, Tilman and {Ferguson}, Cameron and {Harnois-D{\'e}raps}, Joachim and {McCarthy}, Ian G.},
        title = "{Painting with baryons: augmenting N-body simulations with gas using deep generative models}",
      journal = {\mnras},
         year = 2019,
       volume = {487},
       number = {1},
        pages = {L24-L29},
          doi = {10.1093/mnrasl/slz075},
archivePrefix = {arXiv},
       eprint = {1903.12173},
 primaryClass = {astro-ph.CO}
}

@ARTICLE{Thiele-2020b,
       author = {{Thiele}, Leander and {Villaescusa-Navarro}, Francisco and {Spergel}, David N. and {Nelson}, Dylan and {Pillepich}, Annalisa},
        title = "{Teaching neural networks to generate fast Sunyaev-Zel'dovich maps}",
      journal = {\apj},
         year = 2020,
       volume = {902},
       number = {2},
          eid = {129},
        pages = {129},
          doi = {10.3847/1538-4357/abb80f},
archivePrefix = {arXiv},
       eprint = {2007.07267},
 primaryClass = {astro-ph.CO}
}

@ARTICLE{Dai-2021,
       author = {{Dai}, Biwei and {Seljak}, Uro{\v{s}}},
        title = "{Learning effective physical laws for generating cosmological hydrodynamics with Lagrangian deep learning}",
      journal = {Proceedings of the National Academy of Sciences},
         year = 2021,
       volume = {118},
       number = {16},
          eid = {e2020324118},
        pages = {e2020324118},
          doi = {10.1073/pnas.2020324118},
archivePrefix = {arXiv},
       eprint = {2010.02926},
 primaryClass = {astro-ph.CO}
}

@ARTICLE{Chadayammuri-2023,
       author = {{Chadayammuri}, Urmila and {Ntampaka}, Michelle and {ZuHone}, John and {Bogd{\'a}n}, {\'A}kos and {Kraft}, Ralph P.},
        title = "{Painting baryons onto N-body simulations of galaxy clusters with image-to-image deep learning}",
      journal = {\mnras},
         year = 2023,
          doi = {10.1093/mnras/stad2596},
archivePrefix = {arXiv},
       eprint = {2307.16733},
 primaryClass = {astro-ph.CO}
}

@ARTICLE{Liu-2025,
       author = {{Liu}, R. Henry and {Hadzhiyska}, Boryana and {Ferraro}, Simone and others},
        title = "{Fast baryonic field painting for Sunyaev-Zel'dovich analyses: transfer function vs. hybrid effective field theory}",
      journal = {arXiv e-prints},
         year = 2025,
        month = apr,
          eid = {arXiv:2504.11794},
        pages = {arXiv:2504.11794},
          doi = {10.48550/arXiv.2504.11794},
archivePrefix = {arXiv},
       eprint = {2504.11794},
 primaryClass = {astro-ph.CO}
}

@ARTICLE{Huterer-2005,
       author = {{Huterer}, Dragan and {Takada}, Masahiro},
        title = "{Calibrating the nonlinear matter power spectrum: requirements for future weak lensing surveys}",
      journal = {Astroparticle Physics},
         year = 2005,
       volume = {23},
       number = {4},
        pages = {369-376},
          doi = {10.1016/j.astropartphys.2005.02.006},
archivePrefix = {arXiv},
       eprint = {astro-ph/0412142}
}

@ARTICLE{Hearin-2012,
       author = {{Hearin}, Andrew P. and {Zentner}, Andrew R. and {Ma}, Zhaoming},
        title = "{General requirements on matter power spectrum predictions for cosmology with weak lensing tomography}",
      journal = {\jcap},
         year = 2012,
       volume = {2012},
       number = {4},
          eid = {034},
        pages = {034},
          doi = {10.1088/1475-7516/2012/04/034},
archivePrefix = {arXiv},
       eprint = {1111.0052},
 primaryClass = {astro-ph.CO}
}

@ARTICLE{vanDaalen-2011,
       author = {{van Daalen}, Marcel P. and {Schaye}, Joop and {Booth}, C.~M. and {Dalla Vecchia}, Claudio},
        title = "{The effects of galaxy formation on the matter power spectrum: a challenge for precision cosmology}",
      journal = {\mnras},
         year = 2011,
       volume = {415},
       number = {4},
        pages = {3649-3665},
          doi = {10.1111/j.1365-2966.2011.18981.x},
archivePrefix = {arXiv},
       eprint = {1104.1174},
 primaryClass = {astro-ph.CO}
}

@ARTICLE{Dave-2019,
       author = {{Dav{\'e}}, Romeel and {Angl{\'e}s-Alc{\'a}zar}, Daniel and {Narayanan}, Desika and others},
        title = "{SIMBA: cosmological simulations with black hole growth and feedback}",
      journal = {\mnras},
         year = 2019,
       volume = {486},
       number = {2},
        pages = {2827-2849},
          doi = {10.1093/mnras/stz937},
archivePrefix = {arXiv},
       eprint = {1901.10203},
 primaryClass = {astro-ph.GA}
}

@ARTICLE{Lu-2022,
       author = {{Lu}, Tianhuan and {Haiman}, Zolt{\'a}n and {Zorrilla Matilla}, Jos{\'e} Manuel},
        title = "{Simultaneously constraining cosmology and baryonic physics via deep learning from weak lensing}",
      journal = {\mnras},
         year = 2022,
       volume = {511},
       number = {1},
        pages = {1518-1528},
          doi = {10.1093/mnras/stac161},
archivePrefix = {arXiv},
       eprint = {2109.11060},
 primaryClass = {astro-ph.CO}
}

@ARTICLE{Liu-2015,
       author = {{Liu}, Jia and {Petri}, Andrea and {Haiman}, Zolt{\'a}n and {Hui}, Lam and {Kratochvil}, Jan M. and {May}, Morgan},
        title = "{Cosmology constraints from the weak lensing peak counts and the power spectrum in CFHTLenS data}",
      journal = {\prd},
         year = 2015,
       volume = {91},
       number = {6},
          eid = {063507},
        pages = {063507},
          doi = {10.1103/PhysRevD.91.063507},
archivePrefix = {arXiv},
       eprint = {1412.0757},
 primaryClass = {astro-ph.CO}
}

@ARTICLE{Kacprzak-2016,
       author = {{Kacprzak}, T. and {Kirk}, D. and {Friedrich}, O. and others},
        title = "{Cosmology constraints from shear peak statistics in Dark Energy Survey Science Verification data}",
      journal = {\mnras},
         year = 2016,
       volume = {463},
       number = {4},
        pages = {3653-3673},
          doi = {10.1093/mnras/stw2070},
archivePrefix = {arXiv},
       eprint = {1603.05040},
 primaryClass = {astro-ph.CO}
}

@ARTICLE{Martinet-2018,
       author = {{Martinet}, Nicolas and {Schneider}, Peter and {Hildebrandt}, Hendrik and others},
        title = "{KiDS-450: cosmological constraints from weak-lensing peak statistics -- II. Inference from shear peaks using N-body simulations}",
      journal = {\mnras},
         year = 2018,
       volume = {474},
       number = {1},
        pages = {712-730},
          doi = {10.1093/mnras/stx2793},
archivePrefix = {arXiv},
       eprint = {1709.07678},
 primaryClass = {astro-ph.CO}
}

@ARTICLE{Shan-2018,
       author = {{Shan}, HuanYuan and {Liu}, Xiangkun and {Hildebrandt}, Hendrik and others},
        title = "{KiDS-450: cosmological constraints from weak lensing peak statistics -- I. Inference from analytical prediction of high signal-to-noise ratio convergence peaks}",
      journal = {\mnras},
         year = 2018,
       volume = {474},
       number = {1},
        pages = {1116-1134},
          doi = {10.1093/mnras/stx2837},
archivePrefix = {arXiv},
       eprint = {1709.07651},
 primaryClass = {astro-ph.CO}
}

@ARTICLE{Yang-2011,
       author = {{Yang}, Xiuyuan and {Kratochvil}, Jan M. and {Wang}, Sheng and {Lim}, Eugene A. and {Haiman}, Zolt{\'a}n and {May}, Morgan},
        title = "{Cosmological information in weak lensing peaks}",
      journal = {\prd},
         year = 2011,
       volume = {84},
       number = {4},
          eid = {043529},
        pages = {043529},
          doi = {10.1103/PhysRevD.84.043529},
archivePrefix = {arXiv},
       eprint = {1109.6333},
 primaryClass = {astro-ph.CO}
}

@ARTICLE{Liu-Haiman-2016,
       author = {{Liu}, Jia and {Haiman}, Zolt{\'a}n},
        title = "{Origin of weak lensing convergence peaks}",
      journal = {\prd},
         year = 2016,
       volume = {94},
       number = {4},
          eid = {043533},
        pages = {043533},
          doi = {10.1103/PhysRevD.94.043533},
archivePrefix = {arXiv},
       eprint = {1606.01318},
 primaryClass = {astro-ph.CO}
}

@ARTICLE{Lee-2023,
       author = {{Lee}, Max E. and {Lu}, Tianhuan and {Haiman}, Zolt{\'a}n and {Liu}, Jia and {Osato}, Ken},
        title = "{Comparing weak lensing peak counts in baryonic correction models to hydrodynamical simulations}",
      journal = {\mnras},
         year = 2023,
        month = feb,
       volume = {519},
       number = {1},
        pages = {573-584},
          doi = {10.1093/mnras/stac3592},
archivePrefix = {arXiv},
       eprint = {2201.08320},
 primaryClass = {astro-ph.CO},
       adsurl = {https://ui.adsabs.harvard.edu/abs/2023MNRAS.519..573L}
}

@ARTICLE{Lee-2024,
       author = {{Lee}, Max E. and {Genel}, Shy and {Wandelt}, Benjamin D. and {Zhang}, Benjamin and {Delgado}, Ana Maria and {Pandey}, Shivam and {Lau}, Erwin T. and {Carr}, Christopher and {Cook}, Harrison and {Nagai}, Daisuke and {Angl{\'e}s-Alc{\'a}zar}, Daniel and {Villaescusa-Navarro}, Francisco and {Bryan}, Greg L.},
        title = "{Zooming by in the CARPoolGP Lane: New CAMELS-TNG Simulations of Zoomed-in Massive Halos}",
      journal = {\apj},
         year = 2024,
        month = jun,
       volume = {968},
       number = {1},
          eid = {11},
        pages = {11},
          doi = {10.3847/1538-4357/ad3d4a},
archivePrefix = {arXiv},
       eprint = {2403.10609},
 primaryClass = {astro-ph.CO},
       adsurl = {https://ui.adsabs.harvard.edu/abs/2024ApJ...968...11L}
}

@ARTICLE{Lee-2026,
       author = {{Lee}, Max E. and {Haiman}, Zolt{\'a}n and {Pandey}, Shivam and {Genel}, Shy},
        title = "{The Effect of Intrinsic Alignments on Weak-lensing Statistics in Hydrodynamical Simulations}",
      journal = {\apj},
         year = 2026,
        month = jan,
       volume = {996},
       number = {1},
          eid = {36},
        pages = {36},
          doi = {10.3847/1538-4357/ae1ca7},
archivePrefix = {arXiv},
       eprint = {2504.12460},
 primaryClass = {astro-ph.CO},
       adsurl = {https://ui.adsabs.harvard.edu/abs/2026ApJ...996...36L}
}

@ARTICLE{Lee-2026a,
       author = {{Lee}, Max E. and {Haiman}, Zoltan and {Genel}, Shy},
        title = "{The impact of baryons on weak lensing statistics as a function of halo mass and radius}",
      journal = {arXiv e-prints},
         year = 2026,
        month = mar,
          eid = {arXiv:2603.11815},
        pages = {arXiv:2603.11815},
          doi = {10.48550/arXiv.2603.11815},
archivePrefix = {arXiv},
       eprint = {2603.11815},
 primaryClass = {astro-ph.CO},
       adsurl = {https://ui.adsabs.harvard.edu/abs/2026arXiv260311815L}
}

@ARTICLE{Lee-2026b,
       author = {{Lee}, Max E. and {Genel}, Shy and {Haiman}, Zolt{\'a}n and {Bryan}, Greg L.},
        title = "{BIND}",
      journal = {arXiv e-prints},
         year = 2026,
         note = {in preparation}
}

@ARTICLE{Genel-2026,
       author = {{Genel}, Shy and {Jo}, Yongseok and {Oh}, Boon Kiat and {Tillman}, Megan Taylor and {Lee}, Max E. and {Lee}, Jun-Young and {Hern{\'a}ndez-Mart{\'\i}nez}, Elena and {Lovell}, Christopher C. and {Sims}, Xavier and {Burkhart}, Blakesley and {Nagamine}, Kentaro and {Angl{\'e}s-Alc{\'a}zar}, Daniel and {Villaescusa-Navarro}, Francisco},
        title = "{Learning the Universe with the 2nd Generation of CAMELS: Varying 35 parameters of the IllustrisTNG model in (50Mpc/h)\^3 boxes}",
      journal = {arXiv e-prints},
         year = 2026,
        month = jun,
          eid = {arXiv:2606.10038},
        pages = {arXiv:2606.10038},
          doi = {10.48550/arXiv.2606.10038},
archivePrefix = {arXiv},
       eprint = {2606.10038},
 primaryClass = {astro-ph.CO},
       adsurl = {https://ui.adsabs.harvard.edu/abs/2026arXiv260610038G}
}

@ARTICLE{Lipman-2024,
       author = {{Lipman}, Yaron and {Havasi}, Marton and {Holderrieth}, Peter and {Shaul}, Neta and {Le}, Matt and {Karrer}, Brian and {Chen}, Ricky T.~Q. and {Lopez-Paz}, David and {Ben-Hamu}, Heli and {Gat}, Itai},
        title = "{Flow Matching Guide and Code}",
      journal = {arXiv e-prints},
         year = 2024,
        month = dec,
          eid = {arXiv:2412.06264},
        pages = {arXiv:2412.06264},
          doi = {10.48550/arXiv.2412.06264},
archivePrefix = {arXiv},
       eprint = {2412.06264},
 primaryClass = {cs.LG}
}

@ARTICLE{kannan-2025,
       author = {{Kannan}, Sidharth and {Qiu}, Tian and {Cuesta-Lazaro}, Carolina and {Jeong}, Haewon},
        title = "{CosmoFlow: Scale-Aware Representation Learning for Cosmology with Flow Matching}",
      journal = {arXiv e-prints},
         year = 2025,
        month = jul,
          eid = {arXiv:2507.11842},
        pages = {arXiv:2507.11842},
          doi = {10.48550/arXiv.2507.11842},
archivePrefix = {arXiv},
       eprint = {2507.11842},
 primaryClass = {astro-ph.CO}
}

@BOOK{hockney,
       author = {{Hockney}, R.~W. and {Eastwood}, J.~W.},
        title = "{Computer Simulation Using Particles}",
    publisher = {Bristol: Hilger},
         year = 1988,
       adsurl = {https://ui.adsabs.harvard.edu/abs/1988csup.book.....H}
}

@ARTICLE{Springel-2018,
       author = {{Springel}, Volker and {Pakmor}, R{\"u}diger and {Pillepich}, Annalisa and {Weinberger}, Rainer and {Nelson}, Dylan and {Hernquist}, Lars and {Vogelsberger}, Mark and {Genel}, Shy and {Torrey}, Paul and {Marinacci}, Federico and {Naiman}, Jill},
        title = "{First results from the IllustrisTNG simulations: matter and galaxy clustering}",
      journal = {\mnras},
         year = 2018,
        month = mar,
       volume = {475},
       number = {1},
        pages = {676-698},
          doi = {10.1093/mnras/stx3304},
archivePrefix = {arXiv},
       eprint = {1707.03397},
 primaryClass = {astro-ph.GA},
       adsurl = {https://ui.adsabs.harvard.edu/abs/2018MNRAS.475..676S}
}

@ARTICLE{planck-2015,
       author = {{Planck Collaboration} and {Ade}, P.~A.~R. and {Aghanim}, N. and others},
        title = "{Planck 2015 results. XIII. Cosmological parameters}",
      journal = {\aap},
         year = 2016,
        month = sep,
       volume = {594},
          eid = {A13},
        pages = {A13},
          doi = {10.1051/0004-6361/201525830},
archivePrefix = {arXiv},
       eprint = {1502.01589},
 primaryClass = {astro-ph.CO},
       adsurl = {https://ui.adsabs.harvard.edu/abs/2016A&A...594A..13P}
}

@ARTICLE{VillaescusaNavarro-2021,
       author = {{Villaescusa-Navarro}, Francisco and {Angl{\'e}s-Alc{\'a}zar}, Daniel and {Genel}, Shy and others},
        title = "{The CAMELS Project: Cosmology and Astrophysics with Machine-learning Simulations}",
      journal = {\apj},
         year = 2021,
        month = jul,
       volume = {915},
       number = {1},
          eid = {71},
        pages = {71},
          doi = {10.3847/1538-4357/abf7ba},
archivePrefix = {arXiv},
       eprint = {2010.00619},
 primaryClass = {astro-ph.CO},
       adsurl = {https://ui.adsabs.harvard.edu/abs/2021ApJ...915...71V}
}

@ARTICLE{Schaye-2023,
       author = {{Schaye}, Joop and {Kugel}, Roi and {Schaller}, Matthieu and others},
        title = "{The FLAMINGO project: cosmological hydrodynamical simulations for large-scale structure and galaxy cluster surveys}",
      journal = {\mnras},
         year = 2023,
        month = dec,
       volume = {526},
       number = {4},
        pages = {4978-5020},
          doi = {10.1093/mnras/stad2419},
archivePrefix = {arXiv},
       eprint = {2306.04024},
 primaryClass = {astro-ph.CO},
       adsurl = {https://ui.adsabs.harvard.edu/abs/2023MNRAS.526.4978S}
}

@ARTICLE{Jain-2000,
       author = {{Jain}, B. and {Seljak}, U. and {White}, S.},
        title = "{Ray-tracing Simulations of Weak Lensing by Large-Scale Structure}",
      journal = {\apj},
         year = 2000,
        month = feb,
       volume = {530},
       number = {2},
        pages = {547-577},
          doi = {10.1086/308384},
archivePrefix = {arXiv},
       eprint = {astro-ph/9901191},
       adsurl = {https://ui.adsabs.harvard.edu/abs/2000ApJ...530..547J}
}

@ARTICLE{Hilbert-2009,
       author = {{Hilbert}, S. and {Hartlap}, J. and {White}, S.~D.~M. and {Schneider}, P.},
        title = "{Ray-tracing through the Millennium Simulation: Born corrections and lens-lens coupling in cosmic shear and galaxy-galaxy lensing}",
      journal = {\aap},
         year = 2009,
        month = may,
       volume = {499},
       number = {1},
        pages = {31-43},
          doi = {10.1051/0004-6361/200811054},
archivePrefix = {arXiv},
       eprint = {0809.5035},
 primaryClass = {astro-ph},
       adsurl = {https://ui.adsabs.harvard.edu/abs/2009A&A...499...31H}
}

@ARTICLE{Osato-2021,
       author = {{Osato}, Ken and {Liu}, Jia and {Haiman}, Zolt{\'a}n},
        title = "{{\ensuremath{\kappa}}TNG: effect of baryonic processes on weak lensing with IllustrisTNG simulations}",
      journal = {\mnras},
         year = 2021,
        month = apr,
       volume = {502},
       number = {4},
        pages = {5593-5602},
          doi = {10.1093/mnras/stab395},
archivePrefix = {arXiv},
       eprint = {2010.09731},
 primaryClass = {astro-ph.CO},
       adsurl = {https://ui.adsabs.harvard.edu/abs/2021MNRAS.502.5593O}
}

@ARTICLE{Petri-2016,
       author = {{Petri}, Andrea and {Haiman}, Zolt{\'a}n and {May}, Morgan},
        title = "{Sample variance in weak lensing: How many simulations are required?}",
      journal = {\prd},
         year = 2016,
        month = mar,
       volume = {93},
       number = {6},
          eid = {063524},
        pages = {063524},
          doi = {10.1103/PhysRevD.93.063524},
archivePrefix = {arXiv},
       eprint = {1601.06792},
 primaryClass = {astro-ph.CO},
       adsurl = {https://ui.adsabs.harvard.edu/abs/2016PhRvD..93f3524P}
}

@ARTICLE{Petri-2016a,
       author = {{Petri}, Andrea},
        title = "{Mocking the weak lensing universe: The LensTools Python computing package}",
      journal = {Astronomy and Computing},
         year = 2016,
        month = oct,
       volume = {17},
        pages = {73-79},
          doi = {10.1016/j.ascom.2016.06.001},
archivePrefix = {arXiv},
       eprint = {1606.01903},
 primaryClass = {astro-ph.CO},
       adsurl = {https://ui.adsabs.harvard.edu/abs/2016A&C....17...73P}
}

@ARTICLE{Maturi-2010,
       author = {{Maturi}, M. and {Angrick}, C. and {Pace}, F. and {Bartelmann}, M.},
        title = "{An analytic approach to number counts of weak-lensing peak detections}",
      journal = {\aap},
         year = 2010,
        month = sep,
       volume = {519},
          eid = {A23},
        pages = {A23},
          doi = {10.1051/0004-6361/200912866},
archivePrefix = {arXiv},
       eprint = {0907.1849},
 primaryClass = {astro-ph.CO},
       adsurl = {https://ui.adsabs.harvard.edu/abs/2010A&A...519A..23M}
}

@ARTICLE{Coulton-2020,
       author = {{Coulton}, William R. and {Liu}, Jia and {McCarthy}, Ian G. and {Osato}, Ken},
        title = "{Weak lensing minima and peaks: Cosmological constraints and the impact of baryons}",
      journal = {\mnras},
         year = 2020,
        month = jul,
       volume = {495},
       number = {3},
        pages = {2531-2542},
          doi = {10.1093/mnras/staa1098},
archivePrefix = {arXiv},
       eprint = {1910.04171},
 primaryClass = {astro-ph.CO},
       adsurl = {https://ui.adsabs.harvard.edu/abs/2020MNRAS.495.2531C}
}

@ARTICLE{Thiele-2020,
       author = {{Thiele}, Leander and {Hill}, J. Colin and {Smith}, Kendrick M.},
        title = "{Accurate analytic model for the weak lensing convergence one-point probability distribution function and its autocovariance}",
      journal = {\prd},
         year = 2020,
        month = dec,
       volume = {102},
       number = {12},
          eid = {123545},
        pages = {123545},
          doi = {10.1103/PhysRevD.102.123545},
archivePrefix = {arXiv},
       eprint = {2009.06547},
 primaryClass = {astro-ph.CO},
       adsurl = {https://ui.adsabs.harvard.edu/abs/2020PhRvD.102l3545T}
}

@ARTICLE{Kratochvil-2012,
       author = {{Kratochvil}, Jan M. and {Lim}, Eugene A. and {Wang}, Sheng and {Haiman}, Zolt{\'a}n and {May}, Morgan and {Huffenberger}, Kevin},
        title = "{Probing cosmology with weak lensing Minkowski functionals}",
      journal = {\prd},
         year = 2012,
        month = may,
       volume = {85},
       number = {10},
          eid = {103513},
        pages = {103513},
          doi = {10.1103/PhysRevD.85.103513},
archivePrefix = {arXiv},
       eprint = {1109.6334},
 primaryClass = {astro-ph.CO},
       adsurl = {https://ui.adsabs.harvard.edu/abs/2012PhRvD..85j3513K}
}

@ARTICLE{Hartlap-2007,
       author = {{Hartlap}, J. and {Simon}, P. and {Schneider}, P.},
        title = "{Why your model parameter confidences might be too optimistic. Unbiased estimation of the inverse covariance matrix}",
      journal = {\aap},
         year = 2007,
        month = mar,
       volume = {464},
       number = {1},
        pages = {399-404},
          doi = {10.1051/0004-6361:20066170},
archivePrefix = {arXiv},
       eprint = {astro-ph/0608064},
       adsurl = {https://ui.adsabs.harvard.edu/abs/2007A&A...464..399H}
}

@ARTICLE{Zuercher-2021,
       author = {{Z{\"u}rcher}, Dominik and {Fluri}, Janis and {Sgier}, Raphael and {Kacprzak}, Tomasz and {Refregier}, Alexandre},
        title = "{Cosmological forecast for non-Gaussian statistics in large-scale weak lensing surveys}",
      journal = {\jcap},
         year = 2021,
        month = jan,
       volume = {2021},
       number = {1},
          eid = {028},
        pages = {028},
          doi = {10.1088/1475-7516/2021/01/028},
archivePrefix = {arXiv},
       eprint = {2006.12506},
 primaryClass = {astro-ph.CO},
       adsurl = {https://ui.adsabs.harvard.edu/abs/2021JCAP...01..028Z}
}

@ARTICLE{Martinet-2021,
       author = {{Martinet}, Nicolas and {Castro}, Tiago and {Harnois-D{\'e}raps}, Joachim and others},
        title = "{Probing dark energy with tomographic weak-lensing aperture mass statistics}",
      journal = {\aap},
         year = 2021,
        month = feb,
       volume = {646},
          eid = {A62},
        pages = {A62},
          doi = {10.1051/0004-6361/202039679},
archivePrefix = {arXiv},
       eprint = {2010.07376},
 primaryClass = {astro-ph.CO},
       adsurl = {https://ui.adsabs.harvard.edu/abs/2021A&A...646A..62M}
}

@ARTICLE{HarnoisDeraps-2021,
       author = {{Harnois-D{\'e}raps}, Joachim and {Martinet}, Nicolas and {Castro}, Tiago and {Dolag}, Klaus and {Giblin}, Benjamin and {Heymans}, Catherine and {Hildebrandt}, Hendrik and {Xia}, Qianli},
        title = "{Cosmic shear cosmology beyond two-point statistics: a combined peak count and correlation function analysis of DES-Y1}",
      journal = {\mnras},
         year = 2021,
        month = sep,
       volume = {506},
       number = {2},
        pages = {1623-1650},
          doi = {10.1093/mnras/stab1623},
archivePrefix = {arXiv},
       eprint = {2012.02777},
 primaryClass = {astro-ph.CO},
       adsurl = {https://ui.adsabs.harvard.edu/abs/2021MNRAS.506.1623H}
}

@ARTICLE{Gatti-2022,
       author = {{Gatti}, M. and {Jain}, B. and {Chang}, C. and others},
        title = "{Dark Energy Survey Year 3 results: Cosmology with moments of weak lensing mass maps}",
      journal = {\prd},
         year = 2022,
        month = oct,
       volume = {106},
       number = {8},
          eid = {083509},
        pages = {083509},
          doi = {10.1103/PhysRevD.106.083509},
archivePrefix = {arXiv},
       eprint = {2110.10141},
 primaryClass = {astro-ph.CO},
       adsurl = {https://ui.adsabs.harvard.edu/abs/2022PhRvD.106h3509G}
}

@ARTICLE{Zuercher-2022,
       author = {{Z{\"u}rcher}, D. and {Fluri}, J. and {Sgier}, R. and others},
        title = "{Dark energy survey year 3 results: Cosmology with peaks using an emulator approach}",
      journal = {\mnras},
         year = 2022,
        month = apr,
       volume = {511},
       number = {2},
        pages = {2075-2104},
          doi = {10.1093/mnras/stac078},
archivePrefix = {arXiv},
       eprint = {2110.10135},
 primaryClass = {astro-ph.CO},
       adsurl = {https://ui.adsabs.harvard.edu/abs/2022MNRAS.511.2075Z}
}

@ARTICLE{Marques-2024,
       author = {{Marques}, Gabriela A. and {Liu}, Jia and {Shirasaki}, Masato and others},
        title = "{Cosmology from weak lensing peaks and minima with Subaru Hyper Suprime-Cam Survey first-year data}",
      journal = {\mnras},
         year = 2024,
        month = mar,
       volume = {528},
       number = {3},
        pages = {4513-4527},
          doi = {10.1093/mnras/stae098},
archivePrefix = {arXiv},
       eprint = {2308.10866},
 primaryClass = {astro-ph.CO},
       adsurl = {https://ui.adsabs.harvard.edu/abs/2024MNRAS.528.4513M}
}

@ARTICLE{vanDaalen-2020,
       author = {{van Daalen}, Marcel P. and {McCarthy}, Ian G. and {Schaye}, Joop},
        title = "{Exploring the effects of galaxy formation on matter clustering through a library of simulation power spectra}",
      journal = {\mnras},
         year = 2020,
        month = jan,
       volume = {491},
       number = {2},
        pages = {2424-2446},
          doi = {10.1093/mnras/stz3199},
archivePrefix = {arXiv},
       eprint = {1906.00968},
 primaryClass = {astro-ph.CO},
       adsurl = {https://ui.adsabs.harvard.edu/abs/2020MNRAS.491.2424V}
}

@ARTICLE{Broxterman-2024,
       author = {{Broxterman}, Jeger C. and {Schaller}, Matthieu and {Schaye}, Joop and others},
        title = "{The FLAMINGO project: baryonic impact on weak gravitational lensing convergence peak counts}",
      journal = {\mnras},
         year = 2024,
        month = apr,
       volume = {529},
       number = {3},
        pages = {2309-2326},
          doi = {10.1093/mnras/stae698},
archivePrefix = {arXiv},
       eprint = {2312.08450},
 primaryClass = {astro-ph.CO},
       adsurl = {https://ui.adsabs.harvard.edu/abs/2024MNRAS.529.2309B}
}

@ARTICLE{Grandon-2024,
       author = {{Grand{\'o}n}, Daniela and {Marques}, Gabriela A. and {Thiele}, Leander and {Karwal}, Tanvi and {Liu}, Jia and {Jain}, Bhuvnesh},
        title = "{Impact of baryonic feedback on HSC-Y1 weak lensing non-Gaussian statistics}",
      journal = {\prd},
         year = 2024,
        month = nov,
       volume = {110},
       number = {10},
          eid = {103539},
        pages = {103539},
          doi = {10.1103/PhysRevD.110.103539},
archivePrefix = {arXiv},
       eprint = {2403.03807},
 primaryClass = {astro-ph.CO},
       adsurl = {https://ui.adsabs.harvard.edu/abs/2024PhRvD.110j3539G}
}

@ARTICLE{Marinichenko-2025,
       author = {{Marinichenko}, Mariia and {van Daalen}, Marcel P. and {Sellentin}, Elena},
        title = "{FLAMINGO: Baryonic effects on the weak lensing scattering transform}",
      journal = {arXiv e-prints},
         year = 2025,
        month = oct,
          eid = {arXiv:2510.09761},
        pages = {arXiv:2510.09761},
          doi = {10.48550/arXiv.2510.09761},
archivePrefix = {arXiv},
       eprint = {2510.09761},
 primaryClass = {astro-ph.CO},
       adsurl = {https://ui.adsabs.harvard.edu/abs/2025arXiv251009761M}
}

@ARTICLE{Schneider-2015,
       author = {{Schneider}, Aurel and {Teyssier}, Romain},
        title = "{A new method to quantify the effects of baryons on the matter power spectrum}",
      journal = {\jcap},
         year = 2015,
        month = dec,
       volume = {2015},
       number = {12},
          eid = {049},
        pages = {049},
          doi = {10.1088/1475-7516/2015/12/049},
archivePrefix = {arXiv},
       eprint = {1510.06034},
 primaryClass = {astro-ph.CO},
       adsurl = {https://ui.adsabs.harvard.edu/abs/2015JCAP...12..049S}
}

@ARTICLE{Schneider-2019,
       author = {{Schneider}, Aurel and {Teyssier}, Romain and {Stadel}, Joachim and {Chisari}, Nora Elisa and {Le Brun}, Amandine M.~C. and {Amara}, Adam and {Refregier}, Alexandre},
        title = "{Quantifying baryon effects on the matter power spectrum and the weak lensing shear correlation}",
      journal = {\jcap},
         year = 2019,
        month = mar,
       volume = {2019},
       number = {3},
          eid = {020},
        pages = {020},
          doi = {10.1088/1475-7516/2019/03/020},
archivePrefix = {arXiv},
       eprint = {1810.08629},
 primaryClass = {astro-ph.CO},
       adsurl = {https://ui.adsabs.harvard.edu/abs/2019JCAP...03..020S}
}

@ARTICLE{Schneider-2022,
       author = {{Schneider}, Aurel and {Giri}, Sambit K. and {Amodeo}, Stefania and {Refregier}, Alexandre},
        title = "{Constraining baryonic feedback and cosmology with weak-lensing, X-ray, and kinematic Sunyaev-Zeldovich observations}",
      journal = {\mnras},
         year = 2022,
        month = aug,
       volume = {514},
       number = {3},
        pages = {3802-3814},
          doi = {10.1093/mnras/stac1493},
archivePrefix = {arXiv},
       eprint = {2110.02228},
 primaryClass = {astro-ph.CO},
       adsurl = {https://ui.adsabs.harvard.edu/abs/2022MNRAS.514.3802S}
}

@ARTICLE{Schneider-2025,
       author = {{Schneider}, Aurel and {Kova{\v{c}}}, Michael and {Bucko}, Jozef and {Nicola}, Andrina and {Reischke}, Robert and {Giri}, Sambit K. and {Teyssier}, Romain and {Tr{\"o}ster}, Tilman and {Refregier}, Alexandre and {Schaller}, Matthieu and {Schaye}, Joop},
        title = "{Baryonification: an alternative to hydrodynamical simulations for cosmological studies}",
      journal = {\jcap},
         year = 2025,
        month = dec,
       volume = {2025},
       number = {12},
          eid = {043},
        pages = {043},
          doi = {10.1088/1475-7516/2025/12/043},
archivePrefix = {arXiv},
       eprint = {2507.07892},
 primaryClass = {astro-ph.CO},
       adsurl = {https://ui.adsabs.harvard.edu/abs/2025JCAP...12..043S}
}

@ARTICLE{Anbajagane-2024,
       author = {{Anbajagane}, Dhayaa and {Pandey}, Shivam and {Chang}, Chihway},
        title = "{Map-level baryonification: Efficient modelling of higher-order correlations in the weak lensing and thermal Sunyaev-Zeldovich fields}",
      journal = {The Open Journal of Astrophysics},
         year = 2024,
        month = dec,
       volume = {7},
          eid = {108},
        pages = {108},
          doi = {10.33232/001c.126788},
archivePrefix = {arXiv},
       eprint = {2409.03822},
 primaryClass = {astro-ph.CO},
       adsurl = {https://ui.adsabs.harvard.edu/abs/2024OJAp....7E.108A}
}

@ARTICLE{Zhou-2025,
       author = {{Zhou}, Alan Junzhe and {Gatti}, Marco and {Anbajagane}, Dhayaa and {Dodelson}, Scott and {Schaller}, Matthieu and {Schaye}, Joop},
        title = "{Map-level baryonification: unified treatment of weak lensing two-point and higher-order statistics}",
      journal = {\jcap},
         year = 2025,
        month = sep,
       volume = {2025},
       number = {9},
          eid = {073},
        pages = {073},
          doi = {10.1088/1475-7516/2025/09/073},
archivePrefix = {arXiv},
       eprint = {2505.07949},
 primaryClass = {astro-ph.CO},
       adsurl = {https://ui.adsabs.harvard.edu/abs/2025arXiv250507949Z}
}

@ARTICLE{Arico-2021,
       author = {{Aric{\`o}}, Giovanni and {Angulo}, Raul E. and {Contreras}, Sergio and {Ondaro-Mallea}, Lurdes and {Pellejero-Iba{\~n}ez}, Marcos and {Zennaro}, Matteo},
        title = "{The BACCO simulation project: a baryonification emulator with neural networks}",
      journal = {\mnras},
         year = 2021,
        month = sep,
       volume = {506},
       number = {3},
        pages = {4070-4082},
          doi = {10.1093/mnras/stab1911},
archivePrefix = {arXiv},
       eprint = {2011.15018},
 primaryClass = {astro-ph.CO},
       adsurl = {https://ui.adsabs.harvard.edu/abs/2021MNRAS.506.4070A}
}

@ARTICLE{Sharma-2024,
       author = {{Sharma}, Divij and {Dai}, Biwei and {Villaescusa-Navarro}, Francisco and {Seljak}, Uro{\v{s}}},
        title = "{A field-level emulator for modelling baryonic effects across hydrodynamic simulations}",
      journal = {\mnras},
         year = 2025,
        month = apr,
       volume = {538},
       number = {3},
        pages = {1415-1426},
          doi = {10.1093/mnras/staf294},
archivePrefix = {arXiv},
       eprint = {2401.15891},
 primaryClass = {astro-ph.CO},
       adsurl = {https://ui.adsabs.harvard.edu/abs/2025MNRAS.538.1415S}
}

@ARTICLE{Hadzhiyska-2024,
       author = {{Hadzhiyska}, Boryana and {Ferraro}, Simone and {Ried Guachalla}, Bernardita and others},
        title = "{Evidence for large baryonic feedback at low and intermediate redshifts from kinematic Sunyaev-Zel'dovich observations with ACT and DESI photometric galaxies}",
      journal = {arXiv e-prints},
         year = 2024,
        month = jul,
          eid = {arXiv:2407.07152},
        pages = {arXiv:2407.07152},
          doi = {10.48550/arXiv.2407.07152},
archivePrefix = {arXiv},
       eprint = {2407.07152},
 primaryClass = {astro-ph.CO},
       adsurl = {https://ui.adsabs.harvard.edu/abs/2024arXiv240707152H}
}

@ARTICLE{Bigwood-2024,
       author = {{Bigwood}, L. and {Amon}, A. and {Schneider}, A. and others},
        title = "{Weak lensing combined with the kinetic Sunyaev-Zel'dovich effect: a study of baryonic feedback}",
      journal = {\mnras},
         year = 2024,
        month = oct,
       volume = {534},
       number = {1},
        pages = {655-682},
          doi = {10.1093/mnras/stae2100},
archivePrefix = {arXiv},
       eprint = {2404.06098},
 primaryClass = {astro-ph.CO},
       adsurl = {https://ui.adsabs.harvard.edu/abs/2024MNRAS.534..655B}
}

@ARTICLE{Bigwood-2025,
       author = {{Bigwood}, Leah and {Amon}, Alexandra and {McCarthy}, Ian G. and others},
        title = "{The kinetic Sunyaev Zeldovich effect as a benchmark for AGN feedback models in hydrodynamical simulations: insights from DESI + ACT}",
      journal = {arXiv e-prints},
         year = 2025,
        month = oct,
          eid = {arXiv:2510.15822},
        pages = {arXiv:2510.15822},
          doi = {10.48550/arXiv.2510.15822},
archivePrefix = {arXiv},
       eprint = {2510.15822},
 primaryClass = {astro-ph.CO},
       adsurl = {https://ui.adsabs.harvard.edu/abs/2025arXiv251015822B}
}

@ARTICLE{McCarthy-2024,
       author = {{McCarthy}, Ian G. and {Amon}, Alexandra and {Schaye}, Joop and others},
        title = "{FLAMINGO: combining kinetic SZ effect and galaxy-galaxy lensing measurements to gauge the impact of feedback on large-scale structure}",
      journal = {arXiv e-prints},
         year = 2024,
        month = oct,
          eid = {arXiv:2410.19905},
        pages = {arXiv:2410.19905},
          doi = {10.48550/arXiv.2410.19905},
archivePrefix = {arXiv},
       eprint = {2410.19905},
 primaryClass = {astro-ph.CO},
       adsurl = {https://ui.adsabs.harvard.edu/abs/2024arXiv241019905M}
}

@ARTICLE{Siegel-2025,
       author = {{Siegel}, Jared and {Amon}, Alexandra and {McCarthy}, Ian G. and {Bigwood}, Leah and {Yamamoto}, Masaya and others},
        title = "{Joint X-ray, kinetic Sunyaev-Zeldovich, and weak lensing measurements: toward a consensus picture of efficient gas expulsion from groups and clusters}",
      journal = {arXiv e-prints},
         year = 2025,
        month = sep,
          eid = {arXiv:2509.10455},
        pages = {arXiv:2509.10455},
          doi = {10.48550/arXiv.2509.10455},
archivePrefix = {arXiv},
       eprint = {2509.10455},
 primaryClass = {astro-ph.CO},
       adsurl = {https://ui.adsabs.harvard.edu/abs/2025arXiv250910455S}
}

@ARTICLE{Siegel-2025b,
       author = {{Siegel}, Jared and {Bigwood}, Leah and {Amon}, Alexandra and others},
        title = "{The suppression of the matter power spectrum: strong feedback from X-ray gas mass fractions, kSZ effect profiles, and galaxy-galaxy lensing}",
      journal = {arXiv e-prints},
         year = 2025,
        month = dec,
          eid = {arXiv:2512.02954},
        pages = {arXiv:2512.02954},
          doi = {10.48550/arXiv.2512.02954},
archivePrefix = {arXiv},
       eprint = {2512.02954},
 primaryClass = {astro-ph.CO},
       adsurl = {https://ui.adsabs.harvard.edu/abs/2025arXiv251202954S}
}

@ARTICLE{Pandey-2022,
       author = {{Pandey}, S. and {Gatti}, M. and {Baxter}, E. and others},
        title = "{Cross-correlation of Dark Energy Survey Year 3 lensing data with ACT and Planck thermal Sunyaev-Zel'dovich effect observations. II. Modeling and constraints on halo pressure profiles}",
      journal = {\prd},
         year = 2022,
        month = jun,
       volume = {105},
       number = {12},
          eid = {123526},
        pages = {123526},
          doi = {10.1103/PhysRevD.105.123526},
archivePrefix = {arXiv},
       eprint = {2108.01601},
 primaryClass = {astro-ph.CO},
       adsurl = {https://ui.adsabs.harvard.edu/abs/2022PhRvD.105l3526P}
}

@ARTICLE{Pandey-2025,
       author = {{Pandey}, S. and {Hill}, J.~C. and {Alarcon}, A. and others},
        title = "{Constraints on cosmology and baryonic feedback with joint analysis of Dark Energy Survey Year 3 lensing data and ACT DR6 thermal Sunyaev-Zel'dovich effect observations}",
      journal = {arXiv e-prints},
         year = 2025,
        month = jun,
          eid = {arXiv:2506.07432},
        pages = {arXiv:2506.07432},
          doi = {10.48550/arXiv.2506.07432},
archivePrefix = {arXiv},
       eprint = {2506.07432},
 primaryClass = {astro-ph.CO},
       adsurl = {https://ui.adsabs.harvard.edu/abs/2025arXiv250607432P}
}

@ARTICLE{Amon-2022,
       author = {{Amon}, Alexandra and {Efstathiou}, George},
        title = "{A non-linear solution to the S$_{8}$ tension?}",
      journal = {\mnras},
         year = 2022,
        month = nov,
       volume = {516},
       number = {4},
        pages = {5355-5366},
          doi = {10.1093/mnras/stac2429},
archivePrefix = {arXiv},
       eprint = {2206.11794},
 primaryClass = {astro-ph.CO},
       adsurl = {https://ui.adsabs.harvard.edu/abs/2022MNRAS.516.5355A}
}

@ARTICLE{Preston-2023,
       author = {{Preston}, Calvin and {Amon}, Alexandra and {Efstathiou}, George},
        title = "{A non-linear solution to the S$_{8}$ tension - II. Analysis of DES Year 3 cosmic shear}",
      journal = {\mnras},
         year = 2023,
        month = nov,
       volume = {525},
       number = {4},
        pages = {5554-5564},
          doi = {10.1093/mnras/stad2573},
archivePrefix = {arXiv},
       eprint = {2305.09827},
 primaryClass = {astro-ph.CO},
       adsurl = {https://ui.adsabs.harvard.edu/abs/2023MNRAS.525.5554P}
}

@ARTICLE{Bigwood-2025b,
       author = {{Bigwood}, Leah and {McCullough}, Jamie and {Siegel}, Jared and {Amon}, Alexandra and {Efstathiou}, George and others},
        title = "{Confronting cosmic shear astrophysical uncertainties: DES Year 3 revisited}",
      journal = {arXiv e-prints},
         year = 2025,
        month = dec,
          eid = {arXiv:2512.04209},
        pages = {arXiv:2512.04209},
          doi = {10.48550/arXiv.2512.04209},
archivePrefix = {arXiv},
       eprint = {2512.04209},
 primaryClass = {astro-ph.CO},
       adsurl = {https://ui.adsabs.harvard.edu/abs/2025arXiv251204209B}
}

@ARTICLE{2022ApJ...935..167A,
       author = {{Astropy Collaboration} and {Price-Whelan}, Adrian M. and {Lim}, Pey Lian and others},
        title = "{The Astropy Project: Sustaining and Growing a Community-oriented Open-source Project and the Latest Major Release (v5.0) of the Core Package}",
      journal = {\apj},
         year = 2022,
        month = aug,
       volume = {935},
       number = {2},
          eid = {167},
        pages = {167},
          doi = {10.3847/1538-4357/ac7c74},
archivePrefix = {arXiv},
       eprint = {2206.14220},
 primaryClass = {astro-ph.IM},
       adsurl = {https://ui.adsabs.harvard.edu/abs/2022ApJ...935..167A}
}

@ARTICLE{2018AJ....156..123A,
       author = {{Astropy Collaboration} and {Price-Whelan}, A.~M. and {Sip{\H{o}}cz}, B.~M. and others},
        title = "{The Astropy Project: Building an Open-science Project and Status of the v2.0 Core Package}",
      journal = {\aj},
         year = 2018,
        month = sep,
       volume = {156},
       number = {3},
          eid = {123},
        pages = {123},
          doi = {10.3847/1538-3881/aabc4f},
archivePrefix = {arXiv},
       eprint = {1801.02634},
 primaryClass = {astro-ph.IM},
       adsurl = {https://ui.adsabs.harvard.edu/abs/2018AJ....156..123A}
}

@ARTICLE{2013A&A...558A..33A,
       author = {{Astropy Collaboration} and {Robitaille}, Thomas P. and {Tollerud}, Erik J. and others},
        title = "{Astropy: A community Python package for astronomy}",
      journal = {\aap},
         year = 2013,
        month = oct,
       volume = {558},
          eid = {A33},
        pages = {A33},
          doi = {10.1051/0004-6361/201322068},
archivePrefix = {arXiv},
       eprint = {1307.6212},
 primaryClass = {astro-ph.IM},
       adsurl = {https://ui.adsabs.harvard.edu/abs/2013A&A...558A..33A}
}

@ARTICLE{Semboloni-2011,
       author = {{Semboloni}, Elisabetta and {Hoekstra}, Henk and {Schaye}, Joop and {van Daalen}, Marcel P. and {McCarthy}, Ian G.},
        title = "{Quantifying the effect of baryon physics on weak lensing tomography}",
      journal = {\mnras},
         year = 2011,
        month = nov,
       volume = {417},
       number = {3},
        pages = {2020-2035},
          doi = {10.1111/j.1365-2966.2011.19385.x},
archivePrefix = {arXiv},
       eprint = {1105.1075},
 primaryClass = {astro-ph.CO},
       adsurl = {https://ui.adsabs.harvard.edu/abs/2011MNRAS.417.2020S}
}

@ARTICLE{Schneider-2020,
       author = {{Schneider}, Aurel and {Stoira}, Nicola and {Refregier}, Alexandre and {Weiss}, Andreas J. and {Knabenhans}, Mischa and {Stadel}, Joachim and {Teyssier}, Romain},
        title = "{Baryonic effects for weak lensing. Part I. Power spectrum and covariance matrix}",
      journal = {\jcap},
         year = 2020,
        month = apr,
       volume = {2020},
       number = {04},
        pages = {019},
          doi = {10.1088/1475-7516/2020/04/019},
archivePrefix = {arXiv},
       eprint = {1910.11357},
 primaryClass = {astro-ph.CO},
       adsurl = {https://ui.adsabs.harvard.edu/abs/2020JCAP...04..019S}
}

@ARTICLE{Chisari-2019,
       author = {{Chisari}, Nora Elisa and {Mead}, Alexander J. and {Joudaki}, Shahab and {Ferreira}, Pedro G. and {Schneider}, Aurel and {Mohr}, Joseph and {Tr{\"o}ster}, Tilman and {Alonso}, David and {McCarthy}, Ian G. and {Martin-Alvarez}, Sergio and {Devriendt}, Julien and {Slyz}, Adrianne and {van Daalen}, Marcel P.},
        title = "{Modelling baryonic feedback for survey cosmology}",
      journal = {The Open Journal of Astrophysics},
         year = 2019,
        month = jun,
       volume = {2},
       number = {1},
          eid = {4},
        pages = {4},
          doi = {10.21105/astro.1905.06082},
archivePrefix = {arXiv},
       eprint = {1905.06082},
 primaryClass = {astro-ph.CO},
       adsurl = {https://ui.adsabs.harvard.edu/abs/2019OJAp....2E...4C}
}

@ARTICLE{EuclidHOS-2023,
       author = {{Euclid Collaboration: Ajani}, V. and {Baldi}, M. and {Barthelemy}, A. and {Boyle}, A. and others},
        title = "{Euclid preparation. XXVIII. Forecasts for ten different higher-order weak lensing statistics}",
      journal = {\aap},
         year = 2023,
        month = jul,
       volume = {675},
          eid = {A120},
        pages = {A120},
          doi = {10.1051/0004-6361/202346017},
archivePrefix = {arXiv},
       eprint = {2301.12890},
 primaryClass = {astro-ph.CO},
       adsurl = {https://ui.adsabs.harvard.edu/abs/2023A&A...675A.120E}
}

@ARTICLE{Popesso-2026,
       author = {{Popesso}, P. and {Biviano}, A. and {Marini}, I. and {Dolag}, K. and others},
        title = "{The hot gas mass fraction in halos. From Milky Way-like groups to massive clusters}",
      journal = {\aap},
         year = 2026,
       volume = {707},
          eid = {A362},
        pages = {A362},
          doi = {10.1051/0004-6361/202453256},
archivePrefix = {arXiv},
       eprint = {2411.16555},
 primaryClass = {astro-ph.CO},
       adsurl = {https://ui.adsabs.harvard.edu/abs/2026A&A...707A.362P}
}

@ARTICLE{Qu-2026,
       author = {{Qu}, F.~J. and {Ried Guachalla}, B. and {Schaan}, E. and {Hadzhiyska}, B. and others},
        title = "{Precision Kinematic Sunyaev-Zel'dovich Measurements Across Halo Mass and Redshift with DESI DR2 and ACT DR6: Part I. Luminous Red Galaxies}",
      journal = {arXiv e-prints},
         year = 2026,
          eid = {arXiv:2604.19744},
        pages = {arXiv:2604.19744},
          doi = {10.48550/arXiv.2604.19744},
archivePrefix = {arXiv},
       eprint = {2604.19744},
 primaryClass = {astro-ph.CO},
       adsurl = {https://ui.adsabs.harvard.edu/abs/2026arXiv260419744Q}
}

@ARTICLE{Wright-2025,
       author = {{Wright}, Angus H. and {St{\"o}lzner}, Benjamin and {Asgari}, Marika and others},
        title = "{KiDS-Legacy: Cosmological constraints from cosmic shear with the complete Kilo-Degree Survey}",
      journal = {\aap},
         year = 2025,
        month = nov,
       volume = {703},
          eid = {A158},
        pages = {A158},
          doi = {10.1051/0004-6361/202554908},
archivePrefix = {arXiv},
       eprint = {2503.19441},
 primaryClass = {astro-ph.CO},
       adsurl = {https://ui.adsabs.harvard.edu/abs/2025A&A...703A.158W}
}

@ARTICLE{McQuinn-2016,
       author = {{McQuinn}, Matthew},
        title = "{The Evolution of the Intergalactic Medium}",
      journal = {\araa},
         year = 2016,
        month = sep,
       volume = {54},
        pages = {313-362},
          doi = {10.1146/annurev-astro-082214-122355},
archivePrefix = {arXiv},
       eprint = {1512.00086},
 primaryClass = {astro-ph.CO},
       adsurl = {https://ui.adsabs.harvard.edu/abs/2016ARA&A..54..313M}
}

@ARTICLE{Martizzi-2019,
       author = {{Martizzi}, Davide and {Vogelsberger}, Mark and {Artale}, Maria Celeste and {Haider}, Markus and {Torrey}, Paul and {Marinacci}, Federico and {Nelson}, Dylan and {Pillepich}, Annalisa and {Weinberger}, Rainer and {Hernquist}, Lars and {Naiman}, Jill and {Springel}, Volker},
        title = "{Baryons in the Cosmic Web of IllustrisTNG - I: gas in knots, filaments, sheets, and voids}",
      journal = {\mnras},
         year = 2019,
        month = jul,
       volume = {486},
       number = {3},
        pages = {3766-3787},
          doi = {10.1093/mnras/stz1106},
archivePrefix = {arXiv},
       eprint = {1810.01883},
 primaryClass = {astro-ph.CO},
       adsurl = {https://ui.adsabs.harvard.edu/abs/2019MNRAS.486.3766M}
}

@ARTICLE{Leung-2025,
       author = {{Leung}, Calvin and {Borrow}, Josh and {Masui}, Kiyoshi W. and {Andrew}, Shion and {Chen}, Kai-Feng and {Schaye}, Joop and {Schaller}, Matthieu},
        title = "{Nulling baryonic feedback in weak lensing surveys using cross-correlations with fast radio bursts}",
      journal = {arXiv e-prints},
         year = 2025,
        month = sep,
          eid = {arXiv:2509.19514},
        pages = {arXiv:2509.19514},
          doi = {10.48550/arXiv.2509.19514},
archivePrefix = {arXiv},
       eprint = {2509.19514},
 primaryClass = {astro-ph.CO},
       adsurl = {https://ui.adsabs.harvard.edu/abs/2025arXiv250919514L}
}

@ARTICLE{Hadzhiyska-2025,
       author = {{Hadzhiyska}, Boryana and {Ferraro}, Simone and {Farren}, Gerrit S. and {Sailer}, Noah and {Zhou}, Rongpu},
        title = "{Missing baryons recovered: a measurement of the gas fraction in galaxies and groups with the kinematic Sunyaev-Zel'dovich effect and CMB lensing}",
      journal = {arXiv e-prints},
         year = 2025,
        month = jul,
          eid = {arXiv:2507.14136},
        pages = {arXiv:2507.14136},
          doi = {10.48550/arXiv.2507.14136},
archivePrefix = {arXiv},
       eprint = {2507.14136},
 primaryClass = {astro-ph.CO},
       adsurl = {https://ui.adsabs.harvard.edu/abs/2025arXiv250714136H}
}

@ARTICLE{Hadzhiyska-2026a,
       author = {{Hadzhiyska}, B. and {Ferraro}, S. and {Qu}, F.~J. and {Ried Guachalla}, B. and {Schaan}, E. and others},
        title = "{Precision Kinematic Sunyaev-Zel'dovich Measurements Across Halo Mass and Redshift with DESI DR2 and ACT DR6: Part II. Bright Galaxy Survey and Emission-Line Galaxies}",
      journal = {arXiv e-prints},
         year = 2026,
        month = apr,
          eid = {arXiv:2604.19745},
        pages = {arXiv:2604.19745},
          doi = {10.48550/arXiv.2604.19745},
archivePrefix = {arXiv},
       eprint = {2604.19745},
 primaryClass = {astro-ph.CO},
       adsurl = {https://ui.adsabs.harvard.edu/abs/2026arXiv260419745H}
}

@ARTICLE{Hadzhiyska-2026b,
       author = {{Hadzhiyska}, Boryana},
        title = "{Shear-kSZ: A New Estimator for the Matter-Electron Power Spectrum from kSZ Tomography and Weak Lensing}",
      journal = {arXiv e-prints},
         year = 2026,
        month = jul,
          eid = {arXiv:2607.27149},
        pages = {arXiv:2607.27149},
          doi = {10.48550/arXiv.2607.27149},
archivePrefix = {arXiv},
       eprint = {2607.27149},
 primaryClass = {astro-ph.CO},
       adsurl = {https://ui.adsabs.harvard.edu/abs/2026arXiv260727149H}
}

@ARTICLE{Angulo-2010,
       author = {{Angulo}, R.~E. and {White}, S.~D.~M.},
        title = "{One simulation to fit them all - changing the background parameters of a cosmological N-body simulation}",
      journal = {\mnras},
         year = 2010,
        month = jun,
       volume = {405},
       number = {1},
        pages = {143-154},
          doi = {10.1111/j.1365-2966.2010.16459.x},
archivePrefix = {arXiv},
       eprint = {0912.4277},
 primaryClass = {astro-ph.CO},
       adsurl = {https://ui.adsabs.harvard.edu/abs/2010MNRAS.405..143A}
}

@ARTICLE{Zennaro-2019,
       author = {{Zennaro}, Matteo and {Angulo}, Ra{\'u}l E. and {Aric{\`o}}, Giovanni and {Contreras}, Sergio and {Pellejero-Iba{\~n}ez}, Marcos},
        title = "{How to add massive neutrinos to your {$\Lambda$}CDM simulation - extending cosmology rescaling algorithms}",
      journal = {\mnras},
         year = 2019,
        month = nov,
       volume = {489},
       number = {4},
        pages = {5938-5951},
          doi = {10.1093/mnras/stz2612},
archivePrefix = {arXiv},
       eprint = {1905.08696},
 primaryClass = {astro-ph.CO},
       adsurl = {https://ui.adsabs.harvard.edu/abs/2019MNRAS.489.5938Z}
}
\bibliographystyle{aasjournalv7}



\end{document}